\documentclass[10pt, oneside]{article}   	
\usepackage[margin=1.5in, headheight=22.54448pt]{geometry}
\usepackage{geometry}                		
\usepackage{graphicx}				
\usepackage{amssymb}
\usepackage{bbm}
\usepackage{lscape}
\usepackage{float}
\usepackage{xcolor}	
\usepackage{amsthm} 

\usepackage{mathtools}
\usepackage{amsmath} 
\usepackage[utf8]{inputenc}
\usepackage[english]{babel}
\usepackage{setspace}
\usepackage{threeparttable}
\usepackage[title]{appendix}
\usepackage{booktabs}
\usepackage{bm}
\usepackage{multirow}
\usepackage{hyperref}
\usepackage[longnamesfirst]{natbib}
\usepackage{titling}
\usepackage{comment}
\usepackage{cleveref}
\setcitestyle{authoryear,open={(},close={)}}

\newcommand{\norm}[1]{\left\lVert#1\right\rVert}
    
\allowdisplaybreaks

\newtheorem{theorem}{Theorem}[section]
\newtheorem{proposition}{Proposition}[section]

\newtheorem{assumption}{Assumption}
\newtheorem{corollary}{Corollary}[section]
\newtheorem{lemma}[theorem]{Lemma}

\newtheorem{remark}{Remark}
\newtheorem*{remark*}{Remark}
\Crefname{remark}{Remark}{Remarks}
\crefname{remark}{remark}{Remarks}

\newtheorem{definition}{Definition}

\newtheorem{assumptionalt}{Assumption}[assumption]

\newenvironment{assumptionp}[1]{
  \renewcommand\theassumptionalt{#1}
  \assumptionalt
}{\endassumptionalt}

\makeatletter
\def\@seccntformat#1{\@ifundefined{#1@cntformat}%
   {\csname the#1\endcsname\quad}
   {\csname #1@cntformat\endcsname}
}

\newsavebox\myboxA
\newsavebox\myboxB
\newlength\mylenA

\newcommand*\xoverline[2][0.75]{%
    \sbox{\myboxA}{$\m@th#2$}%
    \setbox\myboxB\null
    \ht\myboxB=\ht\myboxA%
    \dp\myboxB=\dp\myboxA%
    \wd\myboxB=#1\wd\myboxA
    \sbox\myboxB{$\m@th\overline{\copy\myboxB}$}
    \setlength\mylenA{\the\wd\myboxA}
    \addtolength\mylenA{-\the\wd\myboxB}%
    \ifdim\wd\myboxB<\wd\myboxA%
       \rlap{\hskip 0.5\mylenA\usebox\myboxB}{\usebox\myboxA}%
    \else
        \hskip -0.5\mylenA\rlap{\usebox\myboxA}{\hskip 0.5\mylenA\usebox\myboxB}%
    \fi}
\makeatother
\newcommand{\indep}{\rotatebox[origin=c]{90}{$\models$}}
\newcommand{\E}{\mathbb{E}}

\title{Clustering with Potential Multidimensionality: \\Inference and Practice \thanks{This paper supersedes papers previously circulated with the working titles ``Design-Based Multiway Clustering" and ``Asymptotic Properties of M-estimators with Finite Populations under Cluster Sampling and Cluster Assignment".}} 
\author{Ruonan Xu\thanks{ruonan.xu@rutgers.edu}\\
Rutgers University \and 
Luther Yap\thanks{lyap@princeton.edu}\\
Princeton University}

\date{\today}							

\begin{document}

\maketitle

\begin{abstract}
\singlespacing
We show how clustering standard errors in one or more dimensions can be justified in M-estimation when there is sampling or assignment uncertainty. 
Since existing procedures for variance estimation are either conservative or invalid, we propose a variance estimator that refines a conservative procedure and remains valid.
We then interpret environments where clustering is frequently employed in empirical work from our design-based perspective and provide insights on their estimands and inference procedures. 

\bigskip

\noindent \textbf{Keywords}: \textit{Finite population inference; M-estimation; Cluster-robust inference; Two-way clustering; Design-based inference; Potential outcomes; Triple differences}

\bigskip 

\noindent \textbf{JEL classification}: \textit{C21, C23}

\end{abstract}

\section{Introduction}

In our survey of articles published in \emph{American Economic Review} in the years 2021 and 2022, 70\% of 133 articles containing empirical specifications reported some cluster-robust standard errors. Among these papers, 12 reported two-way clustered standard errors. Despite the common use of cluster-robust inference in empirical work, little guidance has been provided on the motivation for clustered standard errors and the level of clustering for linear regressions and nonlinear estimators. 
Even less is known about the formal reasoning underlying two-way clustered standard errors. In this paper, we establish asymptotic properties of M-estimators under finite populations with potentially multiway cluster dependence, allowing for unbalanced and unbounded cluster sizes in the limit. 

We use the finite population framework for the choice of appropriate inference because we can combine sampling-based uncertainty that arises from possibly not observing the entire population with design-based uncertainty caused by the stochastic assignment of treatment or policy variables. Following \cite{abadie2023should}, we distinguish between two situations that justify computing clustered standard errors: \romannumeral 1) cluster sampling induced by random sampling of groups of units, and \romannumeral 2) cluster assignment caused by correlated assignment of ``treatment” within the same group. While \cite{abadie2023should} justify cluster-robust standard errors for the difference-in-means estimator with one-way clustering, we generalize their setup both in considering general M-estimators and in allowing for multiway clustering.

We show that for conducting inference with general M-estimators, one-way clustering is only necessary when there is either cluster sampling or cluster assignment, or both on nested or same dimensions. Multiway clustering can be justified when clustered assignment and clustered sampling occur on different dimensions, or when either sampling or assignment is multiway clustered. The same results are also shown for functions of M-estimators with the estimator of the average partial effect (APE) as a leading example. In the special case of linear regression on a binary treatment variable, one-way clustered standard errors on the assignment dimension is sufficient under homogeneous treatment effects even if the sampling and assignment dimensions are non-nested. 

Until recently, accounting for the large sample behavior of design-based settings with multiway clustering has been a difficult problem. Asymptotic theory for variables that have multi-dimensional dependence has thus far relied on separate exchangeability (e.g., \cite{davezies2018asymptotic}). Separate exchangeability implies that the marginal distributions of clusters are exchangeable \citep{mackinnon2021wild}. However, by construction, separate exchangeability is violated within a design-based framework. Consider a binary assignment variable $X_i$. Since the error term $u_i = X_i u_i(1) + (1 - X_i) u_i(0)$ depends on the nonstochastic potential error $u_i(x)$, even if treatments $X_i$ are identically distributed across clusters, the marginal distribution of $u_i$ differs because $X_i$ is weighted differently. Hence, a limit theory that accommodates heterogeneity of clusters and observations is required. By building on the central limit theorem in \cite{yap2023general}, we derive results on large-sample behavior of standard estimators in this environment.

When estimating the variance-covariance matrix, the usual variance estimators are typically overly conservative for the finite population variance-covariance matrix in one-way clustering. However, we find that the two-way clustered variance estimator as proposed in \cite{cameron2011robust}, henceforth CGM, can be anticonservative.
In response to the anticonservativeness of the CGM estimator in design-based settings, there are two approaches that empirical researchers may take. The first approach is to make an assumption on how the individual treatment effects are correlated within the same cluster: CGM is conservative when the correlation is positive, which is reasonable in most applications. The second approach is to remain agnostic and to use CGM2, a more conservative version of the CGM variance estimator proposed by \cite{davezies2018asymptotic}. Since simulations show CGM2 is often unnecessarily conservative, we propose a simple shrinkage variance estimator relying on adjustments using covariates. The probability limit of our adjusted variance estimator is guaranteed to be no smaller than the finite population variance matrix and provides a smaller upper bound than CGM2. 

We discuss several practical settings involving clustering and justify the validity of cluster-robust inference within a design-based framework. These examples include a standard difference-in-means estimator, fixed effects regressions, a triple difference estimator, and a linear regression on two assignment variables clustered on different dimensions. We also show the performance of our proposed shrinkage variance estimators in simulations and an empirical illustration. In particular, our adjusted standard errors can be substantially smaller than CGM2 while still maintaining correct coverage.

Within the growing literature on design-based inference (e.g., \citet{abadie2020sampling, xu2020potential, athey2022design, abadie2023should, de2023level}), there are at least four nuances unique to our environment that we find from our theory and from analyzing applications of cluster-robust inference. 
First, there are special cases where two-way clustering reduces to one-way clustering in a way that one-way clustering does not reduce to a heteroskedasticity-robust variance-covariance matrix.
Second, CGM can be anticonservative while standard one-way robust variances are conservative.
Third, beyond the special case of \citet{abadie2023should}, estimands from a fixed effects regression cannot be interpreted as the average treatment effect in general.
A broader lesson here is that cluster dependence not only affects variance estimation in inference; it can also affect the interpretation of estimands.
Fourth, with multiple assignment variables on different clustering dimensions, we find that in certain cases it suffices to use one-way cluster variances on the respective dimensions. In contrast, this setting is not allowed or cannot be discussed in one-way clustering.

Our paper contributes to at least three strands of literature that have gained recent attention. First, we contribute to the literature on design-based inference (cited and summarized above). Second, we contribute to the literature on multiway clustering (e.g., \citet{davezies2018asymptotic, mackinnon2019cluster, menzel2021bootstrap, chiang2023using, yap2023general, chiang2024standard}). 
To the best of our knowledge, we are the first to consider any design-based environment with multiway clustering, which is a difficult problem as separate exchangeability used in most multiway clustering limit theorems does not hold in the general design-based setting. 
Third, we contribute to the literature on causal panel data (e.g., \citet{de2020two, callaway2021difference, sun2021estimating, wooldridge2021two, athey2022design, borusyak2024revisiting, dechaisemartin2024difference, gardner2024two,arkhangelsky2024causal}).
By using a design-based setup, we make an assumption on assignment instead of the potential outcome as most of these papers do with parallel trends. Considering multiway clustered assignment is new relative to the design-based setting of \citet{athey2022design}, and our setting provides new insights into the interpretation of the fixed effect estimands in these designs.

\section{Asymptotic Properties of M-estimators}

\subsection{Setup}
\indent 

Consider a sequence of finite populations indexed by population size $M$, where $M$ diverges to infinity in deriving the asymptotic properties. Suppose there are $G$ mutually exclusive clusters in population $M$ defined as either the primary sampling units in the sampling scheme or the partition in the assignment design, where each cluster has $M^G_g$ units, $g=1,2,\dots,G$. 
Further, suppose the population can also be partitioned into $H$ mutually exclusive clusters according to either the sampling scheme or assignment design on dimensions possibly different from that of $G$ clusters. 
Each cluster $H$ contains $M^H_h$ units, $h=1,2,\dots,H$. We use $\mathcal{N}^C_c$ to denote the set of observations in the $c$th cluster on the $C \in \{ G, H, G\cap H \}$ dimension, and $c(i)$ to denote the cluster that $i$ belongs to on the relevant dimension.
If there is only one way of partitioning, $H$ and $G$ clusters coincide with each other. 

For each unit $i$ within cluster $g$ and cluster $h$, we observe $(X_{iM}, z_{iM}, Y_{iM})$.
The vector $X_{iM}$ is the vector of stochastic assignment variables, $z_{iM}$ is a set of non-stochastic attributes, and $Y_{iM}$ is the realized outcome. 
The categorization of assignments and attributes depends on the empirical question. 
Typically, the key variables of interest in an empirical study could be viewed as assignment variables, and the rest covariates as attribute variables. With the potential outcome framework, there exists a mapping, denoted by the potential outcome function $y_{iM}(x)$, from the assignment variables to the potential outcomes. For example, $y_{iM}(x)=x\theta_{01}+z_{iM}\theta_{02}+e_{iM}$ for continuous outcomes, and $y_{iM}(x)=\mathbbm{1}[x\theta_{01}+z_{iM}\theta_{02}+e_{iM}>0]$ for binary outcomes, where $z_{iM}$ and $e_{iM}$ are observed and unobserved attributes respectively.\footnote{We emphasize $x$ as the argument of the potential outcome function because it is the only stochastic variables in the function.} 
The potential outcome function $y_{iM}(\cdot)$ is non-stochastic.\footnote{This implies that the unobserved attributes are non-stochastic.} 
Nevertheless, the realized potential outcome, $Y_{iM}=y_{iM}(X_{iM})$, is random. 
Hence, the finite population setting can be understood as a setting that conditions on the potential outcomes and attributes of the $M$ units in the population.
We use $W_{iM}:=(X_{iM}, Y_{iM})$ to denote the random vector for brevity.

We study solutions to a population minimization problem, where the estimand of interest is a $k\times 1$ vector denoted by $\theta^*_M$: 
\begin{equation}\label{eq:estimand}
\begin{aligned}
\theta^*_M&=\arg\min_\theta \frac{1}{M}\sum_{g=1}^G\sum^H_{h=1}\sum_{i\in \mathcal{N}^{G\cap H}_{(g,h)}} \mathbb{E}\big[q_{iM}(W_{iM}, \theta)\big]\\
&=\arg\min_\theta \frac{1}{M}\sum_{i=1}^M \mathbb{E}\big[q_{iM}(W_{iM}, \theta)\big].
\end{aligned}
\end{equation}
The expectation $\mathbb{E}$ in (\ref{eq:estimand}) is taken over the distribution of $X$ since $X$ is the source of randomness here. 
The function $q_{iM}(\cdot, \cdot)$ is the objective function for a single unit. Examples include (nonlinear) least squares, weighted least squares, maximum likelihood, and instrumental variables estimation in the just identified case (based on the first-order condition of the minimization problem). The subscripts of the objective function imply its dependence on the non-stochastic attribute variables $z_{iM}$, so covariates are allowed in the model. Interpreting this estimand is context-dependent, so we abstract from this discussion in our general framework.

Let $R_{iM}$ denote the binary sampling indicator, which is equal to one if unit $i$ is sampled. Hence, the sample size is $N=\sum^G_{g=1}\sum^{H}_{h=1}\sum_{i\in \mathcal{N}^{G\cap H}_{(g,h)}}R_{iM}=\sum^M_{i=1}R_{iM}$. The sample size is random unless the sample is population. 
The estimator of $\theta^*_M$ is denoted by $\hat{\theta}_N$, which solves the minimization problem in the sample:
\begin{align*}
\hat{\theta}_N&=\arg\min_\theta\frac{1}{N}\sum_{g=1}^G\sum^H_{h=1}\sum_{i\in \mathcal{N}^{G\cap H}_{(g,h)}} R_{iM}q_{iM}(W_{iM}, \theta)\\
&=\arg\displaystyle \min_\theta\frac{1}{N}\sum_{i=1}^M R_{iM} q_{iM}(W_{iM}, \theta).
\tag{\stepcounter{equation}\theequation}
\end{align*}

Our random variables are two-way clustered in that random variables for indices $i$ and $j$ are independent if $g(i)\ne g(j)$ and $h(i) \ne h(j)$, formalized in the following assumptions. 
\begin{assumption}\label{assump1}
The sampling scheme consists of two steps. In the first step, a subset of clusters is drawn according to Bernoulli sampling at the $G$ clustering dimension with probability $\rho_{gM}>0$; within the sampled clusters, a subset of clusters is drawn again according to Bernoulli sampling at the $H$ clustering dimension with probability $\rho_{hM}>0$. Sampling probabilities $\rho_{gM}$ and $\rho_{hM}$ do not differ across clusters. In the second step, units are independently sampled, according to a Bernoulli trial with probability $\rho_{uM}>0$, from the subpopulation consisting of all the sampled intersections of clusters. The sequence of sampling probabilities $\rho_{gM}$, $\rho_{hM}$, and $\rho_{uM}$ satisfies $\rho_{lM }\to \rho_l \in (0,1]$ for $l\in\{g,h,u\}$, as $M\to\infty$.
\end{assumption}

Various sampling schemes are allowed under Assumption \ref{assump1}. When all sampling probabilities are equal to one, we observe the entire population; when $\rho_{gM}=\rho_{hM}=1$ but $\rho_{uM}<1$, we have independent sampling; if the dimensions of $H$ and $G$ coincide or $H$ is nested within $G$ without loss of generality (e.g., zip code areas nested within counties), $\rho_{gM}<1$ or $\rho_{hM}<1$ implies one-way cluster sampling; lastly, if the dimensions of $H$ and $G$ differ, $\rho_{hM}<1$ and $\rho_{gM}<1$ implies two-way cluster sampling. 
For instance, one can first sample according to occupations and industries and then sample individuals from the chosen intersections of occupations and industries. 

\begin{remark} \label{remark:twoway_vs_intersect}
This sampling scheme is different from one-way cluster sampling at the intersection level. To see this, one-way clustering at the intersection level implies that $R_{iM} \indep R_{jM}$ for $g(i)\ne g(j)$ or $h(i) \ne h(j)$, which implies $R_{iM} \indep R_{jM}$ for some $i,j$ with $g(i)= g(j)$ and $h(i) \ne h(j)$. However, under our sampling scheme $Cov(R_{iM}, R_{jM}) = \mathbb{E}[R_{iM} R_{jM}] - \mathbb{E}[R_{iM}] \mathbb{E}[R_{jM}] = \rho_{gM} (1-\rho_{gM}) \rho_{hM}^2 \rho_{uM}^2 \ne 0$.
\end{remark}

\begin{assumption}\label{assump2}
The assignments are independent if units do not share any cluster, i.e., $c(i)\neq c(j), \forall\ c\in\{g,h\}$, but are allowed to be correlated within clusters. 
\end{assumption}

\begin{assumption}\label{assump3}
The vector of assignments is independent of the vector of sampling indicators.
\end{assumption}

Assumption \ref{assump2} allows assignments to be correlated within clusters on the dimensions of $G,H$, or both. Cluster assignment is another source of within-cluster correlation in addition to cluster sampling. The assignment variables $X_{iM}$ are not necessarily identically distributed, which allows the assignments to depend on the fixed attributes $z_{iM}$. 
Independent sampling and assignment processes imply Assumption \ref{assump3}.

Cluster sizes in our theorems are allowed to be unbalanced and unbounded in the limit. Nevertheless, in order to apply asymptotic theory, we have assumptions that restrict cluster heterogeneity and the growth rate of the cluster sizes relative to the population size and variances. 

\begin{assumption}\label{assump4}
$\frac{\sum\limits^G_{g=1}\left(M^G_g\right)^2}{M^2} \rightarrow 0$ and $\frac{\sum\limits^H_{h=1}\left(M^H_h\right)^2}{M^2} \rightarrow 0$ as $M \rightarrow \infty$. 
\end{assumption}

This assumption implies $G, H \to \infty$, and rules out the case where a particular subset of clusters dominates the population. 
In the results that follow, for matrices $A$ and $B$, when we say $A \geq B$, we mean that $A- B$ is positive semidefinite.

\subsection{Asymptotic Distribution}

\subsubsection{One-way Clustering}
To fix ideas, we start with the simpler case of one-way clustering. Namely, there is only one way of partitioning so that $G$ and $H$ clusters coincide. Without loss of generality, let $\rho_{hM}=1$ in this case. Let $m_{iM}(W_{iM},\theta)$ denote the score function of $q_{iM}(W_{iM}, \theta)$. 
The variance matrix of M-estimators is defined as 
\begin{equation}
\begin{aligned}
V_M:=&L_M(\theta^*_M)^{-1} V_{\Delta M} L_M(\theta^*_M)^{-1},
\end{aligned}
\end{equation}
where
\begin{equation}
V_{\Delta M} := \mathbb{V}\left[ \frac{1}{\sqrt{M}}\sum^M_{i=1}\frac{R_{iM}}{\sqrt{\rho_{gM}\rho_{uM}}}m_{iM} (W_{iM},\theta^*_M) \right]
\end{equation}
and
\begin{equation}
L_M(\theta):=\frac{1}{M}\sum^M_{i=1}\mathbb{E}\big[\nabla_\theta m_{iM}(W_{iM},\theta)\big].
\end{equation}

It can be shown that:
\begin{equation}
    V_{\Delta M} = \Delta_{ehw,M}(\theta^*_M)+\rho_{uM} \Delta_{cluster,M}(\theta^*_M) -\rho_{uM}\rho_{gM}\Delta_{E,M}-\rho_{uM} \rho_{gM} \Delta_{EC,M},
\end{equation}
where
\begin{equation}
\Delta_{ehw,M}(\theta)=\frac{1}{M}\sum_{i=1}^M\mathbb{E}\big[m_{iM}(W_{iM}, \theta) m_{iM}(W_{iM}, \theta)'\big]
\end{equation}
and
\begin{equation}
\Delta_{cluster,M}(\theta)=\frac{1}{M}\sum^{G}_{g=1}\sum_{i \in \mathcal{N}^G_g}\sum_{j \in \mathcal{N}^G_g \backslash \{i\}}\mathbb{E}\big[m_{iM}(W_{iM}, \theta)m_{jM}(W_{jM}, \theta)'\big]
\end{equation}
account for heteroskedasticity (Eicker-Huber-White (EHW)) and within-cluster correlation respectively. The terms 
\begin{equation}
\Delta_{E,M}=\frac{1}{M}\sum_{i=1}^M\mathbb{E}\big[m_{iM}(W_{iM}, \theta^*_M)\big]\mathbb{E}\big[m_{iM}(W_{iM}, \theta^*_M)\big]'
\end{equation}
and
\begin{equation}
\Delta_{EC,M}=\frac{1}{M}\sum^{G}_{g=1}\sum_{i \in \mathcal{N}^G_g}\sum_{j \in \mathcal{N}^G_g \backslash \{i\}}\mathbb{E}\big[m_{iM}(W_{iM}, \theta^*_M)\big]\mathbb{E}\big[m_{jM}(W_{jM}, \theta^*_M)\big]'
\end{equation}
are the finite population counterparts of $\Delta_{ehw,M}(\theta^*_M)$ and $\Delta_{cluster,M}(\theta^*_M)$.

The conventional superpopulation variance matrix is denoted by
\begin{equation}
V_{SM}=L_M(\theta^*_M)^{-1}\big(\Delta_{ehw,M}(\theta^*_M)+\rho_{uM}\Delta_{cluster,M}(\theta^*_M)\big)L_M(\theta^*_M)^{-1}. 
\end{equation}
Notice that the middle part of the sandwich form of $V_M$ is different from that of $V_{SM}$ due to two ``extra" (E) terms $\Delta_{E,M}$ and $\Delta_{EC,M}$ scaled by the composite sampling probability. 

The usual cluster-robust variance estimator (CRVE) that uses the estimator from \citet{liang1986longitudinal} for $V_{\Delta M}$ is given by 
\begin{equation}
\hat{V}_{SN}=\hat{L}_N\big(\hat{\theta}_N)^{-1}\big(\hat{\Delta}_{ehw,N}(\hat{\theta}_N)+\hat{\Delta}_{cluster,N}(\hat{\theta}_N)\big)\hat{L}_N(\hat{\theta}_N)^{-1},
\end{equation}
where
\begin{equation}
\hat{L}_N(\theta)=\frac{1}{N}\sum^M_{i=1}R_{iM}\nabla_\theta m_{iM}(W_{iM},\theta),
\end{equation}
\begin{equation}
\hat{\Delta}_{ehw,N}(\theta)=\frac{1}{N}\sum^M_{i=1}R_{iM}\cdot m_{iM}(W_{iM},\theta)m_{iM}(W_{iM},\theta)',
\end{equation}
and
\begin{equation}
\hat{\Delta}_{cluster,N}(\theta)=\frac{1}{N}\sum^{G}_{g=1}\sum_{i \in \mathcal{N}^G_g}\sum_{j \in \mathcal{N}^G_g \backslash \{i\}}R_{iM}R_{jM}\cdot m_{iM}(W_{iM}, \theta)m_{jM}(W_{jM}, \theta)'.
\end{equation}

For one-way clustering, we use a stronger version of Assumptions \ref{assump4} to enhance interpretability. This assumption ensures convergence at rate $N^{-1/2}$ and follows \citet{hansen2019asymptotic}.
The results are extended to an arbitrary convergence rate and two-way clustering in the next subsection.
\begin{assumptionp}{\ref*{assump4}$'$}
$\frac{\sum\limits^G_{g=1}\left(M^G_g\right)^2}{M}\leq C<\infty$ and $\max\limits_{g\leq G}\frac{\left(M^G_g\right)^2}{M}\to 0$, as $M\to \infty$.
\end{assumptionp}
\begin{theorem}\label{thm:oneway}
Under Assumptions \ref{assump1}-\ref{assump3}, Assumption $4^\prime$, and Assumption \ref{assumpa1} in Appendix A, (1) $V_M^{-1/2}\sqrt{N}(\hat{\theta}_N-\theta^*_M)\overset{d}\to \mathcal{N}(\textbf{0}, I_k)$; (2) $V_{SM}^{-1/2}\hat{V}_{SN}V_{SM}^{-1/2}\overset{p}\to I_k$.
\end{theorem}

Theorem \ref{thm:oneway} shows asymptotic normality with the finite population cluster-robust asymptotic variance (CRAV). In the variance-covariance matrices, the term $\Delta_{cluster,M}(\theta^*_M)$ is scaled by the sampling probability $\rho_{uM}$ because of the two-stage sampling scheme. Nevertheless, the usual CRVE, $\hat{V}_{SN}$, converges to $V_{SM}$, in which the estimation of $\rho_{uM}$ has been accounted for.

\begin{remark}\label{coro:oneway1}
Clustering is necessary if and only if there is cluster sampling ($\rho_{gM}<1$) or cluster assignment ($\Delta_{cluster,M}(\theta^*_M)\neq\Delta_{EC,M}$), or both.
\end{remark}

The term related to clustering in the variance formula, $\Delta_{cluster,M}(\theta^*_M) - \rho_{gM} \Delta_{EC,M}$, is zero if we have both independent sampling and independent assignment. Otherwise, these components in the variance must be accounted for.
Hence, Remark \ref{coro:oneway1} suggests that we should adjust standard errors of M-estimators for clustering at the level of cluster sampling or cluster assignment.
It generalizes the results in \cite{abadie2023should}: they prove the case for the difference-in-means estimator, while the remark above holds for all M-estimators with either continuous or discrete assignment variables.

\begin{remark}\label{coro:oneway2}
The superpopulation CRAV of M-estimators is no less than the finite population CRAV, in the matrix sense.
\end{remark}

Since the sum of the two additional terms, $\Delta_{E,M}+\Delta_{EC,M}$, is positive semidefinite, we reach the conclusion in Remark \ref{coro:oneway2}. Remark \ref{coro:oneway2} together with Theorem \ref{thm:oneway}(2) imply that the usual CRVE is often too conservative. There are exceptions where using the usual CRVE for inference is approximately correct, with the leading scenario summarized in the remark below. 
\begin{remark}
If a relatively small number of clusters is sampled from a large population of clusters, i.e., $\rho_{gM}$ is close to zero, or there is at most one unit sampled from each cluster, i.e., $\rho_{uM}$ is close to zero, then it is approximately correct to use the usual CRVE of M-estimators for inference.
\end{remark}

Another special case for the usual CRVE to be correct for inference is when $\Delta_{E,M}+\Delta_{EC,M}=\textbf{0}$, which is true if either $\mathbb{E}\big[m_{iM}(W_{iM},\theta^*_M)\big]=\textbf{0},\ \forall\ i=1,\dots,M_g^G,\ g=1,\dots,G$ or $\sum_{i \in \mathcal{N}^G_g}\mathbb{E}\big[m_{iM}(W_{iM},\theta^*_M)\big]=\textbf{0},\ \forall\ g=1,\dots,G$. The former is true for the variance of the coefficient estimator on the assignment variables under the sufficient conditions provided by \cite{abadie2020sampling}, including constant treatment effects and other linearity conditions. The latter holds if the finite population is composed of repetitions of the smallest cluster. With this kind of data structure, $\theta^*_M$ that solves $\mathbb{E}\Big[\sum^G_{g=1}\sum_{i \in \mathcal{N}^G_g}m_{iM}(W_{iM},\theta^*_M)\Big]= \textbf{0}$ is also the solution to $\mathbb{E}\Big[\sum_{i \in \mathcal{N}^G_g}m_{iM}(W_{iM},\theta^*_M)\Big]=\textbf{0}$ for each cluster $g$. 
However, these kinds of special cases rarely hold in practice. 

Sometimes, we are interested in the functions of M-estimators rather than M-estimators themselves. Let $f_{iM}(W_{iM},\theta^*_M)$ be a $q \times 1$ function of $W_{iM}$ and $\theta^*_M$. Suppose we wish to estimate $\gamma^*_M=\frac{1}{M}\sum^M_{i=1}\mathbb{E}\big[f_{iM}(W_{iM},\theta^*_M)\big]$. As an example, $\gamma^*_M$ could be the APE from nonlinear models, where $f(\cdot, \cdot)$ is some partial derivative for continuous variables or some difference function for discrete variables. 

Let $\hat{\gamma}_N=\frac{1}{N}\sum^M_{i=1}R_{iM}f_{iM}(W_{iM},\hat{\theta}_N)$ be the estimator of $\gamma^*_M$. 
Denote the finite population variance matrix by
\begin{equation}
V_{f,M}=\Delta^f_{ehw,M}+\rho_{uM}\Delta^f_{cluster,M}-\rho_{uM}\rho_{gM}\Delta^f_{E,M}-\rho_{uM}\rho_{gM}\Delta^f_{EC,M}.
\end{equation}
The superpopulation variance matrix is then $V_{f,SM}=\Delta^f_{ehw,M}+\rho_{uM}\Delta^f_{cluster,M}$. 
And the usual CRVE is denoted by $\hat{V}_{f,SN}=\hat{\Delta}^f_{ehw,N}+\hat{\Delta}^f_{cluster,N}$.
The detailed definition of each term can be found in Appendix A. 

\begin{theorem}\label{thm:ate}
Under Assumptions \ref{assump1}-\ref{assump3}, Assumption $4^\prime$, and Assumptions \ref{assumpa1}-\ref{assumpa2} in Appendix A, (1) $V_{f,M}^{-1/2}\sqrt{N}(\hat{\gamma}_N-\gamma^*_M)\overset{d}\to \mathcal{N}(\textbf{0}, I_q)$; (2) $V_{f,SM}^{-1/2}\cdot \hat{V}_{f,SN}\cdot V_{f,SM}^{-1/2}\overset{p}\to I_q$.
\end{theorem} 

Theorem \ref{thm:ate} shows that the conservative property of the usual CRVE of M-estimators also applies to the usual CRVE of any functions of M-estimators. 

\subsubsection{Two-way Clustering }
Now, suppose the $H$ and $G$ clusters are partitioned on different dimensions. 
The variance matrix of M-estimators is defined as 
\begin{equation}
\begin{aligned}
V_{TWM}:=&L_M(\theta^*_M)^{-1} V_{\Delta TWM} L_M(\theta^*_M)^{-1},
\end{aligned}
\end{equation}
where
\begin{equation}
V_{\Delta TWM} := \mathbb{V}\left[ \frac{1}{\sqrt{M}}\sum^M_{i=1}\frac{R_{iM}}{\sqrt{\rho_{gM}\rho_{hM}\rho_{uM}}}m_{iM} (W_{iM},\theta^*_M) \right].
\end{equation}

By grouping the cross products of score functions into different cases: individual units ($\Delta_{ehw,M}(\theta^*_M)$), units belonging to the intersection of $G$ and $H$ ($\Delta_{(G\cap H),M}(\theta^*_M)$), units belonging to $G$ but in different $H$'s ($\Delta_{G,M}(\theta^*_M)$), units belonging to $H$ but in different $G$'s ($\Delta_{H,M}(\theta^*_M)$), it can be shown that:
\begin{equation}
\begin{aligned}
V_{\Delta TWM} =& \Delta_{ehw,M}(\theta^*_M)+\rho_{uM} \Delta_{(G\cap H),M}(\theta^*_M)\\
&+\rho_{uM}\rho_{hM}\Delta_{G,M}(\theta^*_M)+\rho_{uM}\rho_{gM}\Delta_{H,M}(\theta^*_M)\\
&-\rho_{uM}\rho_{gM}\rho_{hM}\Delta_{E,M}-\rho_{uM} \rho_{gM}\rho_{hM} \Delta_{E(G\cap H),M}\\
&-\rho_{uM}\rho_{gM}\rho_{hM}\Delta_{EG,M}-\rho_{uM}\rho_{gM}\rho_{hM}\Delta_{EH,M}
\end{aligned}
\end{equation}
where
\begin{equation}
\Delta_{(G\cap H),M}(\theta)=\frac{1}{M}\sum^{G}_{g=1}\sum^H_{h=1}\sum_{i\in \mathcal{N}^{G\cap H}_{(g,h)}}\sum_{j\in \mathcal{N}^{G\cap H}_{(g,h)} \backslash \{ i\}}\mathbb{E}\big[m_{iM}(W_{iM}, \theta)m_{jM}(W_{jM}, \theta)'\big],
\end{equation}
\begin{equation}
\Delta_{E(G\cap H),M}=\frac{1}{M}\sum^{G}_{g=1}\sum^H_{h=1}\sum_{i\in \mathcal{N}^{G\cap H}_{(g,h)}}\sum_{j\in \mathcal{N}^{G\cap H}_{(g,h)} \backslash \{ i\}}\mathbb{E}\big[m_{iM}(W_{iM}, \theta^*_M)\big]\mathbb{E}\big[m_{jM}(W_{jM}, \theta^*_M)\big]',
\end{equation}
\begin{equation}
\Delta_{G,M}(\theta)=\frac{1}{M}\sum^{G}_{g=1}\sum^H_{h=1}\sum^H_{h'\neq h}\sum_{i\in \mathcal{N}^{G\cap H}_{(g,h)}}\sum_{j\in \mathcal{N}^{G \cap H}_{(g,h')}}\mathbb{E}\big[m_{iM}(W_{iM}, \theta)m_{jM}(W_{jM}, \theta)'\big],
\end{equation}
\begin{equation}
\Delta_{EG,M}=\frac{1}{M}\sum^{G}_{g=1}\sum^H_{h=1}\sum^H_{h'\neq h}\sum_{i\in \mathcal{N}^{G\cap H}_{(g,h)}}\sum_{j\in \mathcal{N}^{G \cap H}_{(g,h')}}\mathbb{E}\big[m_{iM}(W_{iM}, \theta^*_M)\big]\mathbb{E}\big[m_{jM}(W_{jM}, \theta^*_M)\big]',
\end{equation}
\begin{equation}
\Delta_{H,M}(\theta)=\frac{1}{M}\sum^{H}_{h=1}\sum^G_{g=1}\sum^G_{g'\neq g}\sum_{i\in \mathcal{N}^{G\cap H}_{(g,h)}}\sum_{j\in \mathcal{N}^{G\cap H}_{(g',h)}}\mathbb{E}\big[m_{iM}(W_{iM}, \theta)m_{jM}(W_{jM}, \theta)'\big],
\end{equation}
and
\begin{equation}
\Delta_{EH,M}=\frac{1}{M}\sum^{H}_{h=1}\sum^G_{g=1}\sum^G_{g'\neq g}\sum_{i\in \mathcal{N}^{G\cap H}_{(g,h)}}\sum_{j\in \mathcal{N}^{G\cap H}_{(g',h)}}\mathbb{E}\big[m_{iM}(W_{iM}, \theta^*_M)\big]\mathbb{E}\big[m_{jM}(W_{jM}, \theta^*_M)\big]'.
\end{equation}
As before, $\Delta_{E(G\cap H),M}$, $\Delta_{EG,M}$, and $\Delta_{EH,M}$ are the finite population counterparts of $\Delta_{(G\cap H),M}(\theta^*_M)$, $\Delta_{G,M}(\theta^*_M)$, and $\Delta_{H,M}(\theta^*_M)$, respectively.

\begin{assumption} \label{asmp:clt_reg}
Let $\lambda_{min}(\cdot)$ denote the smallest eigenvalue. With $\lambda_M := M \lambda_{min}(V_{\Delta TWM})$, and for some $C < \infty$, we have $\frac{1}{\lambda_M} \max_g (M^G_g)^2 \rightarrow 0$, $\frac{1}{\lambda_M} \max_h (M^H_h)^2 \rightarrow 0$, $\frac{1}{\lambda_M} \sum_g (M^G_g)^2 \leq C$, and $\frac{1}{\lambda_M} \sum_h (M^H_h)^2 \leq C$.
\end{assumption}

Assumption \ref{asmp:clt_reg} ensures that the overall variance is not driven by a few large clusters.
A stronger way of stating Assumptions \ref{assump4} and \ref{asmp:clt_reg} is that $\frac{1}{M} \sum\limits^G_{g=1}\left(M^G_g\right)^2\leq C<\infty$ and $\max\limits_{g\leq G}\frac{\left(M^G_g\right)^2}{M}\to 0$, as $M\to \infty$ with an analogous condition in the $H$ dimension where $\lambda_M \geq \underline{c} M$ for some $\underline{c} >0$. 
This assumption is more similar to the setting of \citet{hansen2019asymptotic}, but rules out two-way balanced clusters where there is one unit in every intersection: if there are $G$ clusters on both the $G$ and $H$ dimensions, then $M=G^2$ so $\frac{1}{M} \sum\limits^G_{g=1}\left(M^G_g\right)^2= G^3/ G^2 = G \rightarrow \infty$. 
Assumption \ref{assump4} as stated makes no such restriction as $\frac{1}{M^2} \sum\limits^G_{g=1}\left(M^G_g\right)^2 = G^3/ G^4 =1/G \rightarrow 0$.
With more flexible cluster sizes, the convergence rate depends on the variance of the sum, which motivates Assumption \ref{asmp:clt_reg}.
The stronger version of the assumption only allows a convergence rate of $M^{-1/2}$, which is not necessarily true in the weaker version. For instance, the weaker version can allow a slower convergence rate of $(\sum_{g=1}^G (M^G_g)^2/M^2)^{-1/2} = G^{-1/2}$ instead of $M^{-1/2}=G^{-1}$.
Since we can allow for different convergence rates, we use the scale $\lambda_M$ to restrict cluster heterogeneity in Assumption \ref{asmp:clt_reg} to derive the asymptotic distribution.

\begin{theorem}\label{thm:twoway}
Under Assumptions \ref{assump1}-\ref{asmp:clt_reg} and Assumption \ref{assumpa3} in Appendix A, $V_{TWM}^{-1/2}\sqrt{N}(\hat{\theta}_N-\theta^*_M)\overset{d}\to \mathcal{N}(\textbf{0}, I_k)$.
\end{theorem}

The proof of this theorem is largely analogous to Theorem \ref{thm:oneway}, just that we apply the central limit theorem (CLT) from \citet{yap2023general} instead of \citet{hansen2019asymptotic}. 
\citet{chiang2023using} (Table 1) pointed out that the two-way cluster-robust standard errors are usually valid. A notable exception is when the additive components are degenerate, such that the random variable can be written as $D_{it}=\alpha_i \gamma_t$, where $\alpha_i,\gamma_t$ are cluster-specific random variables on the respective dimensions. As noted in Remark 1 of \citet{yap2023general}, this data generating process (earlier pointed out by \citet{menzel2021bootstrap}) is ruled out by our summability condition in Assumption \ref{asmp:clt_reg}.

\begin{remark}\label{coro:twoway}
$V_{TWM}$ reduces to $V_M$ (i.e., the one-way CRAV) if (\romannumeral 1) $\rho_{hM}=1$ and $\Delta_{H,M}(\theta^*_M)=\Delta_{EH,M}$ or (\romannumeral 2) $\rho_{gM}=1$ and $\Delta_{G,M}(\theta^*_M)=\Delta_{EG,M}$.
\end{remark}

Remark \ref{coro:twoway} coupled with Remark \ref{coro:oneway1} imply that two-way clustering is only justified if there is (\romannumeral 1) two-way clustered sampling (i.e., $\rho_{gM}<1$ and $\rho_{hM}<1$); (\romannumeral 2) two-way clustered assignments (i.e., $\Delta_{G,M}(\theta^*_M)\neq \Delta_{EG,M}$ and $\Delta_{H,M}(\theta^*_M)\neq \Delta_{EH,M}$); or (\romannumeral 3) clustered sampling and clustered assignments on different dimensions. For the third case, there could be many combinations of sampling schemes and assignment designs, possibly combining two-way sampling and two-way assignments at the same time.

In a special case, $G$ and $H$ clusters could be partitioned at different but nested levels. Without loss of generality, suppose $H$ clusters are nested in $G$ clusters. The variance-covariance matrix can be simplified to be 
\begin{equation}
\begin{aligned}
V_{TWM}=&L_M(\theta^*_M)^{-1}\big(\Delta_{ehw,M}(\theta^*_M)+\rho_{uM} \Delta_{(G\cap H),M}(\theta^*_M)+\rho_{1M}\Delta_{G,M}(\theta^*_M)\\
&-\rho_{2M}\Delta_{E,M}-\rho_{2M} \Delta_{E(G\cap H),M}-\rho_{2M}\Delta_{EG,M}\big)L_M(\theta^*_M)^{-1},
\end{aligned}
\end{equation}
where (\romannumeral 1) $\rho_{1M}=\rho_{uM}\rho_{hM}$ and $\rho_{2M}=\rho_{uM}\rho_{gM}\rho_{hM}$ for nested sampling; (\romannumeral 2) $\rho_{1M}=\rho_{2M}=\rho_{uM}$ for nested assignment; (\romannumeral 3) $\rho_{1M}=\rho_{uM}$ and $\rho_{2M}=\rho_{uM}\rho_{gM}$ for sampling at the $G$ level and assignment at the $H$ level; (\romannumeral 4) $\rho_{1M}=\rho_{2M}=\rho_{uM}\rho_{hM}$ for assignment at the $G$ level and sampling at the $H$ level. As a consequence, one-way clustering at the higher level $G$ is sufficient. 

\begin{remark} \label{remark:diff_means_clus_unnecessary}
In the special case of the difference-in-means estimator, which is equivalent to linear regression on a constant and a binary treatment variable, the score has mean zero for all observations when there are constant treatment effects. 
If there is clustered sampling on dimension G and clustered assignment on dimension H, where G and H are non-nested, then it suffices to cluster on the assignment dimension H. To see this, since $\E[m_{iM}(W_{iM},\theta)] =0$ for all $i$, the difference between $V_{\Delta TWM}$ and one-way cluster-robust variance on H is $\frac{1}{M}\sum_i \sum_{j \in \mathcal{N}^G_{g(i)} \backslash \mathcal{N}^H_{h(i)}} \E [m_{iM} (W_{iM}, \theta) m_{jM} (W_{jM},\theta)]$. Since $\E [m_{iM} (W_{iM}, \theta) m_{jM} (W_{jM},\theta)] = \E [m_{iM} (W_{iM}, \theta)] \E[m_{jM} (W_{jM},\theta)]$ when $i,j$ do not share a H cluster, the difference is zero. This is also the case where the usual one-way CRVE is no longer conservative. 
\end{remark}

While the usual variance estimators are generally conservative for the finite population variance-covariance matrix in one-way clustering, this is not necessarily true with multiway clustering. 
We denote the usual two-way cluster-robust variance estimator initially proposed in \cite{cameron2011robust} by 
\begin{equation}
\hat{V}_{CGM}=\hat{L}_N\big(\hat{\theta}_N)^{-1}\big(\hat{\Delta}_{ehw,N}(\hat{\theta}_N)+\hat{\Delta}_{G,N}(\hat{\theta}_N)+\hat{\Delta}_{H,N}(\hat{\theta}_N)-\hat{\Delta}_{{G\cap H},N}(\hat{\theta}_N)\big)\hat{L}_N(\hat{\theta}_N)^{-1},
\end{equation}
where
\begin{equation}
\hat{\Delta}_{C,N}(\theta)=\frac{1}{N}\sum_{c}\sum_{i \in \mathcal{N}^C_c}\sum_{j \in \mathcal{N}^C_c \backslash \{ i \}}R_{iM}R_{jM}\cdot m_{iM}(W_{iM}, \theta)m_{jM}(W_{jM}, \theta)',
\end{equation}
for $C\in \{G,H,G\cap H\}$. 
Define the superpopulation two-way CRAV to be 
\begin{equation}
\begin{aligned}
V_{TWSM}=&L_M(\theta^*_M)^{-1}\big(\Delta_{ehw,M}(\theta^*_M)+\rho_{uM} \Delta_{(G\cap H),M}(\theta^*_M)\\
&+\rho_{uM}\rho_{hM}\Delta_{G,M}(\theta^*_M)+\rho_{uM}\rho_{gM}\Delta_{H,M}(\theta^*_M)\big)L_M(\theta^*_M)^{-1}.
\end{aligned}
\end{equation}

For variance estimators to converge to the variance-covariance matrices in the general environment, we impose an additional assumption. 
\begin{assumption} \label{asmp:cross_var}
For $C, C^{\prime} \in \{ G,H \}$, \\ let
    $\lambda^C_M := \lambda_{min} \left(  \sum_{i=1}^{M} \sum_{j \in \mathcal{N}^{C}_{c(i)}} \mathbb{E}[R_{iM} R_{jM} m_{iM}(W_{iM}, \theta^*_M)m_{jM}(W_{jM}, \theta^*_M)'] \right)$. Then, $(\lambda^C_M)^{-1} \max_{c^\prime} (M^{C^\prime}_{c^\prime})^2 = o(1)$ and $(\lambda^C_M)^{-1} \sum_{c^\prime} (M^{C^\prime}_{c^\prime})^2 = O(1)$.
\end{assumption}
The condition in Assumption \ref{asmp:cross_var} is required in the following propositions so that the asymptotic error incurred by using the matrix estimator $\hat{V}$ relative to the true matrix $V$ converges to zero. 
The difference between Assumption \ref{asmp:cross_var} and the existing assumptions is that the previous assumptions defined $\lambda_M$ as the variance of the sum, which includes all two-way clustered terms, but here, $\lambda^C_M$ only includes terms from one of the two dimensions. 
Since the strategy for showing such convergence is similar to \citet{yap2023general}, an analogous summability condition and a condition on the largest cluster having a negligible contribution to the variance are required.
Since Assumption \ref{asmp:cross_var} only accounts for one-way clustering in the denominator, Assumption \ref{asmp:cross_var} is stronger than Assumption \ref{asmp:clt_reg}.\footnote{Assumption \ref{asmp:cross_var} in its present form requires that the cluster correlation on each clustering dimension be of comparable scale. However, this can be moderated in situations where one clustering dimension dominates the other dimension in the estimation of variance matrix.}

\begin{proposition} \label{prop:cgm}
Under Assumptions  \ref{assump1}-\ref{assump4}, Assumption \ref{asmp:cross_var}, and Assumption \ref{assumpa3} in Appendix A, $V_{TWSM}^{-1/2}\hat{V}_{CGM}V_{TWSM}^{-1/2}\overset{p}\to I_k$. However, $V_{TWSM}$ could be smaller than $V_{TWM}$ in the matrix sense. 
\end{proposition} 

The anticonservativeness results from the subtraction of the correlation terms within intersection clusters $\hat{\Delta}_{{G\cap H},N}(\hat{\theta}_N)$ so that the difference in the meat of the variance sandwich is $\rho_{uM} \rho_{gM} \rho_{hM} (\Delta_{E,M} + \Delta_{E(G\cap H),M} + \Delta_{EG,M} + \Delta_{EH,M})$, which can be positive or negative in general. 
In Example B.1 in Appendix B, we give an example where $\hat{V}_{CGM}$ is anticonservative. Nevertheless, if all within-cluster correlation of $E[m_{iM}(W_{iM},\theta^*_M)]$ is positive, then $\hat{V}_{CGM}$ is still a conservative variance estimator. 
\cite{davezies2018asymptotic} propose an alternative variance estimator that does not adjust for double counting the intersection clusters. 
Let 
\begin{equation}
\hat{V}_{CGM2}=\hat{L}_N\big(\hat{\theta}_N)^{-1}\big(2\hat{\Delta}_{ehw,N}(\hat{\theta}_N)+\hat{\Delta}_{G,N}(\hat{\theta}_N)+\hat{\Delta}_{H,N}(\hat{\theta}_N)\big)\hat{L}_N(\hat{\theta}_N)^{-1}
\end{equation}
and 
\begin{equation}
\begin{aligned}
V_{TWSM2}=&L_M(\theta^*_M)^{-1}\big(2\Delta_{ehw,M}(\theta^*_M)+2\rho_{uM} \Delta_{(G\cap H),M}(\theta^*_M)\\
&+\rho_{uM}\rho_{hM}\Delta_{G,M}(\theta^*_M)+\rho_{uM}\rho_{gM}\Delta_{H,M}(\theta^*_M)\big)L_M(\theta^*_M)^{-1}.
\end{aligned}
\end{equation}

\begin{proposition} \label{prop:cgm2}
Under Assumptions \ref{assump1}-\ref{assump4}, Assumption \ref{asmp:cross_var}, and Assumption \ref{assumpa3} in Appendix A, $V_{TWSM2}^{-1/2}\hat{V}_{CGM2}V_{TWSM2}^{-1/2}\overset{p}\to I_k$. $V_{TWSM2}$ is guaranteed to be no smaller than $V_{TWM}$ in the matrix sense. 
\end{proposition} 

Hence, in contrast to CGM, CGM2 is asymptotically conservative. 

\section{Proposed Variance Estimation}
Even though we restore conservativeness of the usual variance estimator by using $\hat{V}_{CGM2}$, it can be too conservative. The terms in the usual CRAV can be estimated in the standard way. Taking one-way clustering as an example, it is more challenging to estimate the two extra terms, $\Delta_{E,M}$ and $\Delta_{EC,M}$, because  $\mathbb{E}\big[m_{iM}(W_{iM},\theta^*_M)\big]$ is generally non-identifiable due to the missing data problem of the potential outcome framework. For instance, with a binary assignment variable, $\mathbb{E}\big[m_{iM}(W_{iM},\theta^*_M)\big]=P(X_{iM}=1)\cdot m_{iM}\big((1,y_{iM}(1)),\theta^*_M\big)+P(X_{iM}=0)\cdot m_{iM}\big((0,y_{iM}(0)),\theta^*_M\big)$, and we do not observe both $y_{iM}(0)$ and $y_{iM}(1)$ at the same time. 
This observation motivates a simple method to estimate a bound on these extra terms such that the corrected variance estimators are still conservative, but are smaller than the one-way CRVE or CGM2. 

The variance estimator depends on the sampling and assignment schemes. If the cluster variance matrix is purely induced by sampling, then $\Delta_{EC,M}=\Delta_{cluster,M}$, which can be consistently estimated, as $\hat{\Delta}_{cluster,N}$ consistently estimates $\rho_{uM} \Delta_{cluster,M}$.
Then, it remains to identify a lower bound of $\Delta_{E,M}$. 
On the other hand, if there is cluster assignment, we have to identify a lower bound of $\Delta_{E,M}+\Delta_{EC,M}$ jointly. 
Similarly, if there is cluster sampling on the $G$ dimension but cluster assignment on the $H$ dimension, we only need to adjust for the $H$ dimension. If there are multiway cluster assignments, we need to adjust for both dimensions. 

We first discuss estimating the $\Delta_{E,M}$ component. We can remove part of $\Delta_{E,M}$ using the regression-based approach below, by eliminating the variation that is linearly predictable from the covariates $z_{iM}$.
Consider the estimator, 
\begin{equation}
\hat{\Delta}^Z_N=\frac{1}{N}\sum\limits^M_{i=1}R_{iM}\hat{K}_N'z_{iM}'z_{iM}\hat{K}_N, 
\end{equation}
where $\hat{K}_N=\Big(\sum\limits^M_{i=1}R_{iM}z_{iM}'z_{iM}\Big)^{-1}\bigg[\sum\limits^M_{i=1}R_{iM}z_{iM}'m_{iM}(W_{iM},\hat{\theta}_N)'\bigg]$. With clustered data, we can include cluster dummies as regressors in the linear projection of $m_{iM}(W_{iM},\hat{\theta}_N)$ onto the fixed attributes. 

\begin{theorem} \label{thm:unit_shrink}
In addition to Assumptions \ref{assump1}-\ref{assump4} and Assumption \ref{assumpa3} in Appendix A, assume that (\romannumeral 1) $\frac{1}{M}\sum^M_{i=1}z_{iM}'z_{iM}$ is nonsingular; (\romannumeral 2) $\sup\limits_{i,M}\left\|z_{iM}\right\|<\infty$. Then $\textbf{0}\leq\Delta^Z_M\leq\Delta_{E,M}$, where $\left\|\hat{\Delta}^Z_N-\Delta^Z_M\right\|\overset{p}\to 0$ (all inequalities are in the matrix sense).
\end{theorem}

Under one-way clustered sampling but independent assignment, the estimator for an upper bound of the finite population CRAV is 
\begin{equation}
\hat{\Delta}_{ehw,N}(\hat{\theta}_N)+\bigg(1-\frac{G_N}{G}\bigg)\hat{\Delta}_{cluster,N}(\hat{\theta}_N)-\frac{N}{M}\hat{\Delta}^Z_N,
\end{equation}
where $G_N$ is the number of clusters in the sample. The composite sampling probability $\rho_{uM}\rho_{gM}$ can be estimated by $N/M$, where the population size $M$ is assumed to be known. If the entire population is observed, $\rho_{uM}\rho_{gM}$ is simply one. We ignore the Hessian matrix as it does not affect the discussion here. This estimator is asymptotically conservative because $\Delta^Z_M \leq \Delta_{E,M}$.

Next, we turn to $\Delta_{E,M}+\Delta_{EC,M}$. We could sum $m_{iM}(W_{iM},\hat{\theta}_N)$ within each cluster, and linearly project $\sum_{i \in \mathcal{N}^G_g}R_{iM} m_{iM}(W_{iM},\hat{\theta}_N)$ onto the fixed attributes. The number of observations in the linear projection is the number of clusters in the sample. To reduce the dimensionality of the regressors, the fixed attributes can also be summed within clusters as one way of aggregation. As a result, $\sum_{i \in \mathcal{N}^G_g}\mathbb{E}\big[m_{iM}(W_{iM},\theta^*_M)\big]$ can be partially estimated by its predicted value from the linear projection.
Let 
\begin{equation}
\tilde{z}_{gM}=\sum_{i \in \mathcal{N}^G_g}z_{iM},
\end{equation} 
\begin{equation}
\hat{\tilde{z}}_{gM}=\sum_{i \in \mathcal{N}^G_g} R_{iM} z_{iM},
\end{equation} 
\begin{equation}
\tilde{m}_{gM}(\theta)=\sum_{i \in \mathcal{N}^G_g} m_{iM}(W_{iM},\theta),
\end{equation}
\begin{equation}
\hat{\tilde{m}}_{gM}(\theta)=\sum_{i \in \mathcal{N}^G_g} R_{iM} m_{iM}(W_{iM},\theta),
\end{equation}
and 
\begin{equation}
\hat{P}_N=\bigg(\sum^G_{g=1}\hat{\tilde{z}}_{gM}'\hat{\tilde{z}}_{gM}\bigg)^{-1}\bigg(\sum^G_{g=1}\hat{\tilde{z}}_{gM}'\hat{\tilde{m}}_{gM}(\hat{\theta}_N)'\bigg).
\end{equation}
Estimate $\Delta_{E,M}+\Delta_{EC,M}$ with 
\begin{equation}\label{eqn:shrink}
\hat{\Delta}^Z_{CE,N}=\frac{1}{N }\sum^G_{g=1}\hat{P}_N' \hat{\tilde{z}}_{gM}'\hat{\tilde{z}}_{gM}\hat{P}_N.
\end{equation}

\begin{theorem}\label{thm:shrink_oneway}
In addition to Assumptions \ref{assump1}-\ref{assump3}, Assumption $4^\prime$, and Assumption \ref{assumpa1} in Appendix A, suppose that 
 (\romannumeral 1) $\sum\limits^G_{g=1}\tilde{z}_{gM}'\tilde{z}_{gM}$ is nonsingular; (\romannumeral 2) $\sup\limits_{i,M}\left\|z_{iM}\right\|<\infty$; (\romannumeral 3) $\rho_{uM}=1$. Then $0\leq\Delta^Z_{CE,M}\leq\big(\Delta_{E,M}+\Delta_{EC,M}\big)$, where $\left\|\hat{\Delta}^Z_{CE,N}-\Delta^Z_{CE,M}\right\| \overset{p}\to 0$ (all inequalities are in the matrix sense).
\end{theorem}

Theorem \ref{thm:shrink_oneway} proposes an easy way to partially remove $\Delta_{E,M}+\Delta_{EC,M}$ all at once with one-way clustering, and $\hat{\Delta}^Z_{CE,N}$ is positive semidefinite. In this case, one-way cluster sampling is allowed as long as there is no within-cluster sampling. With large samples, even though the limit of the adjusted finite population CRVE is still conservative (as $\Delta^Z_{CE,M}\leq\big(\Delta_{E,M}+\Delta_{EC,M}\big)$), it is still less conservative than the limit of the usual CRVE.

\begin{table}
    \centering
    \begin{tabular}{c|ccc}
    Case & Cluster Sampling & Cluster Assignment & Variance Estimator \\
    \hline
    1    & $\checkmark$ & $\times$ & $\hat{\Delta}_{ehw,N} + (1-G_N/G)\cdot \hat{\Delta}_{cluster,N} - N/M \cdot \hat{\Delta}^Z_N $ \\
    2    & $\times$ & $\checkmark$ & $\hat{\Delta}_{ehw,N} + \hat{\Delta}_{cluster,N} - N/M \cdot \hat{\Delta}^Z_{CE,N} $ \\
    3    & $\checkmark$ & $\checkmark$ & $\hat{\Delta}_{ehw,N} +  \hat{\Delta}_{cluster,N} - N/M \cdot \hat{\Delta}^Z_{CE,N} $ \\
    4    & $\times$ & $\times$ & $\hat{\Delta}_{ehw,N}  - N/M \cdot \hat{\Delta}^Z_N$ \\
    \end{tabular}
    \caption{Variance Estimators for One-Way Clustering}
\end{table}

We list all possible cases of sampling and assignment and their corresponding adjusted variance estimators in Table 1 for one-way clustering. Details of the derivation are relegated to Appendix B. Case 4 reduces exactly to the approach in \citet{abadie2020sampling} for linear regression. While they take the square of the difference between the score and the predicted score as their estimator, their approach is numerically equivalent to taking the difference of the second moments as we propose.\footnote{Our approach is fundamentally different from the proposal in \citet{abadie2023should}. They split the data into subsamples and directly estimate the finite population variance, while our approach here shrinks the variance using information from covariates. All cluster sizes need to diverge in their approach, which is not required here. We also allow for constant assignment within clusters, which is ruled out by \citet{abadie2023should}.}

A similar shrinkage procedure can be applied to two-way clustering with CGM2 since the additively separable one-way cluster objects can be shown to converge to their limit even with multiway dependence. The variance estimator after adjustment is still conservative for the finite population two-way CRAV. To be precise, let:
\begin{equation}
\hat{P}_{G,N}=\bigg(\sum^G_{g=1}\tilde{z}_{gM}'\tilde{z}_{gM}\bigg)^{-1}\bigg(\sum^G_{g=1}\tilde{z}_{gM}'\tilde{m}_{gM}(\hat{\theta}_N)'\bigg),
\end{equation}
\begin{equation}
\hat{P}_{H,N}=\bigg(\sum^H_{h=1}\tilde{z}_{hM}'\tilde{z}_{hM}\bigg)^{-1}\bigg(\sum^H_{h=1}\tilde{z}_{hM}'\tilde{m}_{hM}(\hat{\theta}_N)'\bigg),
\end{equation}
\begin{equation}
\hat{\Delta}^Z_{GE,N}=\frac{1}{M}\sum^G_{g=1}\hat{P}_{G,N}' \tilde{z}_{gM}'\tilde{z}_{gM}\hat{P}_{G,N},
\end{equation}
and
\begin{equation}
\hat{\Delta}^Z_{HE,N}=\frac{1}{M} \sum^H_{h=1}\hat{P}_{H,N}' \tilde{z}_{hM}'\tilde{z}_{hM}\hat{P}_{H,N}.
\end{equation}

\begin{theorem}\label{thm:shrink}
In addition to  Assumptions \ref{assump1}-\ref{assump4}, Assumption \ref{asmp:cross_var}, and Assumption \ref{assumpa3} in Appendix A, suppose that (\romannumeral 1) $\sum\limits^G_{g=1}\tilde{z}_{gM}'\tilde{z}_{gM}$ is nonsingular; (\romannumeral 2) $\sup\limits_{i,M}\left\|z_{iM}\right\|<\infty$; (\romannumeral 3) $\rho_{uM}= \rho_{gM} = \rho_{hM} = 1$; (\romannumeral 4) the variance order condition (\ref{eqn:zm_order}) in Appendix C holds. Then, for $C \in \{ G,H \}$, $0\leq\Delta^Z_{CE,M}\leq\big(\Delta_{E,M}+\Delta_{EC,M}+\Delta_{E(G\cap H),M}\big)$, where $ \left\| (\Delta^Z_{CE,M})^{-1} \left(\hat{\Delta}^Z_{CE,N}-\Delta^Z_{CE,M} \right)\right\| \overset{p}\to 0$ (all inequalities are in the matrix sense). 
Further, either $$ \left\| (\rho_{uM} \rho_{hM} \Delta_{G,M} + \rho_{uM} \Delta_{G \cap H,M} + \Delta_{ehw,M})^{-1} \left(\hat{\Delta}_{G \cap H,N} (\hat{\theta}) + \hat{\Delta}_{ehw,N} (\hat{\theta}) \right)\right\| \overset{p}\to 0 \text{ or }$$ 
$$ \left\| (\rho_{uM} \Delta_{G \cap H,M}  + \Delta_{ehw,M})^{-1} \left(\hat{\Delta}_{G \cap H,N} + \hat{\Delta}_{ehw,N} (\theta) - \rho_{uM}\Delta_{G \cap H,M} (\hat{\theta}) - \Delta_{ehw,M}(\hat{\theta}) \right)\right\| \overset{p}\to 0.$$
\end{theorem}

When adjusting the variance matrix estimator with two-way clustering, we use the entire population. Theorem \ref{thm:shrink} also provides formal guarantees that the cluster estimators for the intersection of $G$ and $H$ either converge to their estimand or are negligible based on the scale of the cluster correlation on different clustering dimensions.

\begin{table}
    \centering
    \begin{tabular}{c|cc}
    Case & Environment & Variance Estimator \\
    \hline
    \multirow{3}{*}{1}    & \multirow{3}{*}{Both Sampling Only} & $\hat{\Delta}_{ehw,N}+(1-G_N/G)\cdot (\hat{\Delta}_{G,N}-\hat{\Delta}_{G\cap H,N})$\\
    & & $+(1-H_N/H)\cdot (\hat{\Delta}_{H,N}-\hat{\Delta}_{G\cap H,N})$\\
    & & $+(1-G_N/G\cdot H_N/H)\cdot \hat{\Delta}_{G\cap H,N} -N/M \cdot \hat{\Delta}^Z_N $ \\
    2    & Both Assignment  & $2\hat{\Delta}_{ehw,N}+\hat{\Delta}_{G,N}+\hat{\Delta}_{H,N}- N/M \cdot \left( \hat{\Delta}^Z_{GE,N} + \hat{\Delta}^Z_{HE,N} \right)$  \\
    \end{tabular}
    \caption{Variance Estimators for Two-Way Clustering}
    
\end{table}

Table 2 summarizes our proposed variance estimators for multiway sampling or multiway assignment (Hessian matrix is ignored here). 
$H_N$ and $G_N$ are the number of clusters in the sample on the dimensions of $H$ and $G$ respectively. We expect cases 2 and 3 in Table 1 and case 2 in Table 2 to be the leading cases in empirical practice. 

\begin{remark}
If all cluster sizes are unbounded, we can allow for sampling. The proof would be constructed first showing that the within-cluster sample average of $z_{iM}$ converges to its within-cluster population average. The adjusted cluster variance estimator $\hat{\Delta}^Z_{CE,N}$ in (\ref{eqn:shrink}) must then be further divided by $\rho_{uM}\rho_{hM}\rho_{gM}$. 
Since our assumptions do not imply that all cluster sizes are unbounded in general, we omit this case from our theorem. Nonetheless, its adjustment is presented in Table 2, where two-way clustered assignment can be combined with sampling.
\end{remark}

In the context of doing inference for functions of M-estimators, we can also apply the same techniques to estimate the two extra terms, $\Delta^f_{E,M}$ and $\Delta^f_{EC,M}$. The only difference is that the dependent variables in the regression-based approach would be the cluster sum of 
\begin{equation}
f_{iM}(W_{iM},\hat{\theta}_N)-\hat{\gamma}_N-\hat{F}_N(\hat{\theta}_N)\hat{L}_N(\hat{\theta}_N)^{-1}m_{iM}(W_{iM},\hat{\theta}_N)
\end{equation} 
rather than $m_{iM}(W_{iM},\hat{\theta}_N)$ alone. (See the details of the notation in Appendix A.)
The results for multiway clustering can be derived in a similar fashion and hence are omitted here.

\section{Simulation}
In this section, we compare the Monte Carlo standard deviation of the coefficient estimator and the APE estimator of the assignment variable in a binary response model with a set of different standard errors. We are mainly interested in the finite sample performance of the proposed shrinkage variance estimators. As a leading case in empirical practice, we focus on (multiway) clustered assignment with the entire population observed.  

In the population generating process, there is a single assignment variable $X_{iM}\in\{0,1\}$ and a single attribute variable $z_{iM}=z_{g(i)M}+z_{h(i)M}$, where $z_{hM}=\pm 1$ with equal probability and $z_{gM}=\pm 1$ with equal probability in the design of two-way clustered assignment, and $z_{gM}=\pm 2$ with equal probability in the design of one-way clustered assignment. The potential outcome of a binary response is generated as 
\begin{equation}
y_{iM}(x)=\mathbbm{1}[x+2z_{iM}\cdot x+e_{iM}>0].
\end{equation}
The idiosyncratic unobservable $e_{iM}$ is the residual from regressing random realization of a standard normal distribution on $z_{iM}$. The data of $z_{iM}$ and $e_{iM}$ are generated once and kept fixed in the population $M$. 

We partition the population units into 50 clusters each on the two dimensions $G$ and $H$ with one unit for every $(g,h)$ cluster pair. As a result, the population size is 2,500. The results for 100 clusters on both dimensions are similar and hence omitted to save space.  

The assignment variable $X_{iM}=A_{g(i)}B_{h(i)}$, where $A_g$ and $B_h$ are binary cluster assignment variables drawn independently with $P(A_g=1)=P(B_h=1)=1/2$, $\forall\ g=1,2,\dots, G$ and $h=1,2,\dots,H$. Therefore, the assignments are clustered at both the $G$ and $H$ dimensions. For the case of one-way cluster assignment, we fix $B_h=1$. There are 10,000 replications for both designs. For each replication, $X_{iM}$ is re-assigned according to the assignment rules above.

Estimates from the pooled probit regression of $Y_{iM}$ on 1, $X_{iM}$, and $z_{iM}$ are displayed in Table \ref{tab:sim} below. 
Columns (1) and (2) collect results for one-way cluster assignment and columns (3) and (4) show results for two-way cluster assignment.

The first two rows of Table \ref{tab:sim} report the Monte Carlo standard deviation of the point estimates and the coverage rate of the 95\% confidence interval based on the oracle standard error, i.e., Monte Carlo standard deviation. The oracle coverage rates are very close to the nominal level of 95\%. Thus, normal approximation seems to work well in finite samples. The next two rows report the superpopulation EHW standard errors and the corresponding coverage rate of the 95\% confidence interval.\footnote{The $97.5^{th}$ percentile of $t(G-1)$ is used as the critical value in constructing the confidence intervals.} The EHW standard errors are too small and the confidence interval undercovers as expected. 

For one-way clustered assignment, we focus on one-way cluster-robust standard errors at the level $G$. Here, we demonstrate how our shrinkage variance estimators work in finite samples. 
For two-way clustered assignment, we report results on both one-way cluster-robust and two-way cluster-robust standard errors. 

The adjusted one-way clustered standard errors at the level $G$ are more than half smaller than the superpopulation one-way clustered standard errors, though still above the Monte Carlo standard deviation under one-way clustered assignment. Switching to two-way clustered assignment, the adjusted one-way clustered standard errors are both too small. The two-way clustered standard errors for both CGM and CGM2 estimators work well in this population generating process. There is slight downward bias of the adjusted CGM standard errors. The adjusted CGM2 standard errors are larger but are guaranteed to be conservative. 

We conclude from the simulation results that the usual superpopulation one-way cluster-robust standard errors and the two-way CGM2 cluster-robust standard errors are overly conservative. When there are fixed attributes available, they can be used to estimate an upper bound of the finite population CRAV. Although the adjusted finite population cluster-robust standard error is still conservative, it often improves over the usual cluster-robust standard error. In particular, our adjusted standard errors are 40\% smaller than CGM while still maintaining correct coverage.

\begin{table}[htbp]
  \centering
  \caption{Standard Errors and Coverage Rates for Pooled Probit}
  \begin{threeparttable}
    \begin{tabular}{lcccc}
    \toprule
               & \multicolumn{2}{c}{One-way Assignment} & \multicolumn{2}{c}{Two-way Assignment} \\
          \cmidrule(lr){2-3} \cmidrule(lr){4-5} 
              & APE   & Coefficient & APE   & Coefficient \\
              & (1) & (2) & (3) & (4) \\
          \hline
 SD    & 0.0225 & 0.0643 & 0.0397 & 0.1065 \\
           Cov (Oracle) & (0.953) & (0.953) & (0.952) & (0.952) \\
           EHW   & 0.0185 & 0.0531 & 0.0220 & 0.0581 \\
           Cov (EHW) & (0.918) & (0.918) & (0.755) & (0.748) \\
           One-way (G) & 0.0591 & 0.1716 & 0.0446 & 0.1187 \\
           Cov (G) & (1.000) & (1.000) & (0.972) & (0.973) \\
           One-way adj (G) & 0.0260 & 0.0752 & 0.0294 & 0.0792 \\
           Cov (G, adj) & (0.993) & (0.994) & (0.876) & (0.879) \\
           One-way (H) & -     & -     & 0.0481 & 0.1284 \\
           Cov (H) & -     & -     & (0.982) & (0.983) \\
           One-way adj (H) & -     & -     & 0.0321 & 0.0867 \\
           Cov (H, adj) & -     & -     & (0.909) & (0.912) \\
           Two-way CGM & -     & -     & 0.0620 & 0.1655 \\
           Cov (CGM) & -     & -     & (0.997) & (0.998) \\
           CGM adj & -     & -     & 0.0377 & 0.1022 \\
           Cov (CGM, adj) & -     & -     & (0.955) & (0.957) \\
           Two-way CGM2  & -     & -     & 0.0658 & 0.1756 \\
           Cov (CGM2) & -     & -     & (0.999) & (0.999) \\
           CGM2 adj & -     & -     & 0.0437 & 0.1180 \\
           Cov (CGM2, adj) & -     & -     & (0.980) & (0.981) \\
          \bottomrule
    \end{tabular}%
    \begin{tablenotes}
     \footnotesize
\item[1] Columns (1) and (2) collect superpopulation and adjusted finite population standard errors and the coverage rates of the 95\% confidence interval based on these standard errors for the APE and coefficient estimators on $X$ under one-way clustered assignment; Columns (3) and (4) report the same set of statistics under two-way clustered assignment. 
\item[2] ``SD" stands for the standard deviation of the point estimates across 10,000 replications; ``Oracle" stands for the coverage rate of the 95\% confidence interval based on the Monte Carlo standard deviation; ``EHW" stands for the superpopulation heteroskedasticity-robust standard errors; ``One-way (C), $C\in\{G,H\}$" stands for the superpopulation one-way cluster-robust standard errors clustered at the level $C$; ``Two-way" stands for the two-way cluster-robust standard errors at both levels of $G$ and $H$, using either CGM or CGM2 estimator; ``adj" stands for the adjusted finite population standard errors; the columns below each standard error report the corresponding coverage rates.  
\end{tablenotes}
\end{threeparttable}
  \label{tab:sim}%
\end{table}%

\section{Clustering in Practice}
In this section, we first apply our general theorems to the difference-in-means estimator, which has been studied thoroughly in the literature. Next, we investigate the treatment effect estimand that a cluster fixed effect regression targets. Lastly, we discuss some empirical settings where two-way clustered standard errors have been considered. 

\subsection{Application to Difference-in-Means} \label{sec:diff_means} 
To make a direct comparison with \citet{abadie2023should}, we consider a difference-in-means estimator for a binary assignment variable without covariates with multiway clustering.
For treatment variable $X \in \{ 0,1 \}$, we denote the nonstochastic potential outcome as $y_{iM} (x)$. We are interested in the population average treatment effect (ATE): 
\begin{equation}
\tau_M = \frac{1}{M} \sum_{i=1}^{M} (y_{iM} (1) - y_{iM} (0))
\end{equation}

Multiway assignment is treated in the following way. Data is generated by independently drawing $A_{g}\in[0,1]$, $B_{h}\in [0,1]$ and $e_{i}\sim U[0,1]$, with $X_{iM}=1\left\{ e_{i}<A_{g(i)}B_{h(i)}\right\}$. The random variables $A_{g}$ and $B_{h}$ have means $\mu_{A}, \mu_{B} > 0$ and variances $\sigma_{A}^2, \sigma_{B}^2$ respectively. This process nests several cases. If assignment is one-way clustered, then we can simply set $A_{g}=1$. Another case is where assignment occurs at the intersection level, and we need both dimensions $G$ and $H$ to be assigned treatment for the unit to be treated. Then, $A_{g}, B_{h} \in \{0,1 \}$. 
If $\mu_{A}$ and $\mu_{B}$ are both non-zero, then even though $X_{iM} = A_{g(i)} B_{h(i)}$ is an interaction model, we do not need to be concerned about non-normality.\footnote{
Let $A_g = \mu_A + \epsilon^A_g$ and $B_h = \mu_B + \epsilon^B_h$ where $\mathbb{E}[\epsilon_g] = \mathbb{E}[\epsilon_h]=0$. Then, $\sum_{i=1}^M X_{iM} = \sum_{i=1}^M (\mu_A + \epsilon^A_{g(i)}) (\mu_B + \epsilon^B_{h(i)})  = \sum_{i=1}^M (\mu_A \mu_B+ \epsilon^A_{g(i)} \mu_B + \mu_A \epsilon^B_{h(i)}+ \epsilon^A_{g(i)}\epsilon^B_{h(i)})$ so the dominant stochastic terms are $\epsilon^A_{g(i)} \mu_B $ and $ \mu_A \epsilon^B_{h(i)}$ instead of $ \epsilon^A_{g(i)}\epsilon^B_{h(i)}$.
}

With $\alpha_M := (1/M)\sum_{i=1}^{M} y_{iM}(0)$
and error $U_{iM} := Y_{iM} - \alpha_M - \tau_M X_{iM}$, 
the potential errors are denoted:
\begin{align*}
    u_{iM}(1) &= y_{iM}(1) - (\alpha_M + \tau_M) \\
    u_{iM}(0) &= y_{iM}(0) - \alpha_M
\end{align*}

For $R_{iM}=1$, we observe $\{ Y_{iM}, X_{iM} \}$, with $Y_{iM}= X_{iM} y_{iM}(1) + (1-X_{iM})y_{iM}(0)$. Let $b_{1}=\mathbb{E}\left[R_{iM} X_{iM}\right]$, $b_{0}=\mathbb{E}\left[R_{iM} (1-X_{iM})\right]$. $b_{1}$ is the probability that an individual is observed and treated; $b_{0}$ is the probability that an individual is observed and untreated. 
$N_{1} := \sum_{i=1}^{M} R_{iM} X_{iM}$ and $N_{0} := \sum_{i=1}^{M} R_{iM} (1-X_{iM})$. The least squares estimator is:
\begin{equation}
\hat{\tau}_M = \frac{1}{N_{1}} \sum_{i=1}^{M} R_{iM} X_{iM} Y_{iM} - \frac{1}{N_{0}} \sum_{i=1}^{M} R_{iM} (1-X_{iM}) Y_{iM}.
\end{equation}

To state the main result, we first define a few terms. Let $\mathcal{N}_i$ denote the neighborhood of $i$, which is the set of observations that are plausibly correlated with $i$. 
The score is 
\begin{equation}
\eta_{iM}:=R_{iM}\left(\frac{X_{iM}}{b_{1}}-\frac{1-X_{iM}}{b_{0}}\right)U_{iM}.
\end{equation}
In this context, $\mathbb{E}[\eta_{iM}]=u_{iM}(1) - u_{iM}(0) =: \tau_{iM}-\tau_M$ is not zero in general, but $\sum_i \mathbb{E}[\eta_{iM}]=0$. Hence, we define $\xi_{iM}$ as the demeaned residual for individual $i$ that features in the variance of $\hat{\tau}_M$:
\begin{equation}
\xi_{iM} := \frac{1}{b_{1}}\left(R_{iM}X_{iM}-b_{1}\right)u_{iM}(1)-\frac{1}{b_{0}}\left(R_{iM}\left(1-X_{iM}\right)-b_{0}\right)u_{iM}(0).
\end{equation}

\begin{corollary} \label{thm:tau_distr}
    Under Assumptions \ref{assump1} to \ref{asmp:clt_reg},
    \begin{equation}
    \frac{\sqrt{N} (\hat{\tau}_M - \tau_M)}{\sqrt{v_M}} \xrightarrow{d} N(0,1),
    \end{equation}   
    where 
\begin{equation}
    v_M := \frac{N}{M^2} \sum_{i=1}^{M} \sum_{j \in \mathcal{N}_i} \mathbb{E}[\xi_{iM} \xi_{jM}].
\end{equation}
\end{corollary}

This result is a corollary of the existing normality results and law of large numbers.
Comparing this context to our framework, $\eta_{iM}= m_{iM}$,  $\xi_{iM} = m_{iM} - \mathbbm{E}[m_{iM}]$, and $v_M = V_{TWM}$.
Using our framework, we can answer questions on whether multiway clustering matters and whether multiway clustering is appropriate. Using the one-way CRVE on dimension $G$ (without loss of generality) yields the following estimand:
\begin{equation}
    V_{GM} = L_{M}\left(\theta_{M}^{*}\right)^{-1}\left(\Delta_{ehw,M}\left(\theta_{M}^{*}\right)+\rho_{uM} \Delta_{(G\cap H),M}\left(\theta_{M}^{*}\right)+\rho_{uM}\rho_{hM}\Delta_{G,M}\left(\theta_{M}^{*}\right)\right)L_{M}\left(\theta_{M}^{*}\right)^{-1}
\end{equation}
Then, answering the question on whether two-way clustering matters involves comparing $V_{GM}$ with $V_{TWSM}$ and answering the question on whether two-way clustering is appropriate involves comparing $V_{GM}$ with $V_{TWM}$. 
The comparisons yield:
\begin{equation}
\begin{split}
&V_{TWM} - V_{GM} \\
&= L_{M}\left(\theta_{M}^{*}\right)^{-1} \left(\rho_{uM}\rho_{gM}\Delta_{H,M}\left(\theta_{M}^{*}\right) - \rho_{uM}\rho_{gM}\rho_{hM}\left(\Delta_{E,M}+\Delta_{E(G\cap H),M}+\Delta_{EG,M}+\Delta_{EH,M}\right) \right) L_{M}\left(\theta_{M}^{*}\right)^{-1}
\end{split}
\end{equation}
\begin{equation} \label{eqn:V_compare}
V_{TWSM} - V_{GM} = L_{M}\left(\theta_{M}^{*}\right)^{-1} \left(\rho_{uM}\rho_{gM}\Delta_{H,M}\left(\theta_{M}^{*}\right) \right) L_{M}\left(\theta_{M}^{*}\right)^{-1}
\end{equation}

Consider the environment with constant treatment effects.
Then, $u_{iM}(1) - u_{iM}(0) = 0$ and hence $\mathbbm{E}[m_{iM}] =0$ and all the $\Delta_E$ terms in the equation above are 0, so the two comparisons become identical.
If there is multi-way clustered assignment only or if there is clustered assignment on $H$ and clustered sampling on $G$, then (\ref{eqn:V_compare}) cannot be simplified further.
However, if there is multiway clustered sampling only or if there is clustered assignment on $G$ and clustered sampling on $H$, then $\Delta_{H,M}(\theta^*_M) =0$ in (\ref{eqn:V_compare}) so $V_{TWSM} = V_{TWM} = V_{GM}$.
This result implies that, under constant treatment effects, it suffices to cluster on the assignment dimension and the usual CRVE is no longer conservative.\footnote{This result complements Corollary 1 of \citet{abadie2017should}: they find that one-way clustering is unnecessary if there is both constant treatment effects and no clustering in the assignment. }

\subsection{Fixed Effects Estimand under Clustered Assignment} \label{sec:fe}
With clustered data, empirical papers often include fixed effects (FE) in a regression. A first-order question is: What are these FE estimators approaching to in the limit within our finite population framework? In particular, when can FE estimands be interpreted as the ATE? 
We focus on an environment where assignment $X_{iM}$ is binary and two-way clustered with $A_g, B_h \in \{ 0,1 \}$ as described in Section \ref{sec:diff_means}, and we observe the entire population, i.e., $R_{iM}=1$ for all $i$.\footnote{Due to how fixed effects estimators transform variables using within-cluster means, any form of sampling leads to random sample sizes within each cluster which introduces additional technical complications that is beyond the immediate scope of our theory, so it is left for future research.} 
Nonetheless, since we allow for two-way clustered assignment and study estimands for both one-way and two-way fixed effects, the results in this section are new relative to \citet{abadie2023should} and \citet{athey2022design}: \citet{abadie2023should} study one-way fixed effects with one-way clustering, and \citet{athey2022design} require a random adoption date in the two-way fixed effects environment with one-way clustered assignment.

We start with one-way fixed effect. Let the ATE be defined in the following way. 
\begin{align*}
    \tau_{M}&:=\frac{1}{M}\sum_{i=1}^M\left(y_{iM}(1)-y_{iM}(0)\right)
    \tag{\stepcounter{equation}\theequation}
\end{align*}

In a one-way FE estimator, let
\begin{align*}
\bar{X}_{gM} & :=\frac{\sum_{i\in\mathcal{N}_{g}^{G}} X_{iM}}{ M^G_{g(i)} }, \text{ and }\\
\hat{\tau}_{OWFE} & :=\frac{\sum_{i=1}^M Y_{iM}\left(X_{iM}-\bar{X}_{g(i)M}\right)}{\sum_{i=1}^M X_{iM}\left(X_{iM}-\bar{X}_{g(i)M}\right)}. \tag{\stepcounter{equation}\theequation}
\end{align*}

\begin{proposition} \label{lem:OWFE}
Under Assumptions \ref{assump1} to \ref{assump4}, $\hat{\tau}_{OWFE} \xrightarrow{P} \tau_{OWFE}$, where
\begin{align*}
\tau_{OWFE}	&:=\sum_{i}\frac{\omega_{i}}{\sum_j \omega_j} \left(y_{iM}(1)-y_{iM}(0)\right), \text{ and } \\
\omega_{i}	&=\mu_{A}\mu_{B}\left(1-\frac{M_{(g(i),h(i))}^{G\cap H}+\left(M_{g(i)}^{G}-M_{(g(i),h(i))}^{G\cap H}\right)\mu_{B}}{M_{g(i)}^{G}}\right).
\end{align*}
\end{proposition}

Since $\mu_B \in [0, 1]$, all weights $\omega_i$ are positive, so the estimand is a weighted average of treatment effects. In the special case where we have perfectly balanced clusters $M_{(g(i),h(i))}^{G\cap H}=k$ and hence $M_{g(i)}^{G}=Hk$, then $\tau_{OWFE} = \tau_M$. 
This result suggests that, unlike the difference-in-means example where clustered dependence only affects the variance, clustered dependence here also affects the interpretation of the estimand.
Mechanically, the difference is that the estimator here contains products $X_{iM} X_{jM}$ for $i \ne j$ that is not present in Section \ref{sec:diff_means}, so correlation in $X$ affects the estimand.

Moving on to two-way fixed effect (TWFE), we focus on the case with multiway assignment where a unit is treated if and only if both its clusters are treated. 
This setup is different from one-way assignment at the intersection level (see \Cref{remark:twoway_vs_intersect}). 
With a linear model
\begin{equation}
Y_{i}=\tau X_{i}+\eta_{g(i)}+\gamma_{h(i)}+\varepsilon_{i},
\end{equation}
the TWFE estimator is
\begin{equation} \label{eqn:TWFE_equation}
\hat{\tau}_{TWFE} :=\frac{\sum_{i=1}^M\tilde{X}_{iM}Y_{iM}}{\sum_{i=1}^M\tilde{X}_{iM}X_{iM}},
\end{equation}
where $\tilde{X}_{iM}$ is the residual when regressing $X$ on the fixed effects. 
Due to \citet{baltagi2008econometric} Chapter 3, 
\begin{equation} \label{eqn:TWFE_transform}
\tilde{X}_{iM}=X_{iM}-\frac{\sum_{j\in\mathcal{N}_{g}^{G}}R_{jM}X_{jM}}{M^G_{g(i)}}-\frac{\sum_{j\in\mathcal{N}_{h}^{H}}R_{jM}X_{jM}}{M^H_{h(i)} }+\frac{\sum_{i=1}^M X_{iM}}{M}
\end{equation}

\begin{proposition} \label{lem:TWFE}
Under Assumptions \ref{assump1} to \ref{assump4}, if $M_{(g,h)}^{G\cap H}=k,M_{g}^{G}=Hk,M_{h}^{H}=Gk$ (i.e., all cluster intersections are balanced), then $\hat{\tau}_{TWFE} \xrightarrow{P} \tau_{TWFE}$, where 
\begin{align*}
\tau_{TWFE} = \frac{\sum_{i=1}^M\E\left[\tilde{X}_{iM}Y_{iM}\right]}{\sum_{i=1}^M\E\left[\tilde{X}_{iM}X_{iM}\right]} & =\tau_{M}.
\end{align*}
\end{proposition}

This result is fairly weak in that it requires balanced clusters, but in this idealized situation, the TWFE estimand is the ATE. 
Unlike Proposition \ref{lem:OWFE}, we cannot interpret the estimand as a weighted average of treatment effects in general, but this result is not surprising considering the large difference-in-differences (DD) literature on how TWFE cannot be interpreted as a weighted average of treatment effects, even with parallel trends.

\subsection{Two-way Clustering v.s. Spatiotemporal Correlation}
One scenario in which two-way clustering is often employed is with panel data, where one clusters at the cross-sectional entity level and time level to adjust for both cross-sectional dependence and serial correlation. Clustering at the entity level can be justified by cluster sampling if cross-sectional entities are independently sampled from a finite population and each entity comes with its serial observations. If the entire cross-sectional population is observed, clustering at the entity level can be justified by serially correlated assignments for each entity across time. A leading example is the standard difference-in-differences, where either before or after the initial treatment period, the treatment assignments within entities are perfectly positively correlated. Meanwhile, treatments across the initial treatment period are negatively correlated. Justification for clustering at the time level is more ambiguous. One leading argument is to account for spatial correlation and hence two-way clustering has been used as a substitute for spatiotemporal-correlation robust inference.

Let us ignore the Hessian matrix for now, as it does not matter for the comparison of different variance matrix estimators. For spatial data, we often work with the entire population and hence sampling may not be a particularly important aspect here. Let us suppose that all sampling probabilities are one. Since our asymptotic theory requires that the number of clusters at both dimensions goes to infinity in the sequence of populations, we work with long panel data where $(|D_M|,T)\to\infty$. Notice that $D_M$ denotes lattices in $\mathbb{R}^d$ and $|D_M|$ stands for the cross-sectional population size. Adapting the argument from \cite{xu2022design}, the finite population spatiotemporal variance matrix is approximately 
\begin{equation}
\begin{aligned}
V_{PM}=&\frac{1}{T|D_M|}\sum_{t=1}^T\sum_{i\in D_M}\big[\mathbb{E}(m_{it} m_{it}')-\mathbb{E}(m_{it})\mathbb{E}(m_{it})'\big]\\
&+\frac{1}{T|D_M|}\sum^T_{t=1}\sum_{i\in D_M}\sum_{j\in D_M, j\neq i}\omega\bigg(\frac{\nu(i,j)}{b_M}\bigg)\big[\mathbb{E}(m_{it} m_{jt}')-\mathbb{E}(m_{it})\mathbb{E}(m_{jt})'\big]\\
&+\frac{1}{T|D_M|}\sum_{i\in D_M}\sum^T_{t=1}\sum^T_{s\neq t}\omega\bigg(\frac{\nu(t,s)}{b_T}\bigg)\big[\mathbb{E}(m_{it} m_{is}')-\mathbb{E}(m_{it})\mathbb{E}(m_{is})'\big]\\
&+\frac{1}{T|D_M|}\sum^T_{t=1}\sum^T_{s\neq t}\sum_{i\in D_M}\sum_{j\in D_M, j\neq i}\omega\bigg(\frac{\nu(i,j)}{b_M}\bigg)\omega\bigg(\frac{\nu(t,s)}{b_T}\bigg)\big[\mathbb{E}(m_{it} m_{js}')-\mathbb{E}(m_{it})\mathbb{E}(m_{js})'\big],
\end{aligned}
\end{equation} 
where $m_{it}=m_{itM}(W_{itM},\theta^*_M)$ to simplify notation. $\nu(i,j)$ measures the cross-sectional distance and $\nu(t,s)$ measures the time lag. $b_M$ and $b_T$ are the cross-sectional and time dimension bandwidth respectively. $\omega(\cdot)$ is the kernel weighting, which satisfies: $\omega (0)=1$; $\omega\Big(\frac{\nu(\cdot,\cdot)}{b}\Big)=0$ for any $\nu(\cdot,\cdot)>b$; $\Big|\omega\Big(\frac{\nu(\cdot,\cdot)}{b}\Big)\Big|\leq 1$, $\forall$ $\nu(\cdot,\cdot)$. 

If the dependence within clusters is weak, the usual CRVE is still consistent for one-way clustering within the superpopulation framework.  Although without downweighting, the CRVE will have slower rates of convergence than HAC type variance estimators (see \cite{hansen2007asymptotic}). On the other hand, if there is spatiotemporal correlation in assignments, two-way clustered variance matrix accounts for the first three terms in $V_{PM}$ but ignores cross-cluster correlation in different time periods for different units. The magnitude of two-way clustering and spatiotemporal robust inference cannot be ordered without additional information on the correlation pattern of assignment variables. 

\subsection{Two Assignment Variables Clustered on Different Dimensions}

There are instances where 
assignment on multiple variables could be clustered at differing dimensions. Suppose there are two assignment variables $X_{1iM}$ and $X_{2iM}$. The assignment of $X_{1iM}$ is clustered on the dimension of $G$, whereas the assignment of $X_{2iM}$ is clustered on the dimension of $H$. As an empirical example, \cite{hersch1998compensating} studies wage and injury risk trade-off for men and women. The two assignment variables are injury rates for individual $i$'s industry and for $i$'s occupation respectively. Hence, one assignment variable is clustered at the industry level and the other assignment variable clustered at the occupation level. 

When two assignment variables are independent of each other, the Frisch–Waugh theorem implies that we only need to cluster the standard errors on one dimension for each of the assignment variables in a linear regression. We conduct a simple simulation to compare two-way clustered standard errors with one-way clustered standard errors on each dimension for the coefficient estimator on both assignment variables. 

The potential outcome function is given below.
\[
y_{iM}(x_{g(i)},x_{h(i)})=\tau_{1i} x_{g(i)}+\tau_{2i} x_{h(i)}+e_{iM},
\]
where $e_{iM}$ is the nonstochastic individual unobservable. The cluster assignment variables $X_g$ and $X_h$ are binary with probabilities $P(X_g=1)=P(X_h=1)=1/2$.
In the first design, we impose heterogeneous treatment effect on cross dimensions. Namely, $\tau_{1i}=\tau_{h(i)}=\pm 1$ with equal probability; $\tau_{2i}=\tau_{g(i)}=\pm 1$ with equal probability. In the second design, we set $\tau_{1i}=\tau_{g(i)}$ and $\tau_{2i}=\tau_{h(i)}$. We regress $Y_{iM}$ on 1, $X_{g(i)}$, and $X_{h(i)}$ and report different superpopulation standard errors as upper bounds for the finite population ones.

\begin{table}[htbp]
  \centering
  \caption{Standard Errors for Two Assignment Variables}
  \begin{threeparttable}
    \begin{tabular}{lcccc}
    \toprule
          & \multicolumn{2}{c}{Design 1} & \multicolumn{2}{c}{Design 2} \\
           \cmidrule(lr){2-3} \cmidrule(lr){4-5} 
          & \multicolumn{1}{c}{$X_g$} & \multicolumn{1}{c}{$X_h$} & \multicolumn{1}{c}{$X_g$} & \multicolumn{1}{c}{$X_h$} \\
          \hline
    SD    & 0.103 & 0.103 & 0.103 & 0.103 \\
    Coverage (oracle) & (0.950) & (0.953) & (0.950) & (0.954) \\
    One-way s.e. (G) & 0.102 & 0.103 & 0.143 & 0.020 \\
    Coverage (G) & (0.947) & (0.957) & (0.992) & (0.297) \\
    One-way s.e.(H) & 0.103 & 0.102 & 0.020 & 0.143 \\
    Coverage (H) & (0.953) & (0.950) & (0.305) & (0.993) \\
    Two-way s.e. & 0.142 & 0.142 & 0.142 & 0.141 \\
    Coverage (two-way) & (0.993) & (0.995) & (0.992) & (0.993) \\
    \bottomrule
    \end{tabular}%
     \begin{tablenotes}
     \footnotesize
\item[1]  The second and fourth columns collect superpopulation standard errors and the coverage rates of the 95\% confidence interval for the coefficient estimator on $X_g$; the third and fifth columns report the same set of statistics for the coefficient estimator on $X_h$. 
\item[2] ``SD" stands for the standard deviation of the coefficient estimates across 10,000 replications; ``Oracle" stands for the coverage rate of the 95\% confidence interval based on the Monte Carlo standard deviation; ``One-way s.e. (C), $C\in\{G,H\}$" stands for the one-way cluster-robust standard errors clustered at the level $C$; ``Two-way s.e." stands for the two-way cluster-robust standard errors at both levels of $G$ and $H$; the columns below each standard error report the corresponding coverage rates.  
\item[3] We partition the population units into 100 clusters each on the two dimensions $G$ and $H$ with one unit for every $(g,h)$ cluster pair. As a result, the population size is 10,000. We observe the entire population. 
\end{tablenotes}
\end{threeparttable}
  \label{tab:twovar}%
\end{table}%

As we can see from Table \ref{tab:twovar}, indeed one-way clustered standard errors are sufficient for the coefficient estimators on the corresponding assignment variables. Namely, we can report one-way clustered standard errors at the level $G$ for $\hat{\tau}_1$ and one-way clustered standard errors at the level $H$ for $\hat{\tau}_2$. In the second design, using the two-way clustered standard errors is harmless, although both one-way and two-way clustered standard errors are conservative. However, when the heterogeneous treatment effects are more asymmetric in the first design, the two-way clustered standard errors can be a lot larger than the one-way clustered standard errors, making the inference unnecessarily overly conservative.

\subsection{Triple Differences}
Recently, triple differences method has got more attention among empirical researchers. There has been ample discussion on the correct inference on difference-in-differences; see, e.g., \cite{bertrand2004much}. Nevertheless, the dicussion on the inference method for triple differences has been limited. In the simulation of \cite{olden2022triple}, they report one-way cluster-robust standard errors on the treatment level to directly compare with the simulation results in \cite{bertrand2004much}. Recently, \cite{strezhnev2023decomposing} advocates for two-way cluster-robust standard errors for triple difference estimators. Both one-way and two-way clustered standard errors have been reported in empirical research according to our survey. 

Borrowing the notation from \cite{strezhnev2023decomposing}, (\ref{eqn:ddd}) is a common specification for triple differences.
\begin{equation}\label{eqn:ddd}
Y_{ight}=\tau_i D_{ight}+\alpha_{gh}+\gamma_{ht}+ \delta_{gt}+\epsilon_{ight}
\end{equation}
Unit $i$ in stratum $h$ and group $g$ is treated at time $t$ if $D_{ight}=1$. The terms $\alpha_{gh}$, $\gamma_{ht}$, and $\delta_{gt}$ are the group-stratum, stratum-time, and group-time fixed effects respectively. $\epsilon_{ight}$ is the individual idiosyncratic term. Suppose there are multiple strata $h\in\{1,2,\dots,H\}$ and multiple groups $g\in\{1,2,\dots,G\}$. Otherwise, there is not much we can do in clustering the standard errors. 

We can rewrite the treatment variable $D_{ight}$ as an interaction of three variables, $D_{ight}=D_{h(i)} \times D_{g(i)} \times Post$, where $Post$ is a time dummy variable that takes the value of one for periods post the initial treatment period. Consequently, it is tempting to compare one-way clustering with two-way clustering. In our view, other than the time variable, the key boils down to the nature of the other two variables within the triple interaction term. If both grouping indicators are stochastic assignment variables on non-nested dimensions, one would like to report the two-way clustered standard errors. An example is \cite{marchingiglio2019employment}, where $h$ and $g$ represent state and industry respectively. Their assignment variable is the state level adoption of gender-specific minimum wage laws applying to specific industries that employed a larger share of women. Alternatively, if treatment is assigned based on one grouping indicator, and the other grouping indicator is some nonstochastic attribute, one-way clustered standard errors are the most appropriate. As an example, in \cite{bau2021can} $h$ indicates province and $g$ represents ethnicity. Since the policy they study is the roll-out of pension plans at the province level, ethnicity is nonstochastic within our design-based framework. In their context, the pension plan policy affects the ethnicity groups differentially by nature.   

To showcase the differences between one-way and two-way clustered standard errors, we conduct a simple simulation of a triple differences regression based on (\ref{eqn:ddd}). There are two time periods. The treatment is essentially randomly assigned in the second time period. Both $D_g$ and $D_h$ are cluster binary variables with probabilities $P(D_g=1)=P(D_h=1)=1/2$. In the first design, $D_g$ and $D_h$ are stochastic, whereas in the second design, $D_g$ is the nonstochastic attribute variable but $D_h$ remains as the stochastic assignment variable. We construct $\tau_i$ in a way that the parallel trends assumption for triple differences holds.\footnote{Specifically, we construct $\tilde{\tau}_i=\tau_{g(i)}+\tau_{h(i)}$, where $\tau_g=\pm 2$ with equal probability and $\tau_h=\pm 1/2$ with equal probability. In design 1, $\tau_i$ is the demeaned $\tilde{\tau}_i$. In design 2, we average $\tilde{\tau}_i$ across units with $D_{g(i)}=1$ and denote this average by $\bar{\tilde{\tau}}$. $\tau_i=\tilde{\tau}_i-\bar{\tilde{\tau}}$.} We report the adjusted finite population standard errors for $\hat{\tau}$ and the coverage rate of the 95\% confidence interval based on these standard errors.\footnote{We use $\tau_i$ as the attributes in the estimation of the adjusted finite population standard errors.}

\begin{table}[htbp]
  \centering
  \caption{Standard Errors for Triple Differences Estimators}
  \begin{threeparttable}
    \begin{tabular}{lcc}
    \toprule
           & $D_g$ \& $D_h$ stochastic & $D_h$ stochastic \\
           \hline
    SD    & 0.210 & 0.056 \\
    Coverage (oracle) & (0.951) & (0.951) \\
    EHW s.e. & 0.044 & 0.042 \\
    Coverage (EHW) & (0.333) & (0.868) \\
    One-way s.e. (G) & 0.203 & 0.189 \\
    Coverage (G) & (0.945) & (1.000) \\
    One-way s.e.(H) & 0.055 & 0.057 \\
    Coverage (H) & (0.404) & (0.957) \\
    Two-way 2 s.e. & 0.211 & 0.198 \\
    Coverage (two-way 2) & (0.953) & (1.000) \\
    \bottomrule
    \end{tabular}%
     \begin{tablenotes}
     \footnotesize
\item[1] The second column collects adjusted finite population standard errors and the coverage rates of the 95\% confidence interval for the triple differences estimator when both grouping indicators are stochastic; the third column reports the same set of statistics when only one of the grouping indicators is stochastic. 
\item[2] ``SD" stands for the standard deviation of the triple differences estimates across 10,000 replications; ``Oracle" stands for the coverage rate of the 95\% confidence interval based on the Monte Carlo standard deviation; ``EHW s.e." stands for the heteroskedasticity-robust standard errors;``One-way s.e. (C), $C\in\{G,H\}$" stands for the one-way cluster-robust standard errors clustered at the level $C$; ``Two-way 2 s.e." stands for the CGM2 two-way cluster-robust standard errors at both levels of $G$ and $H$; the columns below each standard error report the corresponding coverage rates.  
\item[3] We partition the population units into 100 clusters each on the two dimensions $G$ and $H$ with one unit for every $(g,h)$ cluster pair. As a result, the population size is 10,000. We observe the entire population. 
\end{tablenotes}
\end{threeparttable}
  \label{tab:ddd}%
\end{table}%

As shown in Table \ref{tab:ddd}, the EHW standard errors underestimate as expected. When both grouping indicators are stochastic assignment variables, one-way clustered standard errors are generally not sufficient. By fluke the one-way clustered standard errors could be larger than the standard deviation, but that is because they are conservative within the design-based framework. The adjusted CGM2 standard error works pretty well with coverage rate of the confidence interval close to its nominal level. Switching to the case when only the grouping indicator $D_h$ is stochastically assigned, clustering the standard errors at the level of $H$ suffices. Two-way clustered standard error is overly conservative and can be more than three times larger than the one-way clustered standard errors clustered on $H$.

\section{Empirical Illustration}
The adjusted finite population CRVE proposed in Theorem \ref{thm:shrink} is applied to \cite{antecol2018equal}, who study the effects of tenure clock stopping policies on tenure rates among assistant professors. 
The unique dataset collected by them contains all assistant professor hires at the top-50 Economics departments from 1980-2005 as pooled cross sections, 
resulting in 1,392 observations in total. 
Furthermore, the tenure clock stopping policies are assigned at the university level while the data are collected at the individual level,
implying that we have a setting of observing the entire population with cluster assignment.\footnote{This group of assistant professors is treated as the population.} 
The standard errors in \cite{antecol2018equal} are clustered at the policy university level, which is the correct level to cluster the standard errors as implied by Remark \ref{coro:oneway1}. As a result, there are 49 clusters in total with cluster sizes ranging from 11 to 57. 

Since the dependent variable is a binary response, we analyze the linear probability model (LPM) given in \cite{antecol2018equal} along with an additional probit model given in (\ref{eqn:47}) below, which adopts the same notation from their paper. 
\begin{equation}\label{eqn:47}
\begin{aligned}
P(&Y_{ugit}=1|GN_{ut}, F_{ugit}, E_{ut}, FO_{ut}, X_{ugit}, Z_{ut}, \rho_{gt}, \psi_{ug})=\\
\Phi(&\beta_0+\beta_1 GN_{ut}+\beta_2 GN_{ut}\times F_{ugit}+\beta_3 GN_{ut}\times E_{ut}+\beta_4 GN_{ut}\times E_{ut}\times F_{ugit}\\
&+\beta_5 FO_{ut}+\beta_6 FO_{ut}\times F_{ugit}+\beta_7 FO_{ut}\times E_{ut}+\beta_8 FO_{ut}\times E_{ut}\times F_{ugit}\\
&+X_{ugit}\xi + Z_{ut}\eta+\rho_{gt}+\psi_{ug})
\end{aligned}
\end{equation}
The dependent variable $Y$ is an indicator of obtaining tenure at the university of initial placement. Binary variables $GN$ and $FO$ are indicators of gender-neutral and female-only tenure clock stopping policies respectively. The dummy variable $F$ is the indicator for females. The variable $E$ is an indicator of starting jobs in years zero through three after policy adoption. The vector $X$ contains individual characteristics and the vector $Z$ includes university level controls.\footnote{We refer to \cite{antecol2018equal} for the details of the variables included as controls.} The parameter $\rho$ captures gender-specific time trend and $\psi$ represents gender-specific university heterogeneity. The subscripts, $u$, $g$, $i$, $t$, are indicators for university, gender, individual, and the year the job started, respectively. 

\cite{antecol2018equal} include gender-specific university dummies to capture different unobserved university heterogeneity for males and females. Adding group dummies in the linear model is equivalent to performing fixed effects with clustered data. However, adding group dummies in a nonlinear model may cause the incidental parameter problem. Since the cluster sizes are unbalanced, we use pooled probit with correlated random effects as suggested by \cite{wooldridge2010econometric} to allow correlation between the gender-specific university heterogeneity and the covariates. Using Chamberlain-Mundlak device, the cluster size, the gender-specific university averages of individual and university characteristics and policies, and their interactions with cluster sizes are included as additional controls.

Given the probit model above is a nonlinear ``difference-in-differences'' model, the common trend assumption is imposed on the latent outcome variable following \cite{puhani2012treatment} and \cite{wooldridge2023simple}. The treatment effects are defined as the differences in the probit probabilities when the treatment variables equal one or zero. We report the average of the treatment effect for those actually treated by the specific policy. Since our emphasis in this study is on inference, we adhere to the specification presented in \cite{antecol2018equal} to facilitate a direct comparison with their reported standard errors. 
Assume that $\psi$ conditional on the sufficient statistics (the additional controls included) follows a normal distribution, APEs can be obtained via pooled probit. 

\begin{table}
  \centering
  \caption{Effects of Clock Stopping Policies on the Probability of Tenure at the University of Initial Placement}
\begin{threeparttable}
    \begin{tabular}{lcccccc}
\toprule
          & \multicolumn{3}{c}{LPM} & \multicolumn{3}{c}{Probit} \\
 \cmidrule(lr){2-4} \cmidrule(lr){5-7} 
          & \multirow{2}[0]{*}{APE} & \multicolumn{2}{c}{Standard Error} &  \multirow{2}[0]{*}{APE} & \multicolumn{2}{c}{Standard Error} \\
          \cmidrule(lr){3-4} \cmidrule(lr){6-7}  
        &       & inf pop & finite pop   &       & inf pop & finite pop  \\
 &    (1)   & (2)  & (3)  &    (4)   & (5)  & (6)  \\
 \hline
   Panel A. Policy Effects Years 0-3 &       &       &       &       &       &  \\
    Men FOCS &  -0.0085 & 0.0670 & 0.0616 & -0.0068 & 0.0623 & 0.0558 \\
    Women FOCS &  0.1723 & 0.1405 & 0.1191 & 0.1454 & 0.1750 & 0.1484 \\
    Men GNCS &   0.0511 & 0.0787 & 0.0757 & 0.0446 & 0.0726 & 0.0700 \\
    Women GNCS &   -0.0166 & 0.1071 & 0.0959 & 0.0220 & 0.1031 & 0.0957 \\
         &       &       &       &       &       &  \\
    Panel B. Policy Effects Years 4+ &       &       &       &       &       &  \\
    Men FOCS &  0.0023 & 0.0747 & 0.0701 & -0.0055 & 0.0649 & 0.0606 \\
    Women FOCS &    0.0493 & 0.1015 & 0.0797 & 0.0415 & 0.0902 & 0.0681 \\
    Men GNCS & 0.1757 & 0.0826 & 0.0794 & 0.1537 & 0.0767 & 0.0734 \\
    Women GNCS &    -0.1945 & 0.1057 & 0.0899 & -0.1856 & 0.1041 & 0.0892 \\
   \bottomrule
    \end{tabular}%
     \begin{tablenotes}
     \footnotesize
\item[1]  Standard errors are clustered at the university level. 
\item[2]  Columns (1) and (4) report the APEs under the linear probability model and the correlated random effects probit model, respectively; columns (2) and (5) report the usual infinite population cluster-robust standard errors of the APE estimators (linear functions of the coefficient estimators in the case of the LPM); columns (3) and (6) report the adjusted finite population cluster-robust standard errors of the APE estimators. 
\item[3] We refer to \cite{antecol2018equal} for detailed control variables.  
\end{tablenotes}
\end{threeparttable}
  \label{tab:3}%
\end{table}%

In Table \ref{tab:3}, panel A presents the total effects for men and women hired in years zero through three after policy adoption, and panel B shows the effects for those employed in years four or later.
The left panel (columns (1)-(3)) summarizes the results under the LPM. Columns (1) and (2) report the total effects and the standard errors, as shown in column (1) of Table 2 in \cite{antecol2018equal}, while column (3) reports the adjusted finite population clustered standard errors.
The coefficients (APEs) are interpreted as the policy effect on the tenure attainment of the assistant professors compared with those of the same genders at the same university but without any clock stopping policies.

To estimate the adjusted finite population CRAV, we sum all the estimated score functions and control variables within clusters and apply the variance estimator in Case 2 of Table 1 together with the usual estimator of the Hessian matrix. Since the number of control variables exceeds the number of clusters in the data, we only include university characteristics as the fixed attributes in the linear projection, resulting in a linear regression with 49 observations and eight independent variables. Compared with the usual cluster-robust standard errors, the finite population cluster-robust standard errors shrink by about 4\% to 21\% across the eight treatment groups. In terms of the statistical significance, the effect of gender-neutral policy for women hired three or more years after the policy adoption is significant at the 5\% rather than the 10\% level based on the adjusted finite population cluster-robust standard error. The same result holds when the critical values from $t(48)$ distribution are used. 

In the right panel (columns (4)-(6)), we can see that the APEs from the probit regression are close in magnitudes to those from the linear model. The adjusted finite population CRAV is estimated applying Theorem \ref{thm:shrink} and the delta method. Using the same set of university characteristics as the attribute variables, the reduction from the usual clustered standard errors to the finite population clustered standard errors ranges from 4\% to 25\%. Based on the critical values from $t(48)$, the effect of gender-neutral policy for women hired in later years is significant at the 5\% level rather than the 10\% level when the finite population clustered standard error is adopted.

To sum up, control variables can help shrink the standard errors when the population is treated as finite in both linear and nonlinear models. The empirical evidence suggests that gender-neutral tenure clock stopping policy is beneficial to men in obtaining tenured positions but detrimental to women. 

\section{Conclusion}
This paper develops finite population inference methods for M-estimators with data that is potentially clustered multiway. The takeaway for empirical practice is summarized as follows. One should only adjust standard errors for clustering if there is cluster sampling or cluster assignment. Two-way clustered standard errors are justified if there are two-way cluster sampling or two-way cluster assignments, or cluster sampling and cluster assignment on different dimensions. 
While the standard one-way CRVE from \citet{liang1986longitudinal} is conservative for the true variance under one-way clustering, the standard two-way variance estimator from \citet{cameron2011robust} is no longer conservative. Although a subsequent proposal from \citet{davezies2018asymptotic} is guaranteed to be conservative for two-way clustering, their variance estimator is often too large, so we provide a refinement. 
Our proposed refinement uses control variables, such as baseline characteristics, and ensures that the estimators remain valid for inference.
Evidence from our simulation and empirical illustration suggests that gains from our variance correction can be substantial.

Through a survey of when clustered standard errors are used in empirical work, we offer insights on the appropriateness of clustering in various contexts from our theory on M-estimation with clustering. 
The results apply straightforwardly to a difference-in-means estimator. 
In the context with one-way fixed effects, we find that the estimand is interpretable as an ATE only in special cases, but the estimand is still a weighted average of treatment effects in general.
With two-way fixed effects, the requirements for the estimand to be interpretable as an ATE is even more restrictive.
With spatiotemporal correlation, the magnitude of two-way clustering and spatiotemporal variance estimators cannot be ordered in general.
With two assignment variables clustered on different dimensions, we find that it suffices to apply one-way clustering on the respective dimensions rather than to use two-way clustering in certain cases.
In the estimation of triple differences, we find that the choice between one-way and two-way clustering depends on the nature of the variables in the triple interaction term.

The current paper focuses on the asymptotics as the number of clusters tends to infinity in the limit. For a small number of clusters or wildly unbalanced clusters, the wild cluster bootstrap\footnote{See, for example, \cite{cameron2008bootstrap} and \cite{mackinnon2017wild}.} has been proposed as a better-performing inference method for linear models in the setting of superpopulations. The finite population inference method for few heterogeneous clusters remains an interesting future research topic.

\appendix

\numberwithin{equation}{section}
\numberwithin{assumption}{section}

\section{Notation and Regularity Conditions}
The following notation provides details of the asymptotic variances and the variance estimator for functions of M-estimators: 
\begin{align*}
\Delta^f_{ehw,M}=&\frac{1}{M}\sum^M_{i=1}\mathbb{E}\Big\{\big[f_{iM}(W_{iM},\theta^*_M)-\gamma^*_M-F_M(\theta^*_M)L_M(\theta^*_M)^{-1}m_{iM}(W_{iM},\theta^*_M)\big]\cdot\\
&\big[f_{iM}(W_{iM},\theta^*_M)-\gamma^*_M-F_M(\theta^*_M)L_M(\theta^*_M)^{-1}m_{iM}(W_{iM},\theta^*_M)\big]'\Big\},
\tag{\stepcounter{equation}\theequation}
\end{align*}
\begin{equation}
\begin{aligned}
\Delta^f_{E,M}=&\frac{1}{M}\sum^M_{i=1}\Big\{\mathbb{E}\big[f_{iM}(W_{iM},\theta^*_M)-\gamma^*_M-F_M(\theta^*_M)L_M(\theta^*_M)^{-1}m_{iM}(W_{iM},\theta^*_M)\big]\cdot\\
&\mathbb{E}\big[f_{iM}(W_{iM},\theta^*_M)-\gamma^*_M-F_M(\theta^*_M)L_M(\theta^*_M)^{-1}m_{iM}(W_{iM},\theta^*_M)\big]'\Big\},
\end{aligned}
\end{equation}
\begin{equation}
\begin{aligned}
\Delta^f_{cluster,M}=&\frac{1}{M}\sum^{G}_{g=1}\sum_{i \in \mathcal{N}^G_g}\sum_{j \in \mathcal{N}^G_g \backslash \{i\}}\mathbb{E}\Big\{\big[f_{iM}(W_{iM},\theta^*_M)-\gamma^*_M-F_M(\theta^*_M)L_M(\theta^*_M)^{-1}m_{iM}(W_{iM},\theta^*_M)\big]\cdot\\
&\big[f_{jM}(W_{jM},\theta^*_M)-\gamma^*_M-F_M(\theta^*_M)L_M(\theta^*_M)^{-1}m_{jM}(W_{jM},\theta^*_M)\big]'\Big\},
\end{aligned}
\end{equation}
\begin{equation}
\begin{aligned}
\Delta^f_{EC,M}=&\frac{1}{M}\sum^{G}_{g=1}\sum_{i \in \mathcal{N}^G_g}\sum_{j \in \mathcal{N}^G_g \backslash \{i\}}\Big\{\mathbb{E}\big[f_{iM}(W_{iM},\theta^*_M)-\gamma^*_M-F_M(\theta^*_M)L_M(\theta^*_M)^{-1}m_{iM}(W_{iM},\theta^*_M)\big]\cdot\\
&\mathbb{E}\Big[f_{jM}(W_{jM},\theta^*_M)-\gamma^*_M-F_M(\theta^*_M)L_M(\theta^*_M)^{-1}m_{jM}(W_{jM},\theta^*_M)\big]'\Big\}
\end{aligned}
\end{equation}
with 
\begin{equation}
F_M(\theta)=\frac{1}{M}\sum^M_{i=1}\mathbb{E}\big[\nabla_\theta f_{iM}(W_{iM},\theta)\big].
\end{equation}
The terms $\Delta^f_{ehw,M}$ and $\Delta^f_{cluster,M}$ account for heteroskedasticity and within-cluster correlation, whereas $\Delta^f_{E,M}$ and $\Delta^f_{EC,M}$ are their finite population counterparts. 
The estimators are 
\begin{equation}
\begin{aligned}
\hat{\Delta}^f_{ehw,N}=&\frac{1}{N}\sum^M_{i=1}R_{iM}\big[f_{iM}(W_{iM},\hat{\theta}_N)-\hat{\gamma}_N-\hat{F}_N(\hat{\theta}_N)\hat{L}_N(\hat{\theta}_N)^{-1}m_{iM}(W_{iM},\hat{\theta}_N)\big]\cdot\\
&\big[f_{iM}(W_{iM},\hat{\theta}_N)-\hat{\gamma}_N-\hat{F}_N(\hat{\theta}_N)\hat{L}_N(\hat{\theta}_N)^{-1}m_{iM}(W_{iM},\hat{\theta}_N)\big]'
\end{aligned}
\end{equation}
and 
\begin{equation}
\begin{aligned}
\hat{\Delta}^f_{cluster,N}=&\frac{1}{N}\sum^{G}_{g=1}\sum_{i \in \mathcal{N}^G_g}\sum_{j \in \mathcal{N}^G_g \backslash \{i\}}R_{iM}R_{jM}\big[f_{iM}(W_{iM},\hat{\theta}_N)-\hat{\gamma}_N-\hat{F}_N(\hat{\theta}_N)\hat{L}_N(\hat{\theta}_N)^{-1}m_{iM}(W_{iM},\hat{\theta}_N)\big]\cdot\\
&\big[f_{jM}(W_{jM},\hat{\theta}_N)-\hat{\gamma}_N-\hat{F}_N(\hat{\theta}_N)\hat{L}_N(\hat{\theta}_N)^{-1}m_{jM}(W_{jM},\hat{\theta}_N)\big]',
\end{aligned}
\end{equation}
where
\begin{equation}
\hat{F}_N(\theta)=\frac{1}{N}\sum^M_{i=1}R_{iM}\nabla_\theta f_{iM}(W_{iM},\theta).
\end{equation}

\begin{definition}
The random function $g_{iM}(W_{iM},\theta)$ is said to be Lipschitz in the parameter $\theta$ on $\Theta$ if there is $h(u)\downarrow 0$ as $u\downarrow 0$ and $b(\cdot): \mathcal{W}\to R$ such that $\sup _{i,M}\mathbb{E}\big[|b_{iM}(W_{iM})|\big]<\infty$, and for all $\tilde{\theta},\theta\in\Theta$, $\big|g_{iM}(W_{iM},\tilde{\theta})-g_{iM}(W_{iM},\theta)\big|\leq b_{iM}(W_{iM}) h(\|\tilde{\theta}-\theta\|)$, $\forall\ i,M$.
\end{definition}
 
We impose the following regularity conditions for the theorems in the paper. 
\begin{assumption}\label{assumpa1}
Suppose that $\frac{1}{N}\sum\limits^M_{i=1}R_{iM}\cdot m_{iM}(W_{iM},\hat{\theta}_N)=o_p(N^{-1/2})$ and \\
(\romannumeral 1) Let $Q_M(\theta):=\frac{1}{M}\sum^M_{i=1}\mathbb{E}\big[q_{iM}(W_{iM},\theta)\big]$. $\{Q_M(\theta)\}$ has identifiably unique minimizers $\{\theta_M^*\}$ on $\Theta$ as in Definition 3.2 in \cite{gallant1988unified}; \\
(\romannumeral 2) $\Theta$ is compact; \\
(\romannumeral 3) $\theta_M^*\in int(\Theta)$ uniformly in $M$; \\
(\romannumeral 4) $q_{iM}(w,\theta)$ is twice continuously differentiable on $int(\Theta)$ for all $w$ in the support of $W_{iM}$, $\forall\ i,M$; \\
(\romannumeral 5) $\sup\limits_{i,M}\mathbb{E}\Big[\sup\limits_{\theta\in\Theta}|q_{iM}(W_{iM},\theta)|^r\Big]<\infty$ for some $r>1$; \\
(\romannumeral 6) $q_{iM}(W_{iM},\theta)$ is Lipschitz in $\theta$ on $\Theta$; \\
(\romannumeral 7) $\sup\limits_{i,M}\mathbb{E}\Big[\sup\limits_{\theta\in\Theta}\left\|m_{iM}(W_{iM},\theta)\right\|^r\Big]<\infty$ for some $r>2$; \\
(\romannumeral 8) $\lambda_{min}(V_{\Delta M})>0$, where $\lambda_{min}(\cdot)$ stands for the smallest eigenvalue; \\
(\romannumeral 9) $\sup\limits_{i,M}\mathbb{E}\Big[\sup\limits_{\theta\in\Theta}\left\|\nabla_\theta m_{iM}(W_{iM},\theta)\right\|^r\Big]<\infty$ for some $r>1$; \\
(\romannumeral 10) $\nabla_\theta m_{iM}(W_{iM},\theta)$ is Lipschitz in $\theta$ on $\Theta$; \\
(\romannumeral 11) $L_M(\theta^*_M)$ is nonsingular; \\
(\romannumeral 12) there is $h(u)\downarrow 0$ as $u\downarrow 0$ and $b_1(\cdot):\mathcal{W}\to R$ such that $\sup\limits_{i,M}\mathbb{E}\big[b_{1,iM}(W_{iM})^2\big]<\infty$, and for all $\tilde{\theta}, \theta \in \Theta$, $\left\|m_{iM}(W_{iM},\tilde{\theta})-m_{iM}(W_{iM},\theta)\right\|\leq b_{1,iM}(W_{iM}) h(\|\tilde{\theta}-\theta\|)$. 
\end{assumption}

\begin{assumption}\label{assumpa2}
Suppose that \\
(\romannumeral 1) $f_{iM}(w,\theta)$ is continuously differentiable on $int(\Theta)$ for all $w$ in the support of $W_{iM}$, $\forall\ i,M$; \\
(\romannumeral 2) $\sup\limits_{i,M}\mathbb{E}\Big[\sup\limits_{\theta\in\Theta}\left\|f_{iM}(W_{iM},\theta)\right\|^r\Big]<\infty$ for some $r>2$; \\
(\romannumeral 3) $\lambda_{min}(V_{f,M})>0$; \\
(\romannumeral 4) $\sup\limits_{i,M}\mathbb{E}\Big[\sup\limits_{\theta\in\Theta}\left\|\nabla_\theta f_{iM}(W_{iM},\theta)\right\|^r\Big]<\infty$ for some $r>1$; \\
(\romannumeral 5) $\nabla_\theta f_{iM}(W_{iM},\theta)$ is Lipschitz in $\theta$ on $\Theta$; \\
(\romannumeral 6) there is $h(u)\downarrow 0$ as $u\downarrow 0$ and $b_2(\cdot): \mathcal{W}\to R$ such that $\sup\limits_{i,M}\mathbb{E}\big[b_{2,iM}(W_{iM})^2\big]<\infty$, and for all $\tilde{\theta}, \theta\in\Theta$, $\left\|f_{iM}(W_{iM},\tilde{\theta})-f_{iM}(W_{iM},\theta)\right\|\leq b_{2,iM}(W_{iM})h(\|\tilde{\theta}-\theta\|)$.
\end{assumption}

\begin{assumption}\label{assumpa3}
Suppose that 
we maintain conditions (\romannumeral 1)-(\romannumeral 4), (\romannumeral 6), (\romannumeral 10)-(\romannumeral 12) in Assumption \ref{assumpa1}. In addition, \\
(\romannumeral 5) $\sup\limits_{i,M}\mathbb{E}\Big[\sup\limits_{\theta\in\Theta}|q_{iM}(W_{iM},\theta)|^2\Big]<\infty$; \\
(\romannumeral 7) $\sup\limits_{i,M}\mathbb{E}\Big[\sup\limits_{\theta\in\Theta}\left\|m_{iM}(W_{iM},\theta)\right\|^4\Big]<\infty$; \\
(\romannumeral 8) $\lambda_{min}(V_{\Delta TWM})>0$; \\
(\romannumeral 9) $\sup\limits_{i,M}\mathbb{E}\Big[\sup\limits_{\theta\in\Theta}\left\|\nabla_\theta m_{iM}(W_{iM},\theta)\right\|^2\Big]<\infty$.
\end{assumption}

\section{Supplementary Material}
\bigskip
\noindent
\textbf{Example B.1:}
This section theoretically constructs a counterexample such that the CGM estimator is anticonservative in the setting of Section \ref{sec:diff_means}. The CGM estimand is $V_{TWSM}$, so using our framework, $V_{TWSM} - v_M = (N/ M^2) \sum_i \sum_{j \in \mathcal{N}_i} \mathbb{E}[\eta_{iM}] \mathbb{E}[\eta_{jM}]$, and we have shown that $\mathbb{E}[\eta_{iM}] = \tau_{iM} -\tau_{M}$. Hence, for a counterexample where $V_{CGM} - v_k <0$, it suffices that $\sum_i \sum_{j \in \mathcal{N}_i} (\tau_{iM}- \tau_{M}) (\tau_{jM} - \tau_{M}) <0$.

Let $k_1$ denote an odd number, and $(g,h)$ describe a cluster intersection that is in cluster $g$ on the $G$ dimension and in cluster $h$ on the $H$ dimension. Suppose there are $G$ clusters on both $G$ and $H$ dimensions, where $G$ is even. Let $G_0$ be a fixed number.
In the population, individuals can only belong to cluster intersections of the form $(k_1,k_1)$, $(k_1, k_1 \pm 1)$ and $(k_1 \pm 1, k_1)$. This assumption implies that there cannot be individuals in cluster intersections where both cluster indexes are even, or if their difference is more than one. 
There are $4G_0$ observations in $(k_1,k_1)$ clusters and $G_0$ observations in other cluster intersections that are nonempty. Hence, the total number of observations is $M = 8 G_0G$. 

\begin{table}[H]
    \centering
    \caption{Distribution of population with bounded cluster sizes} \label{tab:tau_distr}
    \begin{tabular}{c|ccc}
Type & Proportion & $(g,h)$ & $\tau_{iM}-\tau_{M}$  \\
\midrule
1 & 1/2  &  $(k_1, k_1)$  &  1 \\
2 & 1/4 & $(k_1, k_1 \pm 1)$ & -1 \\
3 & 1/4 & $(k_1 \pm 1, k_1)$ & -1 \\
\hline
\end{tabular}
\end{table}

The $\tau_{iM}-\tau_{M}$ values for observations belonging to the various cluster intersections are given by Table \ref{tab:tau_distr}. Then, we show how this particular construction results in $\sum_i \sum_{j \in \mathcal{N}_i} (\tau_{iM}- \tau_{M}) (\tau_{jM} - \tau_{M}) <0$.
First, consider an observation $i$ in $(k_1,k_1)$ --- they account for half the observations. Then, $|\mathcal{N}_i|=8G_0$, where $4G_0$ of those observations are in $(k_1,k_1)$ and the remaining $4G_0$ are either in $(k_1\pm1,k_1)$ or $(k_1,k_1\pm1)$. Hence, $\sum_{j \in \mathcal{N}_i} (\tau_{iM}- \tau_{M}) (\tau_{jM} - \tau_{M}) = (1) (4-1-1-1-1) G_0 =0$. 
Next, consider an observation $i$ in $(k_1,k_1\pm1)$. The treatment of $(k_1 \pm 1,k_1)$ is identical. The units in either $(k_1,k_1 \pm 1)$ or $(k_1 \pm1,k_1)$ account for the other half of the units. Here, $|\mathcal{N}_i| = 7G_0$. For instance, for some unit $i$ in $(k_1,k_1+1)$, $4G_0$ of $|\mathcal{N}_i|$ are in $(k_1,k_1)$ intersections, $G_0$ are in $(k_1,k_1+1)$, $G_0$ are in $(k_1,k_1-1)$ and $G_0$ are in $(k_1+2,k_1+1)$. Then, $\sum_{j \in \mathcal{N}_i} (\tau_{iM}- \tau_{M}) (\tau_{jM} - \tau_{M}) = (1) (4-1-1-1) G_0 =-1$.
Combining these results,
\begin{align*}
    \frac{1}{M} \sum_i \sum_{j \in \mathcal{N}_i} (\tau_{iM}- \tau_{M}) (\tau_{jM} - \tau_{M}) &= \frac{1}{2} (1) + \frac{1}{2} (-1) = -1/2 <0.
\end{align*}

\bigskip
\noindent
\textbf{Derivation for Table 1:}

It suffices to work with the estimand, because subsequent theorems show consistency. 
The original cluster variance ($V_{SM} = \Delta_{ehw,M} + \rho_{uM} \Delta_{cluster,M}$) over-estimates the true variance by:
\begin{equation}
V_{SM} - V_{\Delta M} = \rho_{uM} \rho_{gM} \Delta_{E,M} + \rho_{uM} \rho_{gM} \Delta_{EC,M}
\end{equation}

First, consider Case 1. Without cluster assignment, $\Delta_{EC,M} = \Delta_{cluster,M}$. Since $\hat{\Delta}_{cluster,N}$ consistently estimates $\rho_{uM} \Delta_{cluster,M}$, $\rho_{uM} \rho_{gM}$ can be estimated by $N/M$, and $\rho_{gM}$ can be estimated by $G_N/ G$, the proposed adjustment is:
\begin{equation}
\left( \frac{G_N}{G}\hat{\Delta}_{cluster,N} +  \frac{N}{M} \hat{\Delta}^Z_N \right)
\end{equation}
Then, the proposed estimator is:
\begin{equation}
\hat{\Delta}_{ehw,N} +\hat{\Delta}_{cluster,N} - \left( \frac{G_N}{G}\hat{\Delta}_{cluster,N} +  \frac{N}{M} \hat{\Delta}^Z_N \right)
\end{equation}
which is of the form required. 

In Case 2 and Case 3, since $\Delta_{EC,M} \ne \Delta_{cluster,M}$ in general, we can correct for both term simultaneously. The proposed adjustment is hence $(N/M) \hat{\Delta}^Z_{CE,N}$. 

In Case 4, $\rho_{gM} =1$ without clustered sampling and $\Delta_{EC,M} = \Delta_{cluster,M}$ without clustered assignment so that $V_{\Delta M} = \Delta_{ehw,M} - \rho_{uM} \Delta_{EC, M}$. Hence, the proposed variance is as stated. 

\bigskip
\noindent
\textbf{Derivation for Table 2:}

In this derivation, we retain the notation with $\rho_{uM}, \rho_{gM}, \rho_{hM}$ to accommodate both the case where we observe the entire population and the case where sampling is allowed with unbounded cluster sizes. 

Taking the meat of the $V_{TWSM2}$ expression, the estimand of $2\hat{\Delta}_{ehw,N} + \hat{\Delta}_{G,N} + \hat{\Delta}_{H,N}$ is
\begin{equation}
V_{\Delta,CGM2} = 2\Delta_{ehw,M}(\theta^*_M)+2\rho_{uM} \Delta_{(G\cap H),M}(\theta^*_M) +\rho_{uM}\rho_{hM}\Delta_{G,M}(\theta^*_M)+\rho_{uM}\rho_{gM}\Delta_{H,M}(\theta^*_M).
\end{equation}

Hence, the CGM2 estimand over-estimates the true variance by:
\begin{equation}
\begin{aligned}
V_{\Delta,CGM2} - V_{\Delta TWM} &= \Delta_{ehw,M}+\rho_{uM} \Delta_{(G\cap H),M} \\
&+\rho_{uM}\rho_{gM}\rho_{hM}\Delta_{E,M}+\rho_{uM} \rho_{gM}\rho_{hM} \Delta_{E(G\cap H),M}\\
&+\rho_{uM}\rho_{gM}\rho_{hM}\Delta_{EG,M}+\rho_{uM}\rho_{gM}\rho_{hM}\Delta_{EH,M}
\end{aligned}
\end{equation}

As before, $\rho_{uM}\rho_{gM}\rho_{hM}$ can be estimated using $N/M$. Since $\rho_{gM}$ and $\rho_{hM}$ can be estimated using $G_N/G$ and $H_N/H$ respectively, $\rho_{uM}$ can be estimated using $\frac{N}{M}\frac{G}{G_N}\frac{H}{H_N}$.

Consider Case 1. With only clustered sampling on both dimensions, $\hat{\Delta}_{G,N}$ consistently estimates $\rho_{uM}\rho_{hM} \Delta_{G,M} + \rho_{uM} \Delta_{(G\cap H), M}$ and $\hat{\Delta}_{H,N}$ consistently estimates $\rho_{uM} \rho_{gM}\Delta_{H,M} + \rho_{uM} \Delta_{(G\cap H), M}$. 
For the terms to appear, the estimation for $\Delta_{ehw,M}+\rho_{uM} \Delta_{(G\cap H),M}$ uses $\hat{\Delta}_{ehw,N} + \hat{\Delta}_{(G\cap H),N}$. 
We estimate $\rho_{uM} \rho_{gM}\rho_{hM} \Delta_{E(G\cap H),M}$ using $ \frac{G_N}{G} \frac{H_N}{H}\hat{\Delta}_{(G\cap H),N} $. 
We estimate $\rho_{uM}\rho_{gM}\rho_{hM}\Delta_{EG,M}$ using $\frac{G_N}{G} (\hat{\Delta}_{G,N} - \hat{\Delta}_{(G\cap H), N})$ and $\rho_{uM}\rho_{gM}\rho_{hM}\Delta_{EH,M}$ using $\frac{H_N}{H} (\hat{\Delta}_{H,N} - \hat{\Delta}_{(G\cap H), N})$. We adjust for $\rho_{uM}\rho_{gM}\rho_{hM}\Delta_{E,M}$ using $N/M \hat{\Delta}^Z_N$.
Hence, the estimator is:
\begin{equation}
\begin{aligned}
2\hat{\Delta}_{ehw,N} + \hat{\Delta}_{G,N} + \hat{\Delta}_{H,N} - \hat{\Delta}_{ehw,N} - \hat{\Delta}_{(G\cap H),N} - N/M \hat{\Delta}^Z_N  \\ 
- \frac{G_N}{G} \frac{H_N}{H}\hat{\Delta}_{(G\cap H),N} - \frac{G_N}{G}(\hat{\Delta}_{G,N} - \hat{\Delta}_{(G\cap H), N})- \frac{H_N}{H} (\hat{\Delta}_{H,N} - \hat{\Delta}_{(G\cap H), N}),
\end{aligned}
\end{equation}
which simplifies to the expression in Table 2.

Consider Case 2. We use ${\Delta}^Z_{HE,M}$ to correct for $\Delta_{E,M} + \Delta_{E(G\cap H),M} + \Delta_{EH,M}$. We can use $N/M\cdot \Delta^Z_{GE,M}$ to correct for $\Delta_{ehw,M} + \rho_{uM}\Delta_{(G\cap H),M} + \rho_{uM}\rho_{gM}\rho_{hM}\Delta_{EG,M}$. Since $\Delta_{ehw,M} + \rho_{uM}\Delta_{(G\cap H),M} \geq \rho_{uM}\rho_{gM}\rho_{hM}\Delta_{E,M} + \rho_{uM}\rho_{gM}\rho_{hM}\Delta_{E(G\cap H),M}$,
\begin{equation}
\begin{aligned}
& \Delta_{ehw,M} + \rho_{uM} \Delta_{(G\cap H),M} + \rho_{uM}\rho_{gM}\rho_{hM}\Delta_{EG,M} - \rho_{uM}\rho_{gM}\rho_{hM} \Delta^Z_{GE,M}  \\
\geq & \rho_{uM}\rho_{gM}\rho_{hM} (\Delta_{E,M} + \Delta_{E(G\cap H),M} + \Delta_{EG,M} -\Delta^Z_{GE,M} )  \geq 0,
\end{aligned}
\end{equation}
so our adjustment of $N/M \cdot ( \hat{\Delta}^Z_{GE,N} + \hat{\Delta}^Z_{HE,N})$ still makes the variance estimator conservative. 

\section{Proofs}
In the following proofs, $C$ denotes a generic positive constant that may be different under different circumstances.

\subsection{Proofs for Section 2}

\begin{lemma} \label{lem:wlln}
Suppose Assumption \ref{assump4} holds. For any scalar two-way clustered random variable $V_{iM}$ with $\E[V_{iM}]=0$ and $\E[V_{iM}^2] \leq C$, $\frac{1}{M} \sum_{i=1}^M V_{iM} \xrightarrow{p} 0$.
\end{lemma}
\noindent
\textbf{Proof:}
By Chebyshev's inequality, for any $\epsilon >0$,
\begin{equation}
\begin{aligned}
\mathbb{P} \left( \left( \frac{\sum_{i=1}^M V_{iM}}{M} \right)^2 >\epsilon \right) &\leq \frac{1}{\epsilon^2 M^2} \E \left[ \left( \sum_{i=1}^M V_{iM} \right)^2 \right] \\
& \leq C\frac{\sum_g (M^G_g)^2 + \sum_h (M^H_h)^2}{M^2} = o(1).
\end{aligned}
\end{equation}
The second inequality follows from bounded variances and the final $o(1)$ equality follows from Assumption \ref{assump4}.

\begin{lemma}\label{lemma1}
Under Assumptions \ref{assump1} and \ref{assump4}, $\frac{N}{M\rho_{uM}\rho_{gM}\rho_{hM}}\overset{p}\to 1$.
\end{lemma}

\noindent
\textbf{Proof:}
Observe that:
\begin{equation}
\frac{N}{M\rho_{uM}\rho_{gM}\rho_{hM}} - 1 = \frac{\sum_{i=1}^M \left( R_{iM} - E[R_{iM}]\right)}{M\rho_{uM}\rho_{gM}\rho_{hM}}.
\end{equation}
Since $\rho_{uM}, \rho_{gM}, \rho_{hM}$ are nonzero by Assumption \ref{assump1}, Lemma \ref{lem:wlln} yields the result.

\begin{lemma}\label{lemma2}
Under Assumptions \ref{assump1} and \ref{assump3}, suppose (\romannumeral 1) $a_{iM}(W_{iM},\theta)$ is Lipschitz in $\theta$ on $\Theta$; 
(\romannumeral 2) $\sup_{i,M}\mathbb{E}\big[\sup _{\theta\in\Theta}\|a_{iM}(W_{iM},\theta)\|^r\big]<\infty$ for some $r>1$. Then (1) Let $A_N(\theta) := \frac{1}{N}\sum_{i=1}^M R_{iM}a_{iM}(W_{iM},\theta)$. $\left\|A_N(\tilde{\theta})-A_N(\theta)\right\|\leq B_N h(\|\tilde{\theta}-\theta\|)$, where $B_N:= \frac{1}{N}\sum_{i=1}^M R_{iM}\cdot b_{iM}(W_{iM})=O_p(1)$ for $b_{iM}(\cdot)$ in Definition 1;
and (2) $A_M(\theta) =\frac{1}{M}\sum_{i=1}^M\mathbb{E}\big[a_{iM}(W_{iM},\theta)\big]$ is uniformly equicontinuous.
\end{lemma}

\noindent
\textbf{Proof:}

We first show result (1).
\begin{align*}
&\left\|A_N(\tilde{\theta})-A_N(\theta)\right\|\\
=&\left\|\frac{1}{N}\sum^M_{i=1}R_{iM}\big[a_{iM}(W_{iM},\tilde{\theta})-a_{iM}(W_{iM},\theta)\big]\right\|\\
\leq&\frac{1}{N}\sum^M_{i=1}R_{iM}\left\|a_{iM}(W_{iM},\tilde{\theta})-a_{iM}(W_{iM},\theta)\right\|\\
\leq&\frac{1}{N}\sum^M_{i=1}R_{iM}\cdot b_{iM}(W_{iM})h(\|\tilde{\theta}-\theta\|)\\
=&B_N h(\|\tilde{\theta}-\theta\|) \tag{\stepcounter{equation}\theequation}
\end{align*}
\begin{equation}
B_N:=\frac{1}{N}\sum^M_{i=1}R_{iM}\cdot b_{iM}(W_{iM})=\frac{M\rho_{uM}\rho_{gM}\rho_{hM}}{N}\frac{1}{M}\sum^M_{i=1}\frac{R_{iM}}{\rho_{uM}\rho_{gM}\rho_{hM}}b_{iM}(W_{iM})
\end{equation}
Because of Lemma \ref{lemma1} and the continuous mapping theorem, $\frac{M\rho_{uM}\rho_{gM}\rho_{hM}}{N}\overset{p}\to 1$. As a result, it is sufficient to prove $\frac{1}{M}\sum\limits^M_{i=1}\frac{R_{iM}}{\rho_{uM}\rho_{gM}\rho_{hM}}b_{iM}(W_{iM})= O_p(1)$.
For all $\epsilon>0$, let $b_\epsilon=C/\epsilon$ for some $C<\infty$,
\begin{equation}
\begin{aligned}
&\mathbb{P} \bigg(\bigg|\frac{1}{M}\sum^M_{i=1}\frac{R_{iM}}{\rho_{uM}\rho_{gM}\rho_{hM}}b_{iM}(W_{iM})\bigg|\geq b_\epsilon\bigg)\\
\leq&\mathbb{E}\bigg(\bigg|\frac{1}{M}\sum^M_{i=1}\frac{R_{iM}}{\rho_{uM}\rho_{gM}\rho_{hM}}b_{iM}(W_{iM})\bigg|\bigg)/b_\epsilon\\
\leq&\frac{1}{M}\sum^M_{i=1}\mathbb{E}\Big(\frac{R_{iM}}{\rho_{uM}\rho_{gM}\rho_{hM}}\Big)\mathbb{E}\Big[\big|b_{iM}(W_{iM})\big|\Big]/b_\epsilon\\
\leq &\sup\limits_{i,M}\mathbb{E}\Big[\big|b_{iM}(W_{iM})\big|\Big]/b_\epsilon<C/b_\epsilon= \epsilon.
\end{aligned}
\end{equation}
Hence, $B_N=O_p(1)$.

Next, we show $\{A_M(\theta)\}$ is uniformly equicontinuous. The proof is based on slight modification of the proof of Theorem 2 in \cite{jenish2009central}.
\begin{align*}
&\sup_{\theta\in\Theta}\sup_{\tilde{\theta}\in B(\theta,\delta)}\norm{A_M(\tilde{\theta})-A_M(\theta)}\\
\leq&\frac{1}{M}\sum_{i=1}^M\sup_{\theta\in\Theta}\sup_{\tilde{\theta}\in B(\theta,\delta)}\norm{\mathbb{E}\big[a_{iM}(W_{iM},\tilde{\theta})-a_{iM}(W_{iM},\theta)\big]}\\
\leq&\frac{1}{M}\sum_{i=1}^M\mathbb{E}\bigg[\sup_{\theta\in\Theta}\sup_{\tilde{\theta}\in B(\theta,\delta)}\norm{a_{iM}(W_{iM},\tilde{\theta})-a_{iM}(W_{iM},\theta)}\bigg]\\
=&\frac{1}{M}\sum_{i=1}^M \mathbb{E} (Y_{iM}(\delta)),
\tag{\stepcounter{equation}\theequation}
\end{align*}
where $Y_{iM}(\delta) := \sup_{\theta\in\Theta}\sup_{\tilde{\theta}\in B(\theta,\delta)}\|a_{iM}(W_{iM},\tilde{\theta})-a_{iM}(W_{iM},\theta)\|$.\\
Define $l_{iM}=\sup_{\theta\in\Theta}\|a_{iM}(W_{iM},\theta)\|$.
Given condition (\romannumeral 2), there exists $k=k(\epsilon)<\infty$ for some $\epsilon>0$ such that 
\begin{equation}
\limsup_{M\to\infty} \frac{1}{M}\sum_{i=1}^M\mathbb{E}\big[l_{iM}\mathbbm{1}(l_{iM}>k)\big]<\frac{\epsilon}{6}.
\end{equation}
Under condition (\romannumeral 1), $a_{iM}(W_{iM},\theta)$ is $L_0$ stochastically equicontinuous on $\Theta$ by Proposition 1 in \cite{jenish2009central}.
Hence, we can find some $\delta=\delta(\epsilon)$ such that 
\begin{equation}
\limsup_{M\to\infty}\frac{1}{M}\sum_{i=1}^M P(Y_{iM}(\delta)>\epsilon/3)\leq \frac{\epsilon}{6k}.
\end{equation}
\begin{align*}
&\limsup_{M\to\infty}\frac{1}{M}\sum_{i=1}^M \mathbb{E}(Y_{iM}(\delta))\\
\leq &\epsilon/3+\limsup_{M\to\infty}\frac{1}{M}\sum_{i=1}^M \mathbb{E}\big[Y_{iM}(\delta)\mathbbm{1}(Y_{iM}(\delta)>\epsilon/3,l_{iM}>k)\big]\\
&+\limsup_{M\to\infty}\frac{1}{M}\sum_{i=1}^M \mathbb{E}\big[Y_{iM}(\delta)\mathbbm{1}(Y_{iM}(\delta)>\epsilon/3,l_{iM}\leq k)\big]\\
\leq & \epsilon/3+2\limsup_{M\to\infty}\frac{1}{M}\sum_{i=1}^M\mathbb{E}\big[l_{iM}\mathbbm{1}(l_{iM}>k)\big]+2k\frac{1}{M}\sum_{i=1}^M\limsup_{M\to\infty}P(Y_{iM}(\delta)>\epsilon/3) =\epsilon
\tag{\stepcounter{equation}\theequation}
\end{align*}
As a result, $\limsup_{M\to\infty}\sup_{\theta\in\Theta}\sup_{\tilde{\theta}\in B(\theta,\delta)}\norm{A_M(\tilde{\theta})-A_M(\theta)}\to 0$ as $\delta\to 0$.

\bigskip
To show the consistency of variance estimators, we use an intermediate lemma, which is the analog of \citet{hansen2019asymptotic} (62) under two-way clustering.
\begin{lemma} \label{lem:HL19_62}
Suppose Assumptions \ref{assump1}-\ref{assump4} hold.
Additionally, assume that for all $\theta$, $\E \left[ \| f(X_i,\theta) \|^4 \right]$ is bounded, 
and that for $C, C^{\prime} \in \{ G,H \}$, and $\lambda^C_M := \lambda_{min} \left(\sum_{i=1}^{M} \sum_{j \in \mathcal{N}^{C}_{c(i)}} \mathbb{E}[ f\left(X_{i},\theta\right) f\left(X_{j},\theta\right)'] \right)$, we have $(\lambda^C_M)^{-1} \max_c (M^{C^\prime}_c)^2 = o(1)$ and $(\lambda^C_M)^{-1} \sum_c (M^{C^\prime}_c)^2 = O(1)$.
Let $\tilde{f}_{g}(\theta) :=\sum_{j\in\mathcal{N}^G_{g}}f\left(X_{j},\theta\right)$ and $\tilde{\Omega}_M(\theta)=\sum_{g=1}^G\tilde{f}_{g}(\theta)\tilde{f}_{g}(\theta)^{\prime}$. 
Then,
\[
 \left\Vert [\mathbb{E}\tilde{\Omega}_{M}(\theta)]^{-1} [ \tilde{\Omega}_{M}(\theta)-\mathbb{E}\tilde{\Omega}_{M}(\theta) ] \right\Vert \xrightarrow{p}0.
\]
\end{lemma}

\noindent
\textbf{Proof:}

The strategy follows Hansen and Lee (2019) (62) in the proof of their Theorem 6. We can suppress dependence on $\theta$ for notational convenience. Fix any $\theta$ and $\delta>0$. Set $\varepsilon=\left(\delta/C\right)^{2}$. Define
\begin{equation}
\tilde{r}_{g}=\tilde{f}_{g}1\left(||\tilde{f}_{g}||\leq\sqrt{M\varepsilon}\right).
\end{equation}
Let $\lambda^G_M$ denote the rate such that $\frac{1}{\lambda^G_M} \| \E \tilde{\Omega}_M (\theta) \|$ is $O(1)$ but not $o(1)$. Then, 
\begin{align*}
\left\Vert [\mathbb{E}\tilde{\Omega}_{M}(\theta)]^{-1} [ \tilde{\Omega}_{M}(\theta)-\mathbb{E}\tilde{\Omega}_{M}(\theta) ] \right\Vert &\leq \left\Vert [\mathbb{E}\tilde{\Omega}_{M}(\theta)]^{-1} \right\Vert \left\Vert[ \tilde{\Omega}_{M}(\theta)-\mathbb{E}\tilde{\Omega}_{M}(\theta) ] \right\Vert \\
\leq C (\lambda^G_M)^{-1} \left\Vert[ \tilde{\Omega}_{M}(\theta)-\mathbb{E}\tilde{\Omega}_{M}(\theta) ] \right\Vert
\end{align*}
\begin{equation}
\begin{aligned}
&(\lambda^G_M)^{-1} \left\| \tilde{\Omega}_{M}(\theta)-\mathbb{E}\tilde{\Omega}_{M}(\theta) \right\| \\
\leq &(\lambda^G_M)^{-1}\mathbb{E}\left\Vert \sum_{g=1}^G\left(\tilde{r}_{g}\tilde{r}_{g}^{\prime}-\mathbb{E}\tilde{r}_{g}\tilde{r}_{g}^{\prime}\right)\right\Vert  + 2(\lambda^G_M)^{-1}\sum_{g=1}^G\mathbb{E}\left(\left\Vert \tilde{f}_{g}\right\Vert ^{2}1\left(\left\Vert \tilde{f}_{g}\right\Vert >\sqrt{M\varepsilon}\right)\right).
\end{aligned}
\end{equation}

Since we have finite moments, for some $C<\infty$, $\mathbb{E}\left\Vert \tilde{f}_{g}\right\Vert ^{2}\leq C\left(M_{g}^{G}\right)^{2}$ by the Cr inequality. Using Jensen's inequality,
\begin{equation}
(\lambda^G_M)^{-1} \mathbb{E}\left\Vert \sum_{g=1}^G\left(\tilde{r}_{g}\tilde{r}_{g}^{\prime}-\mathbb{E}\tilde{r}_{g}\tilde{r}_{g}^{\prime}\right)\right\Vert \leq\left((\lambda^G_M)^{-2}\mathbb{E}\left\Vert \sum_{g=1}^G\left(\tilde{r}_{g}\tilde{r}_{g}^{\prime}-\mathbb{E}\tilde{r}_{g}\tilde{r}_{g}^{\prime}\right)\right\Vert ^{2}\right)^{1/2}.
\end{equation}
The argument that $(\lambda^G_M)^{-2}\mathbb{E}\left\Vert \sum_{g=1}^G\left(\tilde{r}_{g}\tilde{r}_{g}^{\prime}-\mathbb{E}\tilde{r}_{g}\tilde{r}_{g}^{\prime}\right)\right\Vert ^{2}\leq4\delta$
is similar to the proof strategy of variance consistency in Yap (2023)
Lemma 7. To be precise, due to finite moments, and using $A_{ij}$
to denote an adjacency indicator of whether $i,j$ are dependent,
\begin{equation}
\mathbb{E}\left\Vert \sum_{g=1}^G\left(\tilde{r}_{g}\tilde{r}_{g}^{\prime}-\mathbb{E}\tilde{r}_{g}\tilde{r}_{g}^{\prime}\right)\right\Vert ^{2}\leq C\sum_{g=1}^G\sum_{g^{\prime}}\sum_{i,j\in\mathcal{N}_{g}^{G}}\sum_{k,l\in\mathcal{N}_{g^{\prime}}^{G}}\left(A_{ik}+A_{il}+A_{jk}+A_{jl}\right)
\end{equation}

There are only four adjacency terms: if $\left(i,j\right)$ are independent
of $\left(k,l\right)$, then the demeaning would have removed the
relevant terms. Hence, an additional correlation can only exist if
units are correlated across the $g$ clusters. We make the argument
for the $A_{ik}$ term as the other terms are analogous. 
\begin{align*}
\sum_{g=1}^G\sum_{g^{\prime}}\sum_{i,j\in\mathcal{N}_{g}^{G}}\sum_{k,l\in\mathcal{N}_{g^{\prime}}^{G}}A_{ik} & =\sum_{i=1}^M\sum_{k}\sum_{j\in\mathcal{N}_{g(i)}^{G}}\sum_{l\in\mathcal{N}_{g(k)}^{G}}A_{ik}\\
 & \leq\max_{g}\left(M_{g}^{G}\right)^{2}\sum_{i=1}^M\sum_{k\in\mathcal{N}_{i}}A_{ik}\\
 & \leq\max_{g}\left(M_{g}^{G}\right)^{2}\sum_{i=1}^M\left(\sum_{k\in\mathcal{N}_{g(i)}^{G}}+\sum_{k\in\mathcal{N}_{h(i)}^{H}}\right)A_{ik}\\
 & \leq\max_{g}\left(M_{g}^{G}\right)^{2}\left(\sum_{g=1}^G\left(M_{g}^{G}\right)^{2}+\sum_{h=1}^H\left(M_{h}^{H}\right)^{2}\right) 
 \tag{\stepcounter{equation}\theequation}
\end{align*}
Then,
\begin{align*}
(\lambda^G_M)^{-2} \sum_{g=1}^G\sum_{g^{\prime}}\sum_{i,j\in\mathcal{N}_{g}^{G}}\sum_{k,l\in\mathcal{N}_{g^{\prime}}^{G}}A_{ik} & \leq (\lambda^G_M)^{-2} \max_{g}\left(M_{g}^{G}\right)^{2}\left(\sum_{g=1}^G\left(M_{g}^{G}\right)^{2}+\sum_{h=1}^H\left(M_{h}^{H}\right)^{2}\right).
\tag{\stepcounter{equation}\theequation}
\end{align*}
By assumption in the lemma,
$(\lambda^G_M)^{-1} \max_{g}\left(M_{g}^{G}\right)^{2} = o(1)$, $(\lambda^G_M)^{-1} \sum_{g=1}^G\left(M_{g}^{G}\right)^{2} = O(1) $, and $(\lambda^G_M)^{-1} \sum_{h=1}^H\left(M_{h}^{H}\right)^{2} = O(1)$. Hence, we can pick $M$ large enough so that $(\lambda^G_M)^{-2} \max_g \left(M_{g}^{G}\right)^{2}\left(2C\right)\leq\delta$. 

Next, consider $2 \left( \lambda^G_M \right)^{-1} \sum_{g=1}^G\mathbb{E}\left(\left\Vert \tilde{f}_{g}\right\Vert ^{2}1\left(\left\Vert \tilde{f}_{g}\right\Vert >\sqrt{M\varepsilon}\right)\right)$.
Lemma 1 of \citet{hansen2019asymptotic} implies that $\left\Vert \left(M^G_g\right)^{-1}\tilde{f}_{g}\right\Vert ^{2}$
is uniformly integrable. Their lemma can be applied because it holds
regardless of the covariance structure. This means we can pick $B$
sufficiently large so that:
\begin{equation}
\sup_{g}\mathbb{E}\left(\left\Vert \left(M^G_g\right)^{-1}\tilde{f}_{g}\right\Vert ^{2}1\left(\left\Vert \left(M^G_g\right)^{-1}\tilde{f}_{g}\right\Vert >B\right)\right)\leq\frac{\delta}{C}
\end{equation}

Pick $M$ large enough so that 
\begin{equation}
\left\|(\lambda^G_M)^{-1/2}\max_{g}M^G_g \right\|\leq\frac{\sqrt{\varepsilon}}{B}.
\end{equation}
Then
\begin{align*}
2 \left( \lambda^G_M \right)^{-1} \sum_{g=1}^G\mathbb{E}\left(\left\Vert \tilde{f}_{g}\right\Vert ^{2}1\left(\left\Vert \tilde{f}_{g}\right\Vert >\sqrt{M\varepsilon}\right)\right) & \leq
2 \left( \lambda^G_M \right)^{-1} \sum_{g=1}^G\mathbb{E}\left(\left\Vert \tilde{f}_{g}\right\Vert ^{2}1\left(\left\Vert \left(M^G_{g}\right)^{-1}\tilde{f}_{g}\right\Vert >B\right)\right)\\
 & \leq 2 \left( \lambda^G_M \right)^{-1}\sum_{g=1}^G\left(M^G_g\right)^2\frac{\delta}{C}\leq2\delta. 
 \tag{\stepcounter{equation}\theequation}
\end{align*}

We have shown that $\mathbb{E}\left\Vert \tilde{\Omega}_{M}(\theta)-\mathbb{E}\tilde{\Omega}_{M}(\theta)\right\Vert \leq6\delta$.
Since $\delta$ is arbitrary, by Markov's inequality, we obtain the result. 

\bigskip
\noindent
\textbf{Proof of Theorem \ref{thm:oneway}:}

Let $\rho_{hM}=1$. We first show that $\hat{\theta}_N-\theta^*_M\overset{p}\to \bm{0}$.

Denote $Q_N(\theta):=\frac{1}{N}\sum^M_{i=1}R_{iM}q_{iM}(W_{iM},\theta)$. Note that 
\begin{equation}
Q_N(\theta)=\frac{M\rho_{uM}\rho_{gM}}{N}\frac{1}{M}\sum^M_{i=1}\frac{R_{iM}}{\rho_{uM}\rho_{gM}}q_{iM}(W_{iM},\theta).
\end{equation}
By Lemma \ref{lemma1} and the continuous mapping theorem, $\frac{M\rho_{uM}\rho_{gM}}{N}\overset{p}\to 1$.
Hence, it is sufficient to show that for each $\theta\in\Theta$
\begin{equation}\label{eqn:1}
\left\|\frac{1}{M}\sum^M_{i=1}\frac{R_{iM}}{\rho_{uM}\rho_{gM}}q_{iM}(W_{iM},\theta)-\frac{1}{M}\sum^M_{i=1}\mathbb{E}\big[q_{iM}(W_{iM},\theta)\big]\right\|\overset{p}\to 0.
\end{equation}
Condition (\romannumeral 5) in Assumption \ref{assumpa1} implies $\forall\ \theta\in\Theta$
\begin{equation}
\sup_{i,M}\mathbb{E}\Bigg[\bigg |\frac{R_{iM}}{\rho_{uM}\rho_{gM}}q_{iM}(W_{iM},\theta)\bigg |^r\Bigg]\leq\frac{1}{(\rho_{uM}\rho_{gM})^{r-1}}\sup_{i,M}\mathbb{E}\Big[\sup_{\theta\in\Theta}|q_{iM}(W_{iM},\theta)|^r\Big]<\infty
\end{equation}
for some $r>1$,
which further implies 
\begin{equation}
\lim_{C\to\infty}\sup_{i,M}\Bigg\{\mathbb{E}\bigg[\bigg |\frac{R_{iM}}{\rho_{uM}\rho_{gM}}q_{iM}(W_{iM},\theta)\bigg |\cdot \mathbbm{1}\bigg(\bigg |\frac{R_{iM}}{\rho_{uM}\rho_{gM}}q_{iM}(W_{iM},\theta)\bigg |>C\bigg)\bigg]\Bigg\}=0.
\end{equation}
(\ref{eqn:1}) thus follows by Theorem 1 in \cite{hansen2019asymptotic} under Assumption $4^\prime$.
Next, 
\begin{equation}
\sup_{\theta\in\Theta}|Q_N(\theta)-Q_M(\theta)|=o_p(1)
\end{equation}
follows from Lemma \ref{lemma2} above and Corollary 2.2 in \cite{newey1991uniform} under condition (\romannumeral 6) in Assumption \ref{assumpa1}. As a result, consistency follows, e.g., from Theorem 3.3 in \cite{gallant1988unified}.

To prove asymptotic normality, we start by verifying that 
\begin{equation}
\sum^M_{i=1}\mathbb{E}\big[m_{iM}(W_{iM},\theta^*_M)\big]=\textbf{0},
\end{equation}
which holds by Lemma 3.6 in \cite{newey1994large} and Jensen's inequality under conditions (\romannumeral 4) and (\romannumeral 7) in Assumption \ref{assumpa1}.

By the element-by-element mean value expansion around $\theta^*_M$,
\begin{equation}\label{eqn:4}
\begin{aligned}
&o_p(N^{-1/2})=V_M^{-1/2}\frac{1}{N}\sum^M_{i=1}R_{iM}\cdot m_{iM}(W_{iM},\hat{\theta}_N)\\
=&V_M^{-1/2}\frac{1}{N}\sum^M_{i=1}R_{iM}\cdot m_{iM}(W_{iM},\theta^*_M)+V_M^{-1/2}\frac{1}{N}\sum^M_{i=1}R_{iM} \nabla_\theta m_{iM}(W_{iM},\check{\theta})(\hat{\theta}_N-\theta^*_M),
\end{aligned}
\end{equation}
where $\check{\theta}$ lies on the line segment connecting $\theta^*_M$ and $\hat{\theta}_N$.

We first show 
\begin{equation}\label{eqn:2}
\hat{L}_N(\check{\theta})=L_M(\theta^*_M)\big(I_k+o_p(1)\big).
\end{equation}
Since we can write
\begin{equation}
\hat{L}_N(\check{\theta})=L_M(\theta^*_M)\Big[I_k+L_M(\theta^*_M)^{-1}\big(\hat{L}_N(\check{\theta})-L_M(\theta^*_M)\big)\Big],
\end{equation}
it suffices to show 
\begin{equation}
\left\|L_M(\theta^*_M)^{-1}\big(\hat{L}_N(\check{\theta})-L_M(\theta^*_M)\big)\right\|\overset{p}\to 0.
\end{equation}
We can write
\begin{equation}
\begin{aligned}
\hat{L}_N(\theta)=&\frac{M\rho_{uM}\rho_{gM}}{N}\frac{1}{M}\sum^M_{i=1}\frac{R_{iM}}{\rho_{uM}\rho_{gM}}\nabla_\theta m_{iM}(W_{iM},\theta)\\
=&\big(1+o_p(1)\big)\frac{1}{M}\sum^M_{i=1}\frac{R_{iM}}{\rho_{uM}\rho_{gM}}\nabla_\theta m_{iM}(W_{iM},\theta).
\end{aligned}
\end{equation}
Since $\forall\ \theta\in\Theta$
\begin{equation}
\begin{aligned}
&\sup_{i,M}\mathbb{E}\Bigg[\left\|\frac{R_{iM}}{\rho_{uM}\rho_{gM}}\nabla_\theta m_{iM}(W_{iM},\theta)\right\|^r\Bigg]\\
\leq&\frac{1}{(\rho_{uM}\rho_{gM})^{r-1}}\sup\limits_{i,M}\mathbb{E}\Big[\sup\limits_{\theta\in\Theta}\left\|\nabla_\theta m_{iM}(W_{iM},\theta)\right\|^r\Big]< \infty
\end{aligned}
\end{equation}
for some $r>1$,
\begin{equation}
\left\|\frac{1}{M}\sum^M_{i=1}\frac{R_{iM}}{\rho_{uM}\rho_{gM}}\nabla_\theta m_{iM}(W_{iM},\theta)-L_M(\theta)\right\|\overset{p}\to 0
\end{equation}
by Theorem 1 in \cite{hansen2019asymptotic} under Assumption $4^\prime$ and condition (\romannumeral 9) in Assumption \ref{assumpa1}.
By Corollary 2.2 in \cite{newey1991uniform} and Lemma \ref{lemma2} above,
\begin{align}
&\left\|L_M(\theta^*_M)^{-1}\big(\hat{L}_N(\check{\theta})-L_M(\theta^*_M)\big)\right\|\\
\leq& C\Bigg(\sup_{\theta\in\Theta}\left\|\hat{L}_N(\theta)-L_M(\theta)\right\|+\left\|L_M(\check{\theta})-L_M(\theta^*_M)\right\|\Bigg)\overset{p}\to 0
\tag{\stepcounter{equation}\theequation}
\end{align}
under conditions (\romannumeral 10) and (\romannumeral 11) in Assumption \ref{assumpa1}.

(\ref{eqn:2}) implies
\begin{equation}\label{eqn:3}
\hat{L}_N(\check{\theta})^{-1}=L_M(\theta^*_M)^{-1}(I_k+o_p(1)).
\end{equation}
Using (\ref{eqn:3}), (\ref{eqn:4}) can be written as 
\begin{equation}\label{eqn:6}
\begin{aligned}
V_M^{-1/2}\sqrt{N}(\hat{\theta}_N-\theta^*_M)=&-V_M^{-1/2}L_M(\theta^*_M)^{-1}\frac{1}{\sqrt{N}}\sum^M_{i=1}R_{iM}\cdot m_{iM}(W_{iM},\theta^*_M)\\
&-V_M^{-1/2}L_M(\theta^*_M)^{-1}o_p(1)\frac{1}{\sqrt{N}}\sum^M_{i=1}R_{iM}\cdot m_{iM}(W_{iM},\theta^*_M)+o_p(1).
\end{aligned}
\end{equation}

We can write
\begin{equation}\label{eqn:5}
\begin{aligned}
\frac{1}{\sqrt{N}}\sum^M_{i=1}R_{iM}\cdot m_{iM}(W_{iM},\theta^*_M)&=\sqrt{\frac{M\rho_{uM}\rho_{gM}}{N}}\frac{1}{\sqrt{M}}\sum^M_{i=1}\frac{R_{iM}}{\sqrt{\rho_{uM}\rho_{gM}}}m_{iM}(W_{iM},\theta^*_M)\\
&=\big(1+o_p(1)\big)\frac{1}{\sqrt{M}}\sum^M_{i=1}\frac{R_{iM}}{\sqrt{\rho_{uM}\rho_{gM}}}m_{iM}(W_{iM},\theta^*_M).
\end{aligned}
\end{equation}
Plug (\ref{eqn:5}) into (\ref{eqn:6}), we have 
\begin{align*}
&V_M^{-1/2}\sqrt{N}(\hat{\theta}_N-\theta^*_M)\\
=&-V_M^{-1/2}L_M(\theta^*_M)^{-1}\frac{1}{\sqrt{M}}\sum^M_{i=1}\frac{R_{iM}}{\sqrt{\rho_{uM}\rho_{gM}}}m_{iM}(W_{iM},\theta^*_M)\\
&-V_M^{-1/2}L_M(\theta^*_M)^{-1}\frac{1}{\sqrt{M}}\sum^M_{i=1}\frac{R_{iM}}{\sqrt{\rho_{uM}\rho_{gM}}}m_{iM}(W_{iM},\theta^*_M)\cdot o_p(1)+o_p(1).\tag{\stepcounter{equation}\theequation}
\end{align*}
Since
\begin{align*}
&\mathbb{V}\Bigg(\frac{1}{\sqrt{M}}\sum^M_{i=1}\frac{R_{iM}}{\sqrt{\rho_{uM}\rho_{gM}}}m_{iM}(W_{iM},\theta^*_M)\Bigg)\\
=&\frac{1}{M\rho_{uM}\rho_{gM}}\bigg\{\sum^M_{i=1}\mathbb{V}\big[R_{iM}\cdot m_{iM}(W_{iM},\theta^*_M)\big]\\
&+\sum^{G}_{g=1}\sum_{i \in \mathcal{N}^G_g}\sum_{j \in \mathcal{N}^G_g \backslash \{i\}}\mathbb{COV}\big[R_{iM}\cdot m_{iM}(W_{iM},\theta^*_M),R_{jM}\cdot m_{jM}(W_{jM},\theta^*_M)\big]\bigg\}\\
=&\frac{1}{M\rho_{uM}\rho_{gM}}\bigg\{\sum^M_{i=1}\Big[\mathbb{E}\big(R_{iM}\cdot m_{iM}(W_{iM},\theta^*_M)m_{iM}(W_{iM},\theta^*_M)'\big)\\
&-\mathbb{E}\big(R_{iM}\cdot m_{iM}(W_{iM},\theta^*_M)\big)\mathbb{E}\big(R_{iM}\cdot m_{iM}(W_{iM},\theta^*_M)\big)'\Big]\\
&+\sum^{G}_{g=1}\sum_{i \in \mathcal{N}^G_g}\sum_{j \in \mathcal{N}^G_g \backslash \{i\}}\Big[\mathbb{E}\big(R_{iM}R_{jM}\cdot m_{iM}(W_{iM},\theta^*_M)m_{jM}(W_{jM},\theta^*_M)'\big)\\
&-\mathbb{E}\big(R_{iM}\cdot m_{iM}(W_{iM},\theta^*_M)\big)\mathbb{E}\big(R_{jM}\cdot m_{jM}(W_{jM},\theta^*_M)\big)'\Big]\bigg\}\\
=&\frac{1}{M}\bigg\{\sum^M_{i=1}\Big[\mathbb{E}\big(m_{iM}(W_{iM},\theta^*_M)m_{iM}(W_{iM},\theta^*_M)'\big)\\
&-\rho_{uM}\rho_{gM}\mathbb{E}\big(m_{iM}(W_{iM},\theta^*_M)\big)\mathbb{E}\big(m_{iM}(W_{iM},\theta^*_M)\big)'\Big]\\
&+\sum^{G}_{g=1}\sum_{i \in \mathcal{N}^G_g}\sum_{j \in \mathcal{N}^G_g \backslash \{i\}}\Big[\rho_{uM}\mathbb{E}\big(m_{iM}(W_{iM},\theta^*_M)m_{jM}(W_{jM},\theta^*_M)'\big)\\
&-\rho_{uM}\rho_{gM}\mathbb{E}\big(m_{iM}(W_{iM},\theta^*_M)\big)\mathbb{E}\big(m_{jM}(W_{jM},\theta^*_M)\big)'\Big]\bigg\}\\
&=\Delta_{ehw,M}(\theta^*_M)-\rho_{uM}\rho_{gM}\Delta_{E,M}+\rho_{uM} \Delta_{cluster,M}(\theta^*_M)-\rho_{uM} \rho_{gM} \Delta_{EC,M}, \tag{\stepcounter{equation}\theequation}
\end{align*}
we have 
\begin{equation}
\mathbb{V}\Bigg(V_M^{-1/2}L_M(\theta^*_M)^{-1}\frac{1}{\sqrt{M}}\sum^M_{i=1}\frac{R_{iM}}{\sqrt{\rho_{uM}\rho_{gM}}}m_{iM}(W_{iM},\theta^*_M)\Bigg)=I_k.
\end{equation}
Given $\forall\ \theta\in\Theta$
\begin{equation}\label{eqn:8}
\sup_{i,M}\mathbb{E}\Bigg[\left\|\frac{R_{iM}}{\sqrt{\rho_{uM}\rho_{gM}}}m_{iM}(W_{iM},\theta)\right\|^r\Bigg]\leq\frac{1}{(\rho_{uM}\rho_{gM})^{r/2-1}}\sup_{i,M}\mathbb{E}\Big[\sup_{\theta\in\Theta}\left\|m_{iM}(W_{iM},\theta)\right\|^r\Big]<\infty
\end{equation}
for some $r>2$ under condition (\romannumeral 7) in Assumption \ref{assumpa1},
\begin{equation}\label{eqn:7}
V_M^{-1/2}L_M(\theta^*_M)^{-1}\frac{1}{\sqrt{M}}\sum^M_{i=1}\frac{R_{iM}}{\sqrt{\rho_{uM}\rho_{gM}}}m_{iM}(W_{iM},\theta^*_M)\overset{d}\to \mathcal{N}(\textbf{0},I_k)
\end{equation}
by Theorem 2 in \cite{hansen2019asymptotic} under Assumption $4^\prime$ and condition (\romannumeral 8) in Assumption \ref{assumpa1}.

Because of (\ref{eqn:7}), 
\begin{equation}
\begin{aligned}
V_M^{-1/2}\sqrt{N}(\hat{\theta}_N-\theta^*_M)=&-V_M^{-1/2}L_M(\theta^*_M)^{-1}\frac{1}{\sqrt{M}}\sum^M_{i=1}\frac{R_{iM}}{\sqrt{\rho_{uM}\rho_{gM}}}m_{iM}(W_{iM},\theta^*_M)\\
&+o_p(1)O_p(1)+o_p(1)\overset{d}\to \mathcal{N}(\textbf{0},I_k).
\end{aligned}
\end{equation}

As for Theorem \ref{thm:oneway}(2), it is equivalent to show $\left\|V_{SM}^{-1/2}\hat{V}_{SN}V_{SM}^{-1/2}-I_k\right\|\overset{p}\to 0$.

Since (\ref{eqn:3}) holds by replacing $\check{\theta}$ with $\hat{\theta}_N$,
\begin{equation}\label{eqn:9}
\hat{L}_N(\hat{\theta}_N)^{-1}=L_M(\theta^*_M)^{-1}\big(I_k+o_p(1)\big).
\end{equation}

We can write
\begin{align*}
&\hat{\Delta}_{ehw,N}(\theta)+\hat{\Delta}_{cluster,N}(\theta)\\
=&\frac{1}{N}\sum^G_{g=1}\bigg[\sum_{i \in \mathcal{N}^G_g}R_{iM}\cdot m_{iM}(W_{iM},\theta)\bigg]\bigg[\sum_{i \in \mathcal{N}^G_g}R_{iM}\cdot m_{iM}(W_{iM},\theta)\bigg]'\\
=&\frac{M\rho_{uM}\rho_{gM}}{N}\frac{1}{M}\sum^G_{g=1}\bigg[\sum_{i \in \mathcal{N}^G_g}\frac{R_{iM}}{\sqrt{\rho_{uM}\rho_{gM}}}m_{iM}(W_{iM},\theta)\bigg]\bigg[\sum_{i \in \mathcal{N}^G_g}\frac{R_{iM}}{\sqrt{\rho_{uM}\rho_{gM}}} m_{iM}(W_{iM},\theta)\bigg]'\\
=&\big(1+o_p(1)\big)\frac{1}{M}\sum^G_{g=1}\bigg[\sum_{i \in \mathcal{N}^G_g}\frac{R_{iM}}{\sqrt{\rho_{uM}\rho_{gM}}}m_{iM}(W_{iM},\theta)\bigg]\bigg[\sum_{i \in \mathcal{N}^G_g}\frac{R_{iM}}{\sqrt{\rho_{uM}\rho_{gM}}} m_{iM}(W_{iM},\theta)\bigg]'.\tag{\stepcounter{equation}\theequation}
\end{align*}
Note that
\begin{align*}
&\mathbb{E}\Bigg\{\frac{1}{M}\sum^G_{g=1}\bigg[\sum_{i \in \mathcal{N}^G_g}\frac{R_{iM}}{\sqrt{\rho_{uM}\rho_{gM}}} m_{iM}(W_{iM},\theta)\bigg]\bigg[\sum_{i \in \mathcal{N}^G_g}\frac{R_{iM}}{\sqrt{\rho_{uM}\rho_{gM}}}m_{iM}(W_{iM},\theta)\bigg]'\Bigg\}\\
=&\mathbb{E}\bigg[\frac{1}{M}\sum^M_{i=1}\frac{R_{iM}}{\rho_{uM}\rho_{gM}}m_{iM}(W_{iM},\theta)m_{iM}(W_{iM},\theta)'\bigg]\\
&+\mathbb{E}\bigg[\frac{1}{M}\sum^{G}_{g=1}\sum_{i \in \mathcal{N}^G_g}\sum_{j \in \mathcal{N}^G_g \backslash \{i\}}\frac{R_{iM}R_{jM}}{\rho_{uM}\rho_{gM}}m_{iM}(W_{iM}, \theta)m_{jM}(W_{jM}, \theta)'\bigg]\\
=&\frac{1}{M}\sum^M_{i=1}\mathbb{E}\big[m_{iM}(W_{iM},\theta)m_{iM}(W_{iM},\theta)'\big]\\
&+\frac{1}{M}\sum^{G}_{g=1}\sum_{i \in \mathcal{N}^G_g}\sum_{j \in \mathcal{N}^G_g \backslash \{i\}}\rho_{uM}\mathbb{E}\big[m_{iM}(W_{iM}, \theta)m_{jM}(W_{jM}, \theta)'\big]\\
=&\Delta_{ehw,M}(\theta)+\rho_{uM}\Delta_{cluster,M}(\theta).
\tag{\stepcounter{equation}\theequation}
\end{align*}
Hence, $\forall\ \theta\in\Theta$
\begin{align*}
&\Bigg\Vert \frac{1}{M}\sum^G_{g=1}\bigg[\sum_{i \in \mathcal{N}^G_g}\frac{R_{iM}}{\sqrt{\rho_{uM}\rho_{gM}}} m_{iM}(W_{iM},\theta)\bigg]\bigg[\sum_{i \in \mathcal{N}^G_g}\frac{R_{iM}}{\sqrt{\rho_{uM}\rho_{gM}}}m_{iM}(W_{iM},\theta)\bigg]'\\
&-\big(\Delta_{ehw,M}(\theta)+\rho_{uM}\Delta_{cluster,M}(\theta)\big)\Bigg\Vert \overset{p}\to 0\tag{\stepcounter{equation}\theequation}
\end{align*}
follows by (\ref{eqn:8}) and the same proof of (62) in \cite{hansen2019asymptotic} under Assumption $4^\prime$.
Also, $\Delta_{ehw,M}(\theta)+\rho_{uM}\Delta_{cluster,M}(\theta)$ is continuous in $\theta$ for all $M$ by the dominated convergence theorem (DCT), Jensen's inequality, and Cauchy-Schwarz Inequality under conditions (\romannumeral 4) and (\romannumeral 7) in Assumption \ref{assumpa1}. 

In addition,
\begin{align*}
&\left\|\hat{\Delta}_{ehw,N}(\tilde{\theta})+\hat{\Delta}_{cluster,N}(\tilde{\theta})-\big(\hat{\Delta}_{ehw,N}(\theta)+\hat{\Delta}_{cluster,N}(\theta)\big)\right\|\\
\leq&\frac{1}{N}\sum^G_{g=1}\Bigg\|\bigg[\sum_{i \in \mathcal{N}^G_g}R_{iM}\cdot m_{iM}(W_{iM},\tilde{\theta})\bigg]\bigg[\sum_{i \in \mathcal{N}^G_g}R_{iM}\cdot m_{iM}(W_{iM},\tilde{\theta})\bigg]'\\
&-\bigg[\sum_{i \in \mathcal{N}^G_g}R_{iM}\cdot m_{iM}(W_{iM},\theta)\bigg]\bigg[\sum_{i \in \mathcal{N}^G_g}R_{iM}\cdot m_{iM}(W_{iM},\theta)\bigg]'\Bigg\|\\
\leq&\frac{1}{N}\sum^G_{g=1}2\sup_{\theta\in\Theta}\left\|\sum_{i \in \mathcal{N}^G_g}R_{iM}\cdot m_{iM}(W_{iM},\theta)\right\|\cdot\\
&\left\|\sum_{i \in \mathcal{N}^G_g}R_{iM}\cdot m_{iM}(W_{iM},\tilde{\theta})-\sum_{i \in \mathcal{N}^G_g}R_{iM}\cdot m_{iM}(W_{iM},\theta)\right\|\\
\leq&\frac{2}{N}\sum^G_{g=1}\sup_{\theta\in\Theta}\left\|\sum_{i \in \mathcal{N}^G_g}R_{iM}\cdot m_{iM}(W_{iM},\theta)\right\|\sum_{i \in \mathcal{N}^G_g}R_{iM}b_{1,iM}(W_{iM})h(\|\tilde{\theta}-\theta\|).
\tag{\stepcounter{equation}\theequation}
\end{align*}
under condition (\romannumeral 12) in Assumption \ref{assumpa1}.
Let
\begin{align*}
B^1_N:= &\frac{2}{N}\sum^G_{g=1}\sup_{\theta\in\Theta}\left\|\sum_{i \in \mathcal{N}^G_g}R_{iM}\cdot m_{iM}(W_{iM},\theta)\right\|\sum_{i \in \mathcal{N}^G_g}R_{iM}b_{1,iM}(W_{iM})\\
=&2\frac{M\rho_{uM}\rho_{gM}}{N}\frac{1}{M}\sum^G_{g=1}\sup_{\theta\in\Theta}\left\|\sum_{i \in \mathcal{N}^G_g}\frac{R_{iM}}{\sqrt{\rho_{uM}\rho_{gM}}}m_{iM}(W_{iM},\theta)\right\|\sum_{i \in \mathcal{N}^G_g}\frac{R_{iM}}{\sqrt{\rho_{uM}\rho_{gM}}}b_{1,iM}(W_{iM})\\
=&\big(1+o_p(1)\big)\frac{2}{M}\sum^G_{g=1}\sup_{\theta\in\Theta}\left\|\sum_{i \in \mathcal{N}^G_g}\frac{R_{iM}}{\sqrt{\rho_{uM}\rho_{gM}}}m_{iM}(W_{iM},\theta)\right\|\sum_{i \in \mathcal{N}^G_g}\frac{R_{iM}}{\sqrt{\rho_{uM}\rho_{gM}}}b_{1,iM}(W_{iM}).\tag{\stepcounter{equation}\theequation}
\end{align*}
Since 
\begin{align*}
\mathbb{E}\Bigg[\sup_{\theta\in\Theta}\left\|\sum_{i \in \mathcal{N}^G_g}\frac{R_{iM}}{\sqrt{\rho_{uM}\rho_{gM}}}m_{iM}(W_{iM},\theta)\right\|^2\Bigg]<C\left(M^G_g\right)^2\tag{\stepcounter{equation}\theequation}
\end{align*}
by Cr inequality and Jensen's inequality under condition (\romannumeral 7) in Assumption \ref{assumpa1},
\begin{align*}
&\mathbb{E}\Bigg[\frac{2}{M}\sum^G_{g=1}\sup_{\theta\in\Theta}\left\|\sum_{i \in \mathcal{N}^G_g}\frac{R_{iM}}{\sqrt{\rho_{uM}\rho_{gM}}}m_{iM}(W_{iM},\theta)\right\|\sum_{i \in \mathcal{N}^G_g}\frac{R_{iM}}{\sqrt{\rho_{uM}\rho_{gM}}}b_{1,iM}(W_{iM})\Bigg]\\
\leq&\frac{2}{M}\sum^G_{g=1}\sum_{i \in \mathcal{N}^G_g}\Bigg\{\mathbb{E}\Bigg[\sup_{\theta\in\Theta}\left\|\sum_{i \in \mathcal{N}^G_g}\frac{R_{iM}}{\sqrt{\rho_{uM}\rho_{gM}}}m_{iM}(W_{iM},\theta)\right\|^2\Bigg]\Bigg\}^{1/2}\cdot\\
&\Bigg\{\mathbb{E}\bigg[\frac{R_{iM}}{\rho_{uM}\rho_{gM}}b_{1,iM}(W_{iM})^2\bigg]\Bigg\}^{1/2}\\
\leq&C\frac{1}{M}\sum^G_{g=1}\left(M^G_g\right)^2<\infty
\tag{\stepcounter{equation}\theequation}
\end{align*}
by Cauchy-Schwarz inequality under Assumption $4^\prime$ and condition (\romannumeral 12) in Assumption \ref{assumpa1}.
As a result, $B^1_N=O_p(1)$ by Markov's inequality.
Therefore, given condition (\romannumeral 8) in Assumption \ref{assumpa1},
\begin{align*}
&\Big\|\big[\Delta_{ehw,M}(\theta^*_M)+\rho_{uM}\Delta_{cluster,M}(\theta^*_M)\big]^{-1}\\
&\big[\hat{\Delta}_{ehw,N}(\hat{\theta}_N)+\hat{\Delta}_{cluster,N}(\hat{\theta}_N)-\Delta_{ehw,M}(\theta^*_M)-\rho_{uM}\Delta_{cluster,M}(\theta^*_M)\big]\Big\|\\
\leq&C\bigg(\sup_{\theta\in\Theta}\|\hat{\Delta}_{ehw,N}(\theta)+\hat{\Delta}_{cluster,N}(\theta)-\Delta_{ehw,M}(\theta)-\rho_{uM}\Delta_{cluster,M}(\theta)\|\\
&+\left\|\Delta_{ehw,M}(\hat{\theta}_N)+\rho_{uM}\Delta_{cluster,M}(\hat{\theta}_N)-\Delta_{ehw,M}(\theta^*_M)-\rho_{uM}\Delta_{cluster,M}(\theta^*_M)\right\|\bigg)=o_p(1)\tag{\stepcounter{equation}\theequation}
\end{align*}
by Corollary 2.2 in \cite{newey1991uniform} under $\hat{\theta}_N-\theta^*_M\overset{p}\to \textbf{0}$.
Hence,
\begin{equation}\label{eqn:10}
\begin{aligned}
&\hat{\Delta}_{ehw,N}(\hat{\theta}_N)+\hat{\Delta}_{cluster,N}(\hat{\theta}_N)\\
=&\big(\Delta_{ehw,M}(\theta^*_M)+\rho_{uM}\Delta_{cluster,M}(\theta^*_M)\big)\Big[I_k+\big(\Delta_{ehw,M}(\theta^*_M)+\rho_{uM}\Delta_{cluster,M}(\theta^*_M)\big)^{-1}\\
&\big(\hat{\Delta}_{ehw,N}(\hat{\theta}_N)+\hat{\Delta}_{cluster,N}(\hat{\theta}_N)-\Delta_{ehw,M}(\theta^*_M)-\rho_{uM}\Delta_{cluster,M}(\theta^*_M)\big)\Big]\\
=&\big(\Delta_{ehw,M}(\theta^*_M)+\rho_{uM}\Delta_{cluster,M}(\theta^*_M)\big)\big(I_k+o_p(1)\big).
\end{aligned}
\end{equation}

Using (\ref{eqn:9}) and (\ref{eqn:10}), 
\begin{align*}
&\left\|V_{SM}^{-1/2}\hat{V}_{SN}V_{SM}^{-1/2}-I_k\right\|\\
=&\left\|V_{SM}^{-1/2}\hat{L}_N(\hat{\theta}_N)^{-1}\big(\hat{\Delta}_{ehw,N}(\hat{\theta}_N)+\hat{\Delta}_{cluster,N}(\hat{\theta}_N)\big)\hat{L}_N(\hat{\theta}_N)^{-1}V_{SM}^{-1/2}-I_k\right\|\\
=&\Big\|V_{SM}^{-1/2}L_M(\theta^*_M)^{-1}\big(I_k+o_p(1)\big)\big(\Delta_{ehw,M}(\theta^*_M)+\rho_{uM}\Delta_{cluster,M}(\theta^*_M)\big)\big(I_k+o_p(1)\big)\cdot\\
&L_M(\theta^*_M)^{-1}\big(I_k+o_p(1)\big)V_{SM}^{-1/2}-I_k\Big\|\\
\leq&\left\|V_{SM}^{-1/2}V_{SM}V_{SM}^{-1/2}-I_k\right\|+\left\|V_{SM}^{-1/2}V_{SM}V_{SM}^{-1/2}\right\|o_p(1)\\
=&o_p(1).
\tag{\stepcounter{equation}\theequation}
\end{align*}

\bigskip
\noindent
\textbf{Proof of Theorem \ref{thm:ate}:}

First, using similar arguments in the proof of Theorem \ref{thm:oneway}, 
\begin{equation}
\hat{\gamma}_N-\gamma^*_M\overset{p}\to \textbf{0}
\end{equation}
under conditions (\romannumeral 1), (\romannumeral 2), and (\romannumeral 6) in Assumption \ref{assumpa2}.

By the mean value expansion around $\theta^*_M$,
\begin{equation}\label{eqn:11}
\begin{aligned}
&V_{f,M}^{-1/2}\frac{1}{\sqrt{N}}\sum^M_{i=1}R_{iM}f_{iM}(W_{iM},\hat{\theta}_N)\\
=&V_{f,M}^{-1/2}\frac{1}{\sqrt{N}}\sum^M_{i=1}R_{iM}f_{iM}(W_{iM},\theta^*_M)\\
&+V_{f,M}^{-1/2}\frac{1}{N}\sum^M_{i=1}R_{iM}\nabla_\theta f_{iM}(W_{iM},\check{\theta})\sqrt{N}(\hat{\theta}_N-\theta^*_M),
\end{aligned}
\end{equation}
where $\check{\theta}$ lies on the line segment connecting $\theta^*_M$ and $\hat{\theta}_N$.

Given Theorem \ref{thm:oneway}, $V_M^{-1/2}\sqrt{N}(\hat{\theta}_N-\theta^*_M)=O_p(1)$. 
Further,
\begin{equation}
\hat{F}_N(\check{\theta})=F_M(\theta^*_M)+o_p(1)
\end{equation}
under conditions (\romannumeral 1), (\romannumeral 4), and (\romannumeral 5) in Assumption \ref{assumpa2}.
Therefore,
\begin{equation}\label{eqn:12}
V_{f,M}^{-1/2}\frac{1}{N}\sum^M_{i=1}R_{iM}\nabla_\theta f_{iM}(W_{iM},\check{\theta})\sqrt{N}(\hat{\theta}_N-\theta^*_M)=V_{f,M}^{-1/2}F_M(\theta^*_M)\sqrt{N}(\hat{\theta}_N-\theta^*_M)+o_p(1).
\end{equation}
According to the mean value expansion in the proof of the asymptotic normality of \\$V_M^{-1/2}\sqrt{N}(\hat{\theta}_N-\theta^*_M)$,
\begin{equation}\label{eqn:13}
V_M^{-1/2}\sqrt{N}(\hat{\theta}_N-\theta^*_M)=-V_M^{-1/2}\frac{1}{\sqrt{N}}\sum^M_{i=1}R_{iM}L_M(\theta^*_M)^{-1}m_{iM}(W_{iM},\theta^*_M)+o_p(1).
\end{equation}
Combining (\ref{eqn:11}), (\ref{eqn:12}), and (\ref{eqn:13}),
\begin{equation}\label{eqn:14}
\begin{aligned}
&V_{f,M}^{-1/2}\frac{1}{\sqrt{N}}\sum^M_{i=1}R_{iM}f_{iM}(W_{iM},\hat{\theta}_N)\\
=&V_{f,M}^{-1/2}\frac{1}{\sqrt{N}}\sum^M_{i=1}R_{iM}\big[f_{iM}(W_{iM},\theta^*_M)-F_M(\theta^*_M)L_M(\theta^*_M)^{-1}m_{iM}(W_{iM},\theta^*_M)\big]+o_p(1).
\end{aligned}
\end{equation}

Subtract $V_{f,M}^{-1/2}\sqrt{N}\gamma^*_M$ from both sides of (\ref{eqn:14}).
\begin{align*}
&V_{f,M}^{-1/2}\sqrt{N}\big(\hat{\gamma}-\gamma^*_M\big)\\
=&V_{f,M}^{-1/2}\frac{1}{\sqrt{N}}\sum^M_{i=1}R_{iM}\big[f_{iM}(W_{iM},\theta^*_M)-\gamma^*_M-F_M(\theta^*_M)L_M(\theta^*_M)^{-1}m_{iM}(W_{iM},\theta^*_M)\big]+o_p(1)\\
=&V_{f,M}^{-1/2}\sqrt{\frac{M\rho_{uM}\rho_{gM}}{N}}\frac{1}{\sqrt{M}}\sum^M_{i=1}\frac{R_{iM}}{\sqrt{\rho_{uM}\rho_{gM}}}\big[f_{iM}(W_{iM},\theta^*_M)-\gamma^*_M\\
&-F_M(\theta^*_M)L_M(\theta^*_M)^{-1}m_{iM}(W_{iM},\theta^*_M)\big]+o_p(1)\\
=&\big(1+o_p(1)\big)V_{f,M}^{-1/2}\frac{1}{\sqrt{M}}\sum^M_{i=1}\frac{R_{iM}}{\sqrt{\rho_{uM}\rho_{gM}}}\big[f_{iM}(W_{iM},\theta^*_M)-\gamma^*_M\\
&-F_M(\theta^*_M)L_M(\theta^*_M)^{-1}m_{iM}(W_{iM},\theta^*_M)\big]+o_p(1)\tag{\stepcounter{equation}\theequation}
\end{align*}
Observe that $\forall\ \theta\in\Theta$
\begin{align*}
&\sup_{i,M}\mathbb{E}\Bigg\{\left\|\frac{R_{iM}}{\sqrt{\rho_{uM}\rho_{gM}}}\big[f_{iM}(W_{iM},\theta)-\gamma^*_M-F_M(\theta^*_M)L_M(\theta^*_M)^{-1}m_{iM}(W_{iM},\theta)\big]\right\|^r\Bigg\}\\
\leq&\frac{1}{(\rho_{uM}\rho_{gM})^{r/2-1}}\Bigg\{\bigg[\sup_{i,M}\mathbb{E}\Big(\sup_{\theta\in\Theta}\left\|f_{iM}(W_{iM},\theta)\right\|^r\Big)\bigg]^{1/r}+\left\|\gamma^*_M\right\|\\
&+C\left\|F_M(\theta^*_M)\right\|\bigg[\sup_{i,M}\mathbb{E}\Big(\sup_{\theta\in\Theta}\left\|m_{iM}(W_{iM},\theta)\right\|^r\Big)\bigg]^{1/r}\Bigg\}^r<\infty \tag{\stepcounter{equation}\theequation}
\end{align*}
for some $r>2$ by Minkowski's inequality and Jensen's inequality under conditions (\romannumeral 2) and (\romannumeral 4) in Assumption \ref{assumpa2}, and condition (\romannumeral 7) in Assumption \ref{assumpa1}. 
Also,
\begin{equation}
\begin{aligned}
&\mathbb{V}\Bigg\{\frac{1}{\sqrt{M}}\sum^M_{i=1}\frac{R_{iM}}{\sqrt{\rho_{uM}\rho_{gM}}}\big[f_{iM}(W_{iM},\theta^*_M)-\gamma^*_M-F_M(\theta^*_M)L_M(\theta^*_M)^{-1}m_{iM}(W_{iM},\theta^*_M)\big]\Bigg\}\\
=&\Delta^f_{ehw,M}-\rho_{uM}\rho_{gM}\Delta^f_{E,M}+\rho_{uM}\Delta^f_{cluster,M}-\rho_{uM}\rho_{gM}\Delta^f_{EC,M}.
\end{aligned}
\end{equation}
By Theorem 2 in \cite{hansen2019asymptotic}
\begin{equation}
V_{f,M}^{-1/2}\frac{1}{\sqrt{M}}\sum^M_{i=1}\frac{R_{iM}}{\sqrt{\rho_{uM}\rho_{gM}}}\big[f_{iM}(W_{iM},\theta^*_M)-\gamma^*_M-F_M(\theta^*_M)L_M(\theta^*_M)^{-1}m_{iM}(W_{iM},\theta^*_M)\big]\overset{d}\to \mathcal{N}(\textbf{0},I_q)
\end{equation}
under Assumption $4^\prime$ and condition (\romannumeral 3) in Assumption \ref{assumpa2}.

To show Theorem \ref{thm:ate}(2), 
observe that 
\begin{align*}
&\hat{\Delta}^f_{ehw,N}+\hat{\Delta}^f_{cluster,N}\\
=&\frac{M\rho_{uM}\rho_{gM}}{N}\frac{1}{M}\sum^G_{g=1}\Bigg\{\sum_{i \in \mathcal{N}^G_g}\frac{R_{iM}}{\sqrt{\rho_{uM}\rho_{gM}}}\big[f_{iM}(W_{iM},\hat{\theta}_N)-\hat{\gamma}_N\\
&-\hat{F}_N(\hat{\theta}_N)\hat{L}_N(\hat{\theta}_N)^{-1}m_{iM}(W_{iM},\hat{\theta}_N)\big]\Bigg\}\cdot\\
&\Bigg\{\sum_{i \in \mathcal{N}^G_g}\frac{R_{iM}}{\sqrt{\rho_{uM}\rho_{gM}}}\big[f_{iM}(W_{iM},\hat{\theta}_N)-\hat{\gamma}_N-\hat{F}_N(\hat{\theta}_N)\hat{L}_N(\hat{\theta}_N)^{-1}m_{iM}(W_{iM},\hat{\theta}_N)\big]\Bigg\}'\\
=&\big(1+o_p(1)\big)\frac{1}{M}\sum^G_{g=1}\Bigg\{\sum_{i \in \mathcal{N}^G_g}\frac{R_{iM}}{\sqrt{\rho_{uM}\rho_{gM}}}\Big[f_{iM}(W_{iM},\hat{\theta}_N)-\gamma^*_M+o_p(1)\\
&-\big(F_M(\theta^*_M)+o_p(1)\big)L_M(\theta^*_M)^{-1}\big(I_k+o_p(1)\big)m_{iM}(W_{iM},\hat{\theta}_N)\Big]\Bigg\}\cdot\\
&\Bigg\{\sum_{i \in \mathcal{N}^G_g}\frac{R_{iM}}{\sqrt{\rho_{uM}\rho_{gM}}}\Big[f_{iM}(W_{iM},\hat{\theta}_N)-\gamma^*_M+o_p(1)\\
&-\big(F_M(\theta^*_M)+o_p(1)\big)L_M(\theta^*_M)^{-1}\big(I_k+o_p(1)\big)m_{iM}(W_{iM},\hat{\theta}_N)\Big]\Bigg\}'\\
=&\big(1+o_p(1)\big)\frac{1}{M}\sum^G_{g=1}\Bigg\{\sum_{i \in \mathcal{N}^G_g}\frac{R_{iM}}{\sqrt{\rho_{uM}\rho_{gM}}}\big[f_{iM}(W_{iM},\hat{\theta}_N)-\gamma^*_M\\
&-F_M(\theta^*_M)L_M(\theta^*_M)^{-1}m_{iM}(W_{iM},\hat{\theta}_N)\big]\Bigg\}\cdot\\
&\Bigg\{\sum_{i \in \mathcal{N}^G_g}\frac{R_{iM}}{\sqrt{\rho_{uM}\rho_{gM}}}\big[f_{igM}(W_{iM},\hat{\theta}_N)-\gamma^*_M-F_M(\theta^*_M)L_M(\theta^*_M)^{-1}m_{iM}(W_{iM},\hat{\theta}_N)\big]\Bigg\}'+o_p(1) \tag{\stepcounter{equation}\theequation}
\end{align*}
under condition (\romannumeral 2) in Assumption \ref{assumpa2}, condition (\romannumeral 7) in Assumption \ref{assumpa1}, and Assumption $4^\prime$.
Denote 
\begin{equation}
\begin{aligned}
\tilde{\Delta}(\theta)=&\frac{1}{M}\sum^G_{g=1}\Bigg\{\sum_{i \in \mathcal{N}^G_g}\frac{R_{iM}}{\sqrt{\rho_{uM}\rho_{gM}}}\big[f_{iM}(W_{iM},\theta)-\gamma^*_M-F_M(\theta^*_M)L_M(\theta^*_M)^{-1}m_{iM}(W_{iM},\theta)\big]\Bigg\}\cdot\\
&\Bigg\{\sum_{i \in \mathcal{N}^G_g}\frac{R_{iM}}{\sqrt{\rho_{uM}\rho_{gM}}}\big[f_{iM}(W_{iM},\theta)-\gamma^*_M-F_M(\theta^*_M)L_M(\theta^*_M)^{-1}m_{iM}(W_{iM},\theta)\big]\Bigg\}'.
\end{aligned}
\end{equation}
It suffices to show 
\begin{equation}\label{eqn:15}
\left\|(\Delta^f_{ehw,M}+\rho_{uM}\Delta^f_{cluster,M})^{-1/2}\tilde{\Delta}(\hat{\theta}_N)(\Delta^f_{ehw,M}+\rho_{uM}\Delta^f_{cluster,M})^{-1/2}-I_q\right\|=o_p(1).
\end{equation}

Note that 
\begin{equation}
\begin{aligned}
&\bigg\|\frac{R_{iM}}{\sqrt{\rho_{uM}\rho_{gM}}}\Big\{\big[f_{iM}(W_{iM},\tilde{\theta})-\gamma^*_M-F_M(\theta^*_M)L_M(\theta^*_M)^{-1}m_{iM}(W_{iM},\tilde{\theta})\big]\\
&-\big[f_{iM}(W_{iM},\theta)-\gamma^*_M-F_M(\theta^*_M)L_M(\theta^*_M)^{-1}m_{iM}(W_{iM},\theta)\big]\Big\}\bigg\|\\
\leq&\frac{R_{iM}}{\sqrt{\rho_{uM}\rho_{gM}}}\bigg[\left\|f_{iM}(W_{iM},\tilde{\theta})-f_{iM}(W_{iM},\theta)\right\|\\
&+C\left\|m_{iM}(W_{iM},\tilde{\theta})-m_{iM}(W_{iM},\theta)\right\|\bigg]\\
\leq&\frac{R_{iM}}{\sqrt{\rho_{uM}\rho_{gM}}}\big[b_{2,iM}(W_{iM})h(\|\tilde{\theta}-\theta\|)+C b_{1,iM}(W_{iM})h(\|\tilde{\theta}-\theta\|)\big].
\end{aligned}
\end{equation}
Let
\begin{equation}
b_{3,iM}(W_{iM})=\frac{R_{iM}}{\sqrt{\rho_{uM}\rho_{gM}}}\big[b_{2,iM}(W_{iM})+Cb_{1,iM}(W_{iM})\big].
\end{equation}
Observe that
\begin{equation}
\begin{aligned}
\sup_{i,M}\mathbb{E}\big[b_{3,iM}(W_{iM})^2\big]\leq&\sup_{i,M}\mathbb{E}\big[b_{2,iM}(W_{iM})^2\big]+C\sup_{i,M}\mathbb{E}\big[b_{1,iM}(W_{iM})^2\big]\\
&+C\Big\{\sup_{i,M}\mathbb{E}\big[b_{2,iM}(W_{iM})^2\big]\sup_{i,M}\mathbb{E}\big[b_{1,iM}(W_{iM})^2\big]\Big\}^{1/2}<\infty
\end{aligned}
\end{equation}
by Cauchy-Schwarz inequality under condition (\romannumeral 12) in Assumption \ref{assumpa1} and condition (\romannumeral 6) in Assumption \ref{assumpa2}.
Therefore, (\ref{eqn:15}) follows from similar arguments in the proof of Theorem \ref{thm:oneway}(2).

\bigskip
\noindent
\textbf{Proof of Theorem \ref{thm:twoway}:}

The proof is analogous to Thereom \ref{thm:oneway}.
Let $Q_N(\theta):=\frac{1}{N}\sum^M_{i=1}R_{iM}q_{iM}(W_{iM},\theta)$. Then,
\begin{equation}
\begin{aligned}
Q_N(\theta)&=\frac{M\rho_{uM}\rho_{gM}\rho_{hM}}{N}\frac{1}{M}\sum^M_{i=1}\frac{R_{iM}}{\rho_{uM}\rho_{gM}\rho_{hM}}q_{iM}(W_{iM},\theta)\\
&=\big(1+o_p(1)\big)\frac{1}{M}\sum^M_{i=1}\frac{R_{iM}}{\rho_{uM}\rho_{gM}\rho_{hM}}q_{iM}(W_{iM},\theta).
\end{aligned}
\end{equation}
Hence, it is sufficient to show that for each $\theta\in\Theta$
\begin{equation}\label{eqn:1tw}
\left\|\frac{1}{M}\sum^M_{i=1}\frac{R_{iM}}{\rho_{uM}\rho_{gM}\rho_{hM}}q_{iM}(W_{iM},\theta)-\frac{1}{M}\sum^M_{i=1}\mathbb{E}\big[q_{iM}(W_{iM},\theta)\big]\right\|\overset{p}\to 0.
\end{equation}
Condition (\romannumeral 5) in Assumption \ref{assumpa3} implies $\forall\ \theta\in\Theta$
\begin{equation}
\sup_{i,M}\mathbb{E}\Bigg[\bigg \|\frac{R_{iM}}{\rho_{uM}\rho_{gM}\rho_{hM}}q_{iM}(W_{iM},\theta)\bigg \|^2\Bigg]\leq\frac{1}{(\rho_{uM}\rho_{gM}\rho_{hM})^{3}}\sup_{i,M}\mathbb{E}\Big[\sup_{\theta\in\Theta}\|q_{iM}(W_{iM},\theta)\|^2\Big]<\infty
\end{equation}
(\ref{eqn:1tw}) thus follows by Lemma \ref{lem:wlln}.
By the element-by-element mean value expansion around $\theta^*_M$,
\begin{equation}\label{eqn:4tw}
\begin{aligned}
&o_p(N^{-1/2})=V_{TWM}^{-1/2}\frac{1}{N}\sum^M_{i=1}R_{iM}\cdot m_{iM}(W_{iM},\hat{\theta}_N)\\
=&V_{TWM}^{-1/2}\frac{1}{N}\sum^M_{i=1}R_{iM}\cdot m_{iM}(W_{iM},\theta^*_M)+V_{TWM}^{-1/2}\frac{1}{N}\sum^M_{i=1}R_{iM} \nabla_\theta m_{iM}(W_{iM},\check{\theta})(\hat{\theta}_N-\theta^*_M),
\end{aligned}
\end{equation}
where $\check{\theta}$ lies on the line segment connecting $\theta^*_M$ and $\hat{\theta}_N$.

We first show 
\begin{equation}\label{eqn:2tw}
\hat{L}_N(\check{\theta})=L_M(\theta^*_M)\big(I_k+o_p(1)\big).
\end{equation}
Since we can write
\begin{equation}
\hat{L}_N(\check{\theta})=L_M(\theta^*_M)\Big[I_k+L_M(\theta^*_M)^{-1}\big(\hat{L}_N(\check{\theta})-L_M(\theta^*_M)\big)\Big],
\end{equation}
it suffices to show 
\begin{equation}
\left\|L_M(\theta^*_M)^{-1}\big(\hat{L}_N(\check{\theta})-L_M(\theta^*_M)\big)\right\|\overset{p}\to 0.
\end{equation}
We can write
\begin{equation}
\begin{aligned}
\hat{L}_N(\theta)=&\frac{M\rho_{uM}\rho_{gM}\rho_{hM}}{N}\frac{1}{M}\sum^M_{i=1}\frac{R_{iM}}{\rho_{uM}\rho_{gM}\rho_{hM}}\nabla_\theta m_{iM}(W_{iM},\theta)\\
=&\big(1+o_p(1)\big)\frac{1}{M}\sum^M_{i=1}\frac{R_{iM}}{\rho_{uM}\rho_{gM}\rho_{hM}}\nabla_\theta m_{iM}(W_{iM},\theta).
\end{aligned}
\end{equation}
Since $\forall\ \theta\in\Theta$
\begin{equation}
\begin{aligned}
&\sup_{i,M}\mathbb{E}\Bigg[\left\|\frac{R_{iM}}{\rho_{uM}\rho_{gM}\rho_{hM}}\nabla_\theta m_{iM}(W_{iM},\theta)\right\|^2\Bigg]\\
\leq&\frac{1}{(\rho_{uM}\rho_{gM}\rho_{hM})^3}\sup\limits_{i,M}\mathbb{E}\Big[\sup\limits_{\theta\in\Theta}\left\|\nabla_\theta m_{iM}(W_{iM},\theta)\right\|^2\Big]< \infty,
\end{aligned}
\end{equation}
\begin{equation}
\left\|\frac{1}{M}\sum^M_{i=1}\frac{R_{iM}}{\rho_{uM}\rho_{gM}\rho_{hM}}\nabla_\theta m_{iM}(W_{iM},\theta)-L_M(\theta)\right\|\overset{p}\to 0
\end{equation}
by Lemma \ref{lem:wlln} under Assumption \ref{assump4} and condition (\romannumeral 9) in Assumption \ref{assumpa3}. 
By Corollary 2.2 in \cite{newey1991uniform} and Lemma \ref{lemma2} above,
\begin{equation} \label{eqn:Lconverge_tw}
\begin{aligned}
&\left\|L_M(\theta^*_M)^{-1}\big(\hat{L}_N(\check{\theta})-L_M(\theta^*_M)\big)\right\|\\
\leq& C\Bigg(\sup_{\theta\in\Theta}\left\|\hat{L}_N(\theta)-L_M(\theta)\right\|+\left\|L_M(\check{\theta})-L_M(\theta^*_M)\right\|\Bigg)\overset{p}\to 0
\end{aligned}
\end{equation}
under conditions (\romannumeral 10) and (\romannumeral 11) in Assumption \ref{assumpa3}.

(\ref{eqn:2tw}) implies
\begin{equation}\label{eqn:3tw}
\hat{L}_N(\check{\theta})^{-1}=L_M(\theta^*_M)^{-1}(I_k+o_p(1)).
\end{equation}
Using (\ref{eqn:3tw}), (\ref{eqn:4tw}) can be written as 
\begin{equation}\label{eqn:6tw}
\begin{aligned}
V_{TWM}^{-1/2}\sqrt{N}(\hat{\theta}_N-\theta^*_M)=&-V_{TWM}^{-1/2}L_M(\theta^*_M)^{-1}\frac{1}{\sqrt{N}}\sum^M_{i=1}R_{iM}\cdot m_{iM}(W_{iM},\theta^*_M)\\
&-V_{TWM}^{-1/2}L_M(\theta^*_M)^{-1}o_p(1)\frac{1}{\sqrt{N}}\sum^M_{i=1}R_{iM}\cdot m_{iM}(W_{iM},\theta^*_M)+o_p(1).
\end{aligned}
\end{equation}

We can write
\begin{equation}\label{eqn:5tw}
\begin{aligned}
\frac{1}{\sqrt{N}}\sum^M_{i=1}R_{iM}\cdot m_{iM}(W_{iM},\theta^*_M)&=\sqrt{\frac{M\rho_{uM}\rho_{gM}\rho_{hM}}{N}}\frac{1}{\sqrt{M}}\sum^M_{i=1}\frac{R_{iM}}{\sqrt{\rho_{uM}\rho_{gM}\rho_{hM}}}m_{iM}(W_{iM},\theta^*_M)\\
&=\big(1+o_p(1)\big)\frac{1}{\sqrt{M}}\sum^M_{i=1}\frac{R_{iM}}{\sqrt{\rho_{uM}\rho_{gM}\rho_{hM}}}m_{iM}(W_{iM},\theta^*_M).
\end{aligned}
\end{equation}
Plug (\ref{eqn:5tw}) into (\ref{eqn:6tw}), we have 
\begin{align*}
&V_{TWM}^{-1/2}\sqrt{N}(\hat{\theta}_N-\theta^*_M)\\
=&-V_{TWM}^{-1/2}L_M(\theta^*_M)^{-1}\frac{1}{\sqrt{M}}\sum^M_{i=1}\frac{R_{iM}}{\sqrt{\rho_{uM}\rho_{gM}\rho_{hM}}}m_{iM}(W_{iM},\theta^*_M)\\
&-V_{TWM}^{-1/2}L_M(\theta^*_M)^{-1}\frac{1}{\sqrt{M}}\sum^M_{i=1}\frac{R_{iM}}{\sqrt{\rho_{uM}\rho_{gM}\rho_{hM}}}m_{iM}(W_{iM},\theta^*_M)\cdot o_p(1)+o_p(1).\tag{\stepcounter{equation}\theequation}
\end{align*}

Define:
\begin{align*}
    V_{\Delta TWM} := \mathbb{V} \left( \frac{1}{\sqrt{M}}\sum^M_{i=1}\frac{R_{iM}}{\sqrt{\rho_{uM}\rho_{gM}\rho_{hM}}}m_{iM}(W_{iM},\theta^*_M) \right). \tag{\stepcounter{equation}\theequation}
\end{align*}
Since
\begin{align*}
    V_{TWM} = L_M(\theta^*_M)^{-1} V_{\Delta TWM} L_M(\theta^*_M)^{-1}, \tag{\stepcounter{equation}\theequation}
\end{align*}
we have 
\begin{equation}
\mathbb{V}\Bigg(V_{TWM}^{-1/2}L_M(\theta^*_M)^{-1}\frac{1}{\sqrt{M}}\sum^M_{i=1}\frac{R_{iM}}{\sqrt{\rho_{uM}\rho_{gM}\rho_{hM}}}m_{iM}(W_{iM},\theta^*_M)\Bigg)=I_k.
\end{equation}
Given that, $\forall\ \theta\in\Theta$,
\begin{equation}\label{eqn:8tw}
\sup_{i,M}\mathbb{E}\Bigg[\left\|\frac{R_{iM}}{\sqrt{\rho_{uM}\rho_{gM}\rho_{hM}}}m_{iM}(W_{iM},\theta)\right\|^4\Bigg]\leq\frac{1}{\rho_{uM}\rho_{gM}\rho_{hM}}\sup_{i,M}\mathbb{E}\Big[\sup_{\theta\in\Theta}\left\|m_{iM}(W_{iM},\theta)\right\|^4\Big]<\infty,
\end{equation}
\begin{equation}\label{eqn:7tw}
V_{TWM}^{-1/2}L_M(\theta^*_M)^{-1}\frac{1}{\sqrt{M}}\sum^M_{i=1}\frac{R_{iM}}{\sqrt{\rho_{uM}\rho_{gM}\rho_{hM}}}m_{iM}(W_{iM},\theta^*_M)\overset{d}\to \mathcal{N}(\textbf{0},I_k)
\end{equation}
by Theorem 1 in \cite{yap2023general} under Assumption \ref{asmp:clt_reg}.

Due to (\ref{eqn:7tw}), 
\begin{equation}
\begin{aligned}
V_{TWM}^{-1/2}\sqrt{N}(\hat{\theta}_N-\theta^*_M)=&-V_{TWM}^{-1/2}L_M(\theta^*_M)^{-1}\frac{1}{\sqrt{M}}\sum^M_{i=1}\frac{R_{iM}}{\sqrt{\rho_{uM}\rho_{gM}\rho_{hM}}}m_{iM}(W_{iM},\theta^*_M)\\
&+o_p(1)O_p(1)+o_p(1)\overset{d}\to \mathcal{N}(\textbf{0},I_k).
\end{aligned}
\end{equation}


\bigskip
\noindent
\textbf{Proof of Proposition \ref{prop:cgm}:}
We define:
\begin{equation}
V_{\Delta TWSM} (\theta) := \Delta_{ehw,M} (\theta) + \rho_{uM} \Delta_{(G \cap H),M} (\theta) + \rho_{uM} \rho_{hM} \Delta_{G,M} (\theta) + \rho_{uM} \rho_{gM} \Delta_{H,M} (\theta).
\end{equation}

We first show pointwise convergence for a given $\theta$ in that $V_{\Delta TWSM} (\theta)^{-1} (\hat{V}_{\Delta TWSM} (\theta) - V_{\Delta TWSM} (\theta)) = o_p(1) I_k$. 
By applying Lemma \ref{lem:HL19_62}, 
\begin{align*}
\left(  \Delta_{ehw,M} (\theta^*_M) + \rho_{uM} \Delta_{G \cap H,M} (\theta^*_M) + \rho_{uM} \rho_{hM} \Delta_{G,M} (\theta^*_M)  \right)^{-1} \left( \hat{\Delta}_{ehw,N}(\hat{\theta}_N) + \hat{\Delta}_{G,N} (\hat{\theta}_N) \right) \xrightarrow{p} I_k \\
\left(  \Delta_{ehw,M} (\theta^*_M) + \rho_{uM} \Delta_{G \cap H,M} (\theta^*_M) + \rho_{uM} \rho_{gM} \Delta_{H,M} (\theta^*_M)  \right)^{-1} \left( \hat{\Delta}_{ehw,N}(\hat{\theta}_N) + \hat{\Delta}_{H,N} (\hat{\theta}_N) \right) \xrightarrow{p} I_k 
\end{align*}

It remains to consider the intersection terms. Consider the following condition
\begin{equation} \label{eqn:MGH_condition}
\frac{\lambda^{G \cap H}_M}{\lambda^{G}_M}=o(1).
\end{equation}
We show that when (\ref{eqn:MGH_condition}) fails, the intersection variance can be consistently estimated, i.e.,
\begin{equation} \label{eqn:GH_converge}
(\rho_{uM} \Delta_{G \cap H,M}  + \Delta_{ehw,M})^{-1} \left(\hat{\Delta}_{G \cap H,N} + \hat{\Delta}_{ehw,N} (\theta) - \rho_{uM}\Delta_{G \cap H,M} (\theta) - \Delta_{ehw,M}(\theta) \right)=o_p(1) I_k,
\end{equation}
but when it holds, the intersection term is negligible:
\begin{equation} \label{eqn:GH_negligible}
 (\rho_{uM} \rho_{hM} \Delta_{G,M} + \rho_{uM} \Delta_{G \cap H,M} + \Delta_{ehw,M})^{-1} \left(\hat{\Delta}_{G \cap H,N} (\hat{\theta}) + \hat{\Delta}_{ehw,N} (\hat{\theta}) \right) = o_p(1) I_k.
\end{equation}

If (\ref{eqn:MGH_condition}) fails, then $C\frac{\lambda^{G \cap H}_M}{\lambda^{G}_M}\ne o(1)$, so $\lambda^{G \cap H}_M \geq C\lambda^{G}_M$. Due to Lemma \ref{lem:wlln}, 
\begin{equation}
\hat{\Delta}_{G \cap H,N} + \hat{\Delta}_{ehw,N} = \frac{1 + o_p(1)}{M \rho_{uM} \rho_{gM} \rho_{hM}} \sum_{g,h}\sum_{i \in \mathcal{N}^{G\cap H}_{(g,h)}}\sum_{j \in \mathcal{N}^{G\cap H}_{(g,h)}}R_{iM}R_{jM}\cdot m_{iM}(W_{iM}, \theta)m_{jM}(W_{jM}, \theta)'.
\end{equation}
Hence,
\small
\begin{align*}
& \left\Vert(\rho_{uM} \Delta_{G \cap H,M}  + \Delta_{ehw,M})^{-1} \left(\hat{\Delta}_{G \cap H,N} + \hat{\Delta}_{ehw,N} (\theta) - \rho_{uM}\Delta_{G \cap H,M} (\theta) - \Delta_{ehw,M}(\theta) \right)\right\Vert	\\
&\leq \frac{C (1+o_p(1))}{\lambda^G_M} \left\Vert\sum_{g,h} \sum_{i,j\in\mathcal{N}_{(g,h)}^{G\cap H}} 
 \left(R_{iM} R_{jM} m_{iM}(W_{iM}, \theta) m_{jM}(W_{jM}, \theta)^\prime-\E\left[R_{iM} R_{jM} m_{iM}(W_{iM}, \theta) m_{jM}(W_{jM}, \theta)^\prime\right]\right)\right\Vert
\tag{\stepcounter{equation}\theequation}
\end{align*}
\normalsize
The term then converges in probability as 
\small
\begin{equation} \label{eqn:intersection_converge}
\begin{split}
 \mathbb{P} &\left( \frac{1}{\lambda^G_M} \left\Vert\sum_{g,h} \sum_{i,j\in\mathcal{N}_{(g,h)}^{G\cap H}} 
 \left(R_{iM} R_{jM} m_{iM}(W_{iM}, \theta) m_{jM}(W_{jM}, \theta)^\prime-\E\left[R_{iM} R_{jM} m_{iM}(W_{iM}, \theta) m_{jM}(W_{jM}, \theta)^\prime\right]\right)\right\Vert > \epsilon \right) \\
 &\leq \frac{1}{\epsilon (\lambda^G_M)^2 } \E\left\Vert \sum_{i} \sum_{j\in\mathcal{N}_{(g(i),h(i))}^{G\cap H}} 
 \left(R_{iM} R_{jM} m_{iM}(W_{iM}, \theta) m_{jM}(W_{jM}, \theta)^\prime-\E\left[R_{iM} R_{jM} m_{iM}(W_{iM}, \theta) m_{jM}(W_{jM}, \theta)^\prime \right] \right) \right\Vert^2 \\
 &\leq \frac{1}{\epsilon (\lambda^G_M)^2 } \sum_{g,h} \sum_{g^\prime, h^\prime} \sum_{i,j \in \mathcal{N}^{G \cap H}_{(g,h)}}  \sum_{k,l \in \mathcal{N}^{G \cap H}_{(g^\prime, h^\prime)}} \left( A_{ik} + A_{il} + A_{jk} + A_{jl}\right) \\
 &\leq \frac{1}{\epsilon (\lambda^G_M)^2 } \sum_{g} \sum_{g^\prime} \sum_{i,j \in \mathcal{N}^{G }_{g}}  \sum_{k,l \in \mathcal{N}^{G }_{g^\prime}} \left( A_{ik} + A_{il} + A_{jk} + A_{jl}\right) = o(1)
\end{split}
\end{equation}
\normalsize
due to the same argument as Lemma \ref{lem:HL19_62}.

If (\ref{eqn:MGH_condition}) holds, then, due to Assumption \ref{asmp:cross_var}, pointwise convergence holds as
\begin{align*}
& \left\Vert(\rho_{uM} \rho_{hM} \Delta_{G,M} + \rho_{uM} \Delta_{G \cap H,M} + \Delta_{ehw,M})^{-1} \left(\hat{\Delta}_{G \cap H,N} (\hat{\theta}) + \hat{\Delta}_{ehw,N} (\hat{\theta}) \right)\right\Vert	\\
	\leq &\frac{C (1+o_p(1))}{\lambda^G_M}\left\Vert\sum_{g}\sum_{h}\sum_{i\in\mathcal{N}_{(g,h)}^{G\cap H}}\sum_{j\in\mathcal{N}_{(g,h)}^{G\cap H}} R_{iM} R_{jM} m_{iM}(W_{iM}, \theta) m_{jM}(W_{jM}, \theta)^\prime \right\Vert\\
	\leq &\frac{C (1+o_p(1))}{\lambda^G_M} \left\Vert\sum_{g,h} \sum_{i,j\in\mathcal{N}_{(g,h)}^{G\cap H}} \left(R_{iM} R_{jM} m_{iM}(W_{iM}, \theta) m_{jM}(W_{jM}, \theta)^\prime-\E\left[R_{iM} R_{jM} m_{iM}(W_{iM}, \theta) m_{jM}(W_{jM}, \theta)^\prime\right]\right)\right\Vert \\
 &+ \frac{C (1+o_p(1))}{\lambda^G_M} \left\Vert\sum_{g}\sum_{h}\sum_{i\in\mathcal{N}_{(g,h)}^{G\cap H}}\sum_{j\in\mathcal{N}_{(g,h)}^{G\cap H}}\E\left[R_{iM} R_{jM} m_{iM}(W_{iM}, \theta) m_{jM}(W_{jM}, \theta)^\prime\right]\right\Vert \\
 =& o_p(1) + C (1+o_p(1))\frac{\lambda^{G \cap H}_{M}}{\lambda^G_M} =o_p(1).
 \tag{\stepcounter{equation}\theequation}
\end{align*}
To obtain the convergence in the final line, apply the convergence in probability from (\ref{eqn:intersection_converge}) and condition (\ref{eqn:MGH_condition}). 
By applying the continuous mapping theorem, $\hat{V}_{\Delta TWSM} (\theta)$ converges in probability.
Next, we proceed with uniform convergence. Here, $\frac{M}{\lambda_M}\big(\Delta_{ehw,M}(\theta)+\rho_{uM} \Delta_{(G\cap H),M}(\theta) +\rho_{uM}\rho_{hM}\Delta_{G,M}(\theta)+\rho_{uM}\rho_{gM}\Delta_{H,M}(\theta)\big)$ is continuous in $\theta$ for all $M$ by the dominated convergence theorem (DCT), Jensen's inequality, and Cauchy-Schwarz inequality under conditions (\romannumeral 4) and (\romannumeral 7) in Assumption \ref{assumpa3}.
In addition,
\begin{align*}
& \frac{N}{\lambda^G_M} \left\|\hat{V}_{\Delta TWSM} (\tilde{\theta}) - \hat{V}_{\Delta TWSM} ({\theta}) \right\| \\
\leq & \frac{1}{\lambda^G_M}\sum^G_{g=1}\Bigg\|\bigg[\sum_{i \in \mathcal{N}^G_g}R_{iM}\cdot m_{iM}(W_{iM},\tilde{\theta})\bigg]\bigg[\sum_{i \in \mathcal{N}^G_g}R_{iM}\cdot m_{iM}(W_{iM},\tilde{\theta})\bigg]'\\
&-\bigg[\sum_{i \in \mathcal{N}^G_g}R_{iM}\cdot m_{iM}(W_{iM},\theta)\bigg]\bigg[\sum_{i \in \mathcal{N}^G_g}R_{iM}\cdot m_{iM}(W_{iM},\theta)\bigg]'\Bigg\|\\
&+\frac{1}{\lambda^G_M}\sum^H_{h=1}\Bigg\|\bigg[\sum_{i \in \mathcal{N}^H_h}R_{iM}\cdot m_{iM}(W_{iM},\tilde{\theta})\bigg]\bigg[\sum_{i \in \mathcal{N}^H_h}R_{iM}\cdot m_{iM}(W_{iM},\tilde{\theta})\bigg]'\\
&-\bigg[\sum_{i \in \mathcal{N}^H_h}R_{iM}\cdot m_{iM}(W_{iM},\theta)\bigg]\bigg[\sum_{i \in \mathcal{N}^H_h}R_{iM}\cdot m_{iM}(W_{iM},\theta)\bigg]'\Bigg\|\\
&+\frac{1}{\lambda^G_M}\sum^G_{g=1} \sum^H_{h=1} \Bigg\|\bigg[\sum_{i \in \mathcal{N}^{G\cap H}_{(g,h)}}R_{iM}\cdot m_{iM}(W_{iM},\tilde{\theta})\bigg]\bigg[\sum_{i \in \mathcal{N}^{G\cap H}_{(g,h)}}R_{iM}\cdot m_{iM}(W_{iM},\tilde{\theta})\bigg]'\\
&-\bigg[\sum_{i \in \mathcal{N}^{G\cap H}_{(g,h)}}R_{iM}\cdot m_{iM}(W_{iM},\theta)\bigg]\bigg[\sum_{i \in \mathcal{N}^{G\cap H}_{(g,h)}}R_{iM}\cdot m_{iM}(W_{iM},\theta)\bigg]'\Bigg\|.
\tag{\stepcounter{equation}\theequation}
\end{align*}
By applying the same expansion steps as (C.47)-(C.50) for each of the three terms, we conclude that 
\begin{equation}
\frac{N}{\lambda^G_M} \left\|\hat{V}_{\Delta TWSM} (\tilde{\theta}) - \hat{V}_{\Delta TWSM} ({\theta}) \right\| \leq B^2_N h ( \| \tilde{\theta} - \theta \| ),
\end{equation}
where $ B^2_N = O_p(1)$.

Since the smallest eigenvalue is bounded below from (viii) in Assumption \ref{assumpa3},
\begin{align*}
&\Big\|V_{\Delta TWSM}(\theta^*_M)^{-1} \big[\hat{V}_{\Delta TWSM} (\hat{\theta}_N) - V_{\Delta TWSM} ({\theta^*_M})\big]\Big\|\\
\leq&C\bigg(\sup_{\theta\in\Theta}\|\hat{V}_{\Delta TWSM} (\theta)-V_{\Delta TWSM} (\theta) \| \frac{N}{\lambda^G_M}+ \|V_{\Delta TWSM} (\hat{\theta}_N)-V_{\Delta TWSM} (\theta^*_M) \| \frac{N}{\lambda^G_M}\bigg) =o_p(1)\tag{\stepcounter{equation}\theequation}
\end{align*}
by Corollary 2.2 in \cite{newey1991uniform} under $\hat{\theta}_N-\theta^*_M\overset{p}\to \textbf{0}$. Hence,
\begin{align*}
\hat{V}_{\Delta TWSM} (\hat{\theta}_N) &= V_{\Delta TWSM} (\theta^*_M) \left[ I_k +  V_{\Delta TWSM} (\theta^*_M)^{-1} \left( \hat{V}_{\Delta TWSM} (\hat{\theta}_N) -   V_{\Delta TWSM} (\theta^*_M)\right) \right] \\
&= V_{\Delta TWSM} (\theta^*_M) (I_k + o_p(1) )
\end{align*}
Given the convergence of $\hat{L}_N(\hat{\theta}_N)$ from (\ref{eqn:Lconverge_tw}), the entire object converges using the same argument as in Theorem \ref{thm:oneway}.

Since $\hat{V}_{CGM}$ is consistent for $V_{TWSM}$, it remains to compare $V_{TWSM}$ with $V_{TWM}$.
\begin{align*}
    V_{TWSM} - V_{TWM} =&L_M(\theta^*_M)^{-1} \bigg(  \rho_{uM}\rho_{gM}\rho_{hM}\Delta_{E,M}+\rho_{uM} \rho_{gM}\rho_{hM} \Delta_{E(G\cap H),M}\\
    &+\rho_{uM}\rho_{gM}\rho_{hM}\Delta_{EG,M}+\rho_{uM}\rho_{gM}\rho_{hM}\Delta_{EH,M} \bigg) L_M(\theta^*_M)^{-1} \\
    =&\rho_{uM}\rho_{gM}\rho_{hM} L_M(\theta^*_M)^{-1} \bigg(  \Delta_{E,M}+  \Delta_{E(G\cap H),M} +\Delta_{EG,M}+\Delta_{EH,M} \bigg) L_M(\theta^*_M)^{-1}
    \tag{\stepcounter{equation}\theequation}
\end{align*}
It is possible to construct a data generating process where $\Delta_{E,M}+  \Delta_{E(G\cap H),M} +\Delta_{EG,M}+\Delta_{EH,M} < \bm{0}$, as seen in Example B.1.

\bigskip
\noindent
\textbf{Proof of Proposition \ref{prop:cgm2}:}
Pointwise convergence of the $\Delta$ objects is immediate from applying Lemma \ref{lem:HL19_62} under Assumption \ref{asmp:cross_var} since the estimator and estimand can be written as an additive combination of one-way cluster-robust objects. 
The argument for uniform convergence is similar to that of the previous proposition. 

\begin{align*}
    V_{TWSM2} - V_{TWM} =&L_M(\theta^*_M)^{-1} \bigg( \Delta_{ehw,M}(\theta^*_M) + \rho_{uM} \Delta_{(G\cap H),M} (\theta^*_M) \\
    &+  \rho_{uM}\rho_{gM}\rho_{hM}\Delta_{E,M}+\rho_{uM} \rho_{gM}\rho_{hM} \Delta_{E(G\cap H),M}\\
    &+\rho_{uM}\rho_{gM}\rho_{hM}\Delta_{EG,M}+\rho_{uM}\rho_{gM}\rho_{hM}\Delta_{EH,M} \bigg) L_M(\theta^*_M)^{-1}
    \tag{\stepcounter{equation}\theequation}
\end{align*}
Given 
\begin{align*}
    \Delta_{ehw,M}(\theta^*_M)+\rho_{uM} \Delta_{(G\cap H),M} (\theta^*_M)  \geq \rho_{uM}\rho_{gM}\rho_{hM}\Delta_{E,M}+\rho_{uM} \rho_{gM}\rho_{hM} \Delta_{E(G\cap H),M},
    \tag{\stepcounter{equation}\theequation}
\end{align*}
\begin{align*}
     V_{TWSM2} - V_{TWM} \geq&\rho_{uM}\rho_{gM}\rho_{hM} L_M(\theta^*_M)^{-1} \bigg( 2 \Delta_{E,M}+ 2 \Delta_{E(G\cap H),M} +\Delta_{EG,M}+\Delta_{EH,M} \bigg) L_M(\theta^*_M)^{-1}.
     \tag{\stepcounter{equation}\theequation}
\end{align*}
Since $\Delta_{E,M}+ \Delta_{E(G\cap H),M} +\Delta_{EG,M} \geq 0$ and $\Delta_{E,M}+ \Delta_{E(G\cap H),M} +\Delta_{EH,M} \geq 0$, we have a conservative estimand.

\subsection{Proofs for Section 3}

\bigskip
\noindent
\textbf{Proof of Theorem \ref{thm:unit_shrink}:}

Let
\begin{equation}
K_M=\bigg(\sum\limits^M_{i=1}z_{iM}'z_{iM}\bigg)^{-1}\bigg\{\sum\limits^M_{i=1}z_{iM}'\mathbb{E}\big[m_{iM}(W_{iM},\theta^*_M)\big]'\bigg\}.
\end{equation}
To show $\left\|\hat{K}_N-K_M\right\|\overset{p}\to 0$, we first show 
\begin{equation}\label{eqn:b65}
\left\|\frac{1}{N}\sum^M_{i=1}R_{iM}z_{iM}'z_{iM}-\frac{1}{M}\sum^M_{i=1}z_{iM}'z_{iM}\right\|\overset{p}\to 0.
\end{equation}
We can write
\begin{equation} 
\frac{1}{N}\sum^M_{i=1}R_{iM}z_{iM}'z_{iM}=\frac{M\rho_{uM}\rho_{gM}\rho_{hM}}{N}\frac{1}{M}\sum^M_{i=1}\frac{R_{iM}}{\rho_{uM}\rho_{gM}\rho_{hM}}z_{iM}'z_{iM}.
\end{equation}
Since $\frac{M\rho_{uM}\rho_{gM}\rho_{hM}}{N}\overset{p}\to 1$, it suffices to show
\begin{equation} \label{eq:A9}
\left\|\frac{1}{M}\sum^M_{i=1}\frac{R_{iM}}{\rho_{uM}\rho_{gM}\rho_{hM}}z_{iM}'z_{iM}-\frac{1}{M}\sum^M_{i=1}z_{iM}'z_{iM}\right\|\overset{p}\to 0.
\end{equation}
Given
\begin{equation}
\sup_{i,M}\mathbb{E}\Bigg[\left\|\frac{R_{iM}}{\rho_{uM}\rho_{gM}\rho_{hM}}z_{iM}'z_{iM}\right\|^2\Bigg]=\frac{1}{\rho_{uM}\rho_{gM}\rho_{hM}}\sup_{i,M}\left\|z_{iM}\right\|^4 <\infty
\end{equation}
under condition (\romannumeral 2) in Theorem \ref{thm:unit_shrink}, (\ref{eq:A9}) is implied by Lemma \ref{lem:wlln}. 

Next, we show 
\begin{equation}\label{eqn:b69}
\left\|\frac{1}{N}\sum^M_{i=1}R_{iM}\cdot m_{iM}(W_{iM},\hat{\theta}_N)z_{iM}-\frac{1}{M}\sum\limits^M_{i=1}\mathbb{E}\big[m_{iM}(W_{iM},\theta^*_M)\big]z_{iM}\right\|\overset{p}\to 0.
\end{equation}
Again, we can write
\begin{align*}
\frac{1}{N}\sum^M_{i=1}R_{iM}\cdot m_{iM}(W_{iM},\hat{\theta}_N)z_{iM}=&\frac{M\rho_{uM}\rho_{gM}\rho_{hM}}{N}\frac{1}{M}\sum^M_{i=1}\frac{R_{iM}}{\rho_{uM}\rho_{gM}\rho_{hM}}m_{iM}(W_{iM},\hat{\theta}_N)z_{iM}\\
=&\big(1+o_p(1)\big)\frac{1}{M}\sum^M_{i=1}\frac{R_{iM}}{\rho_{uM}\rho_{gM}\rho_{hM}}m_{iM}(W_{iM},\hat{\theta}_N)z_{iM}.
\tag{\stepcounter{equation}\theequation}
\end{align*}
 We first show $\forall\ \theta\in\Theta$
\begin{equation} \label{eq:A13}
\left\|\frac{1}{M}\sum^M_{i=1}\frac{R_{iM}}{\rho_{uM}\rho_{gM}\rho_{hM}}m_{iM}(W_{iM},\theta)z_{iM}-\frac{1}{M}\sum^M_{i=1}\mathbb{E}\big[m_{iM}(W_{iM},\theta)\big]z_{iM}\right\|\overset{p}\to 0.
\end{equation}
Since $\forall\ \theta\in\Theta$
\begin{equation}
\begin{aligned}
&\sup_{i,M}\mathbb{E}\Bigg[\left\|\frac{R_{iM}}{\rho_{uM}\rho_{gM}\rho_{hM}}m_{iM}(W_{iM},\theta)z_{iM}\right\|^2\Bigg]\\
\leq&\frac{1}{\rho_{uM}\rho_{gM}\rho_{hM}}\bigg\{\sup_{i,M}\mathbb{E}\Big[\sup_{\theta\in\Theta}\left\|m_{iM}(W_{iM},\theta)\right\|^4\Big]\bigg\}^{1/2}\sup_{i,M}\left\|z_{iM}\right\|^2<\infty,
\end{aligned}
\end{equation}
 (\ref{eq:A13}) holds Lemma \ref{lem:wlln}. 
Next, we show the Lipschitz condition. $\forall\ \tilde{\theta},\theta\in\Theta$
\begin{equation}
\begin{aligned}
\left\|m_{iM}(W_{iM},\tilde{\theta})z_{iM}- m_{iM}(W_{iM},\theta)z_{iM}\right\|\leq&\left\|z_{iM}\right\|\cdot\left\|m_{iM}(W_{iM},\tilde{\theta})-m_{iM}(W_{iM},\theta)\right\|\\
\leq &\left\|z_{iM}\right\|b_{1,iM}(W_{iM})h(\|\tilde{\theta}-\theta\|)
\end{aligned}
\end{equation} 
and
\begin{equation}
\sup_{i,M}\mathbb{E}\big[\left\|z_{iM}\right\| b_{1,iM}(W_{iM})\big]\leq\sup_{i,M}\left\|z_{iM}\right\|\sup_{i,M}\Big\{\mathbb{E}\big[b_{1,iM}(W_{iM})^2\big]\Big\}^{1/2}<\infty \end{equation}
by Jensen's inequality.
As a result, 
\begin{align*}
&\left\|\frac{1}{M}\sum^M_{i=1}\frac{R_{iM}}{\rho_{uM}\rho_{gM}\rho_{hM}}m_{iM}(W_{iM},\hat{\theta}_N)z_{iM}-\frac{1}{M}\sum\limits^M_{i=1}\mathbb{E}\big[m_{iM}(W_{iM},\theta^*_M)\big]z_{iM}\right\|\\
\leq&\sup\limits_{\theta\in\Theta}\left\|\frac{1}{M}\sum^M_{i=1}\frac{R_{iM}}{\rho_{uM}\rho_{gM}\rho_{hM}}m_{iM}(W_{iM},\theta)z_{iM}-\frac{1}{M}\sum\limits^M_{i=1}\mathbb{E}\big[m_{iM}(W_{iM},\theta)\big]z_{iM}\right\|\\
&+\left\|\frac{1}{M}\sum\limits^M_{i=1}\mathbb{E}\big[m_{iM}(W_{iM},\hat{\theta}_N)\big]z_{iM}-\frac{1}{M}\sum\limits^M_{i=1}\mathbb{E}\big[m_{iM}(W_{iM},\theta^*_M)\big]z_{iM}\right\|\overset{p}\to 0
\tag{\stepcounter{equation}\theequation}
\end{align*}
by Lemma \ref{lemma2} above and Corollary 2.2 in \cite{newey1991uniform} under $\hat{\theta}_N-\theta^*_M\overset{p}\to \textbf{0}$.
Combining (\ref{eqn:b65}) and (\ref{eqn:b69}), we conclude that $\left\|\hat{K}_N-K_M\right\|\overset{p}\to 0$.

Hence,
\begin{equation}
\hat{\Delta}^Z_N=\frac{1}{N}\sum^M_{i=1}R_{iM}\big(K_M'+o_p(1)\big)z_{iM}'z_{iM}\big(K_M+o_p(1)\big). 
\end{equation}
Let
\begin{equation} 
\Delta^Z_M=\frac{1}{M}\sum^M_{i=1}\mathbb{E}\big[m_{iM}(W_{iM}, \theta^*_M)\big]z_{iM}\bigg(\frac{1}{M}\sum^M_{i=1}z_{iM}'z_{iM}\bigg)^{-1}\frac{1}{M}\sum^M_{i=1}z_{iM}'\mathbb{E}\big[m_{iM}(W_{iM}, \theta^*_M)\big]'.
\end{equation}
\begin{equation}
\left\|\hat{\Delta}^Z_N-\Delta^Z_M\right\|=o_p(1)
\end{equation}
given (\ref{eqn:b65}).

Let $A_M$ and $D_M$ be the matrices with $i$-th rows equal to $\mathbb{E}\big[m_{iM}(W_{iM},\theta^*_M)\big]'/\sqrt{M}$ and $z_{iM}/\sqrt{M}$ respectively. Let $I_M$ be the identity matrix of size $M$. Then, 
\begin{equation} \label{eq:Delta_EM_psd}
\Delta_{E,M}-\Delta^Z_M=A_M'\big(I_M-D_M(D_M'D_M)^{-1}D_M'\big)A_M,
\end{equation}
which is positive semidefinite.
Hence, the result.

\bigskip
\noindent
\textbf{Proof of Theorem \ref{thm:shrink_oneway}:} The proof is almost the same as that for Theorem \ref{thm:shrink} with sampling indicators and hence is omitted here.

\bigskip
\noindent
\textbf{Proof of Theorem \ref{thm:shrink}:}
Recall that 
\begin{equation}
\hat{P}_N=\bigg(\sum^G_{g=1}\tilde{z}_{gM}'\tilde{z}_{gM}\bigg)^{-1}\bigg(\sum^G_{g=1}\tilde{z}_{gM}'\tilde{m}_{gM}(\hat{\theta}_N)'\bigg).
\end{equation}
Let 
\begin{equation}
P_M:=\Bigg[\sum^G_{g=1}\tilde{z}_{gM}'\tilde{z}_{gM}\Bigg]^{-1}\sum^G_{g=1}\tilde{z}_{gM}'\mathbb{E}\big[\tilde{m}_{gM}(\theta^*_M)\big]'.
\end{equation}
To show convergence of these objects, we require the order of variances for $z$ to be similar to that of $m$. Namely, for $C \in \{ G,H \}$ and arbitrary constants $a >0$, $A < \infty$, 
\begin{equation} \label{eqn:zm_order}
\begin{split}
a \leq \frac{1}{\lambda^G_M} \lambda_{min} \left(  \sum_{i=1}^{M} \sum_{j \in \mathcal{N}^{C}_{c(i)}} \mathbb{E}[ z_{iM} m_{jM}(W_{jM}, \theta^*_M)'] \right) \leq A \\
    a \leq \frac{1}{\lambda^G_M} \lambda_{min} \left(  \sum_{i=1}^{M} \sum_{j \in \mathcal{N}^{C}_{c(i)}} \mathbb{E}[ z_{iM} z_{jM}'] \right) \leq A
\end{split}
\end{equation}
$\lambda^G_M$ is the scaling factor such that $\frac{1}{\lambda^G_M} \sum^G_{g=1}\tilde{z}_{gM}'\tilde{z}_{gM} =O(1) $ but not $o(1)$. 
To show $\left\|\hat{P}_N-P_M\right\|\overset{p}\to 0$, we show 
\begin{equation}\label{eq6}
\frac{1}{\lambda^G_M}\left\|\sum^G_{g=1}\tilde{m}_{gM}(\hat{\theta}_N)\tilde{z}_{gM}-\sum^G_{g=1}\mathbb{E}\big[\tilde{m}_{gM}(\theta^*_M)\big]\tilde{z}_{gM}\right\|\overset{p}\to 0.
\end{equation}

As a first step, we show $\forall\ \theta\in\Theta$
\begin{equation}\label{eq4}
\frac{1}{\lambda^G_M} \left\|\sum^G_{g=1}\tilde{m}_{gM}(\theta)\tilde{z}_{gM} -\sum^G_{g=1}\mathbb{E}\big[\tilde{m}_{gM}(\theta)\big]\tilde{z}_{gM}\right\|\overset{p}\to 0.
\end{equation}
Fix $\delta>0$. Set $\epsilon=(\delta/C)^2$. Let 
\begin{equation}
\tilde{l}_{gM}:=\tilde{m}_{gM}(\theta)\tilde{z}_{gM} \mathbbm{1}\bigg(\left\|\tilde{m}_{gM}(\theta)\tilde{z}_{gM} \right\|\leq M\epsilon\bigg).
\end{equation}
Then
\begin{equation}
\begin{aligned}
&\mathbb{E}\Bigg[ \frac{1}{\lambda^G_M}\left\|\sum^G_{g=1}\tilde{m}_{gM}(\theta)\tilde{z}_{gM}-\sum^G_{g=1}\mathbb{E}\big[\tilde{m}_{gM}(\theta)\big]\tilde{z}_{gM}\right\|\Bigg]\\
\leq&\frac{1}{\lambda^G_M}\mathbb{E}\Bigg\{\left\|\sum^G_{g=1}\Big[\tilde{l}_{gM}-\mathbb{E}\big(\tilde{l}_{gM}\big)\Big]\right\|\Bigg\}\\
&+\frac{2}{\lambda^G_M}\sum^G_{g=1}\mathbb{E}\Bigg[\left\|\tilde{m}_{gM}(\theta)\tilde{z}_{gM}\right\|\mathbbm{1}\bigg(\left\|\tilde{m}_{gM}(\theta)\tilde{z}_{gM}\right\|> M\epsilon\bigg)\Bigg].
\end{aligned}
\end{equation}
Observe that
\begin{equation}\label{eq2}
\begin{aligned}
&\frac{1}{\lambda^G_M}\mathbb{E}\Bigg[\left\|\sum^G_{g=1}\Big(\tilde{l}_{gM}-\mathbb{E}\big(\tilde{l}_{gM}\big)\Big)\right\|\Bigg]\\
\leq& \frac{1}{\lambda^G_M}\Bigg\{\mathbb{E}\Bigg[\left\|\sum^G_{g=1}\Big(\tilde{l}_{gM}-\mathbb{E}\big(\tilde{l}_{gM}\big)\Big)\right\|^2\Bigg]\Bigg\}^{1/2}\\
\leq &\frac{1}{\lambda^G_M}\Bigg\{\sum^G_{g=1}\mathbb{E}\Big[\left\|\tilde{l}_{gM}\right\|^2\Big]\Bigg\}^{1/2} \\
\leq &4 \delta
\end{aligned}
\end{equation}
follows by an argument similar to the proof of Lemma \ref{lem:HL19_62}.
Also, 
\begin{equation}
\begin{aligned}
&\sup_{g,M}\mathbb{E}\bigg[\left\|\tilde{m}_{gM}(\theta)\tilde{z}_{gM}\Big/\left(M^G_g\right)^2\right\|^2\bigg]\\
\leq &\bigg\{\sup_{g,M}\mathbb{E}\Big[\sup_{\theta\in\Theta}\left\|\tilde{m}_{gM}(\theta)/\left( M_g^G\right)\right\|^{4}\Big]\bigg\}^{1/2}\bigg\{\sup_{g,M}\mathbb{E}\Big[\left\|\tilde{z}_{gM}/\left( M_g^G\right)\right\|^{4}\Big]\bigg\}^{1/2}\\
\leq & \bigg\{\sup_{g,M}\sum_{i\in \mathcal{N}_g^G}\left(\mathbb{E}\Big[\sup_{\theta\in\Theta}\left\|m_{iM}(W_{iM},\theta)/\left( M_g^G\right)\right\|^{4}\Big]\right)^{1/4}\bigg\}^2\\
&\cdot \bigg\{\sup_{g,M}\sum_{i\in \mathcal{N}_g^G}\left(\mathbb{E}\Big[\left\|z_{iM}/\left( M_g^G\right)\right\|^{4}\Big]\right)^{1/4}\bigg\}^2\\
\leq & \sup_{i,M}\mathbb{E}\left[\sup_{\theta\in\Theta}\left\|m_{iM}(W_{iM},\theta)\right\|^{4}\right]^{1/2}\sup_{i,M}\left\|z_{iM}\right\|^2<\infty
\end{aligned}
\end{equation}
by Jensen's inequality under condition (\romannumeral 7) in Assumption \ref{assumpa3} and condition (\romannumeral 2) in Theorem \ref{thm:shrink}.
Hence, we can pick $B$ sufficiently large so that 
\begin{equation}
\sup_{g,M}\mathbb{E}\Bigg[\left\|\tilde{m}_{gM}(\theta)\tilde{z}_{gM}\Big/\left(M^G_g\right)^2\right\|\mathbbm{1}\bigg(\left\|\tilde{m}_{gM}(\theta)\tilde{z}_{gM}\Big/\left(M^G_g\right)^2\right\|>B\bigg)\Bigg]\leq \frac{\delta}{C}.
\end{equation}
Pick $M$ large enough so that 
\begin{equation}
\max_{g\leq G}\frac{\left(M^G_g\right)^2}{\lambda^G_M}\leq \frac{\epsilon}{B},
\end{equation} 
which is feasible under Assumption \ref{asmp:cross_var}.
Then
\begin{equation}\label{eq3}
\frac{2}{\lambda^G_M}\sum^G_{g=1}\mathbb{E}\Bigg[\left\|\tilde{m}_{gM}(\theta)\tilde{z}_{gM}\right\|\mathbbm{1}\bigg(\left\|\tilde{m}_{gM}(\theta)\tilde{z}_{gM}\right\|>M\epsilon\bigg)\Bigg]\leq \frac{2}{\lambda^G_M}\sum^G_{g=1}\left(M^G_g\right)^2\frac{\delta}{C}\leq 2\delta.
\end{equation}
Combining (\ref{eq2}) and (\ref{eq3}), (\ref{eq4}) holds by Markov's inequality.

Next, 
\begin{align*}
&\left\|\frac{1}{\lambda^G_M}\sum^G_{g=1}\big[\tilde{m}_{gM}(\tilde{\theta})\tilde{z}_{gM}-\tilde{m}_{gM}(\theta)\tilde{z}_{gM}\big]\right\|\\
=&\frac{1}{\lambda^G_M}\sum^G_{g=1}\left\|\sum_{i \in \mathcal{N}^G_g}\sum_{j \in \mathcal{N}^G_g} m_{iM}(W_{iM},\tilde{\theta})z_{jM}-\sum_{i \in \mathcal{N}^G_g}\sum_{j \in \mathcal{N}^G_g} m_{iM}(W_{iM},\theta)z_{jM}\right\|\\
\leq&\frac{1}{\lambda^G_M}\sum^G_{g=1}\sum_{i \in \mathcal{N}^G_g}\sum_{j \in \mathcal{N}^G_g} \left\|m_{iM}(W_{iM},\tilde{\theta})-m_{iM}(W_{iM},\theta)\right\|\cdot\left\|z_{jM}\right\|\\
\leq&\frac{1}{\lambda^G_M}\sum^G_{g=1}\sum_{i \in \mathcal{N}^G_g}\sum_{j \in \mathcal{N}^G_g} b_{1,iM}(W_{iM})\cdot\left\|z_{jM}\right\|\cdot h(\|\tilde{\theta}-\theta\|) \tag{\stepcounter{equation}\theequation}
\end{align*}
Let 
\begin{align*}
B^3_N:=&\frac{1}{\lambda^G_M}\sum^G_{g=1}\sum_{i \in \mathcal{N}^G_g}\sum_{j \in \mathcal{N}^G_g} b_{1,iM}(W_{iM})\cdot\left\|z_{jM}\right\|.
\tag{\stepcounter{equation}\theequation}
\end{align*}
Since 
\begin{equation}
\begin{aligned}
\mathbb{E}(B_N^3)&= \frac{1}{\lambda^G_M}\sum^G_{g=1}\sum_{i \in \mathcal{N}^G_g}\sum_{j \in \mathcal{N}^G_g}  \mathbb{E}\big[b_{1,iM}(W_{iM})\big]\left\|z_{iM}\right\|\\
&\leq\frac{1}{\lambda^G_M}\sum^G_{g=1}\left(M^G_g\right)^2\sup_{i,M}\Big\{\mathbb{E}\big[b_{1,iM}(W_{iM})^2\big]\Big\}^{1/2}\sup_{i,M}\left\|z_{iM}\right\|<\infty
\end{aligned}
\end{equation}
by Jensen's inequality under condition (\romannumeral 12) in Assumption \ref{assumpa3}, condition (\romannumeral 2) in Theorem \ref{thm:shrink}, and Assumption \ref{asmp:cross_var}, $B^3_N=O_p(1)$ by Markov's inequality. Also, $\frac{1}{\lambda^G_M}\sum^G_{g=1}\mathbb{E}\big[\tilde{m}_{gM}(\theta)\big]\tilde{z}_{gM}$ is continuous in $\theta$ for all $M$ by the DCT and Jensen's inequality under Assumption \ref{asmp:cross_var} and conditions (\romannumeral 4) and (\romannumeral 7) in Assumption \ref{assumpa3}. 
As a result,
\begin{align*}
&\frac{1}{\lambda^G_M}\left\|\sum^G_{g=1}\tilde{m}_{gM}(\hat{\theta}_N)\tilde{z}_{gM}-\sum^G_{g=1}\mathbb{E}\big[\tilde{m}_{gM}(\theta^*_M)\big]\tilde{z}_{gM}\right\|\\
\leq&\frac{1}{\lambda^G_M}\sup_{\theta\in\Theta}\left\|\sum^G_{g=1}\tilde{m}_{gM}(\theta)\tilde{z}_{gM}-\sum^G_{g=1}\mathbb{E}\big[\tilde{m}_{gM}(\theta)\big]\tilde{z}_{gM}\right\|\\
&+\frac{1}{\lambda^G_M}\left\|\sum^G_{g=1}\mathbb{E}\big[\tilde{m}_{gM}(\hat{\theta}_N)\big]\tilde{z}_{gM}-\sum^G_{g=1}\mathbb{E}\big[\tilde{m}_{gM}(\theta^*_M)\big]\tilde{z}_{gM}\right\|\overset{p}\to 0. \tag{\stepcounter{equation}\theequation}
\end{align*}
follows by Corollary 2.2 in \cite{newey1991uniform} under $\hat{\theta}_N-\theta^*_M\overset{p}\to \textbf{0}$.

The result $ \left\|\hat{P}_N-P_M\right\|\overset{p}\to 0$ follows immediately under the continuity of inversion and multiplication.

Denote $\Delta^Z_{CE,M}:=\frac{1}{M}\sum\limits^G_{g=1}P_M'\tilde{z}_{gM}'\tilde{z}_{gM}P_M$. 
We can write
\begin{align*}
\hat{\Delta}^Z_{CE,N}=& \frac{1}{M}\sum^G_{g=1} \hat{P}_N'\tilde{z}_{gM}'\tilde{z}_{gM}\hat{P}_N\\
=&\frac{1}{M}\sum^G_{g=1} \big(P_M'+o_p(1)\big)\tilde{z}_{gM}'\tilde{z}_{gM}\big(P_M+o_p(1)\big).
\tag{\stepcounter{equation}\theequation}
\end{align*}
Then,
\begin{equation} \label{eqn:DeltaZCE_uniform}
\left\| (\Delta^Z_{CE,M})^{-1} \left(\hat{\Delta}^Z_{CE,N}-\Delta^Z_{CE,M} \right)\right\|=o_p(1).
\end{equation}

To show the ordering of the variance-covariance matrices in Theorem \ref{thm:shrink}, notice that
\begin{align*}
&\Delta_{E,M}+\Delta_{EC,M}-\Delta^Z_{CE,M}\\
=&\frac{1}{M}\sum^G_{g=1}\mathbb{E}\big[\tilde{m}_{gM}(\theta^*_M)\big]\mathbb{E}\big[\tilde{m}_{gM}(\theta^*_M)\big]'\\
&-\frac{1}{M}\sum^G_{g=1}\mathbb{E}\big[\tilde{m}_{gM}(\theta^*_M)\big]\tilde{z}_{gM}
\Bigg[\frac{1}{M}\sum^G_{g=1}\tilde{z}_{gM}'\tilde{z}_{gM}\Bigg]^{-1}\frac{1}{M}\sum^G_{g=1}\tilde{z}_{gM}'\mathbb{E}\big[\tilde{m}_{gM}(\theta^*_M)\big]',\tag{\stepcounter{equation}\theequation}
\end{align*}
which is positive semidefinite due to the same argument as (\ref{eq:Delta_EM_psd}).

The final part of the theorem follows from previous derivation in (\ref{eqn:GH_converge}) and (\ref{eqn:GH_negligible}).

\bigskip
\subsection{Proofs for Section 5}
\noindent

\begin{lemma}\label{lem:mwclusW} 
Suppose there is multiway clustering in a binary assignment described in Section \ref{sec:diff_means}. Then, if $c(i)=c(j)$, then $\E\left[X_{iM}X_{jM}\right]=\left(\mu_{A}^{2}+\sigma_{A}^{2}\right)\left(\mu_{B}^{2}+\sigma_{B}^{2}\right)$. If $g(i)=g(j),h(i)\ne h(j)$, then $\E\left[X_{iM}X_{jM}\right]=\left(\mu_{A}^{2}+\sigma_{A}^{2}\right)\mu_{B}^{2}$. If $g(i) \ne g(j)$ and $h(i) \ne h(j)$, then $\E\left[X_{iM}X_{jM}\right]=\mu_{A}^{2}\mu_{B}^{2}$. 
\end{lemma}

\noindent
\textbf{Proof:}
\begin{align*}
\E\left[X_{iM}X_{jM}\right] & =\E\left[\E\left[X_{iM}X_{jM}|A_{g},B_{h}\right]\right]\\
 & =\E\left[A_{g}^{2}B_{h}^{2}\right] =\E\left[A_{g}^{2}\right]\E\left[B_{h}^{2}\right] =\left(\mu_{A}^{2}+\sigma_{A}^{2}\right)\left(\mu_{B}^{2}+\sigma_{B}^{2}\right)
 \tag{\stepcounter{equation}\theequation}
\end{align*}
Further, 
\begin{align*}
\E\left[X_{iM}X_{jM}\right] & =\E\left[\E\left[X_{iM}X_{jM}|A_{g},B_{h},B_{h^{\prime}},g(i)=g(j)=g,h(i)=h,h(j)=h^{\prime}\right]\right]\\
 & =\E\left[A_{g}^{2}B_{h}B_{h^{\prime}}\right] =\E\left[A_{g}^{2}\right]\E\left[B_{h}\right]^{2} =\left(\mu_{A}^{2}+\sigma_{A}^{2}\right)\mu_{B}^{2}
 \tag{\stepcounter{equation}\theequation}
\end{align*}

\begin{lemma} \label{lem:OWclusLLN}
Let $V_{iM}$ and $S_{iM}$ be two-way clustered scalar random variables that are potentially correlated that have bounded second moments. Under Assumptions \ref{assump1} to \ref{assump4}, 
\begin{equation}
\frac{1}{M}\sum_{i=1}^{M}S_{iM}\frac{1}{M^G_{g(i)}}\sum_{j\in\mathcal{N}_{g(i)}^{G}}V_{jM}-\frac{1}{M}\sum_{i=1}^{M}\sum_{j\in\mathcal{N}_{g(i)}^{G}}\frac{1}{M^G_{g(i)}}\E\left[S_{iM}V_{jM}\right]\xrightarrow{P}0
\end{equation}
\end{lemma}
\noindent
\textbf{Proof:}
Using Chebyshev's inequality,
\begin{align*}
\mathbb{P}&\left(\left(\frac{1}{M}\sum_{i=1}^{M}\sum_{j\in\mathcal{N}_{g(i)}^{G}}\frac{1}{M^G_{g(i)}}\left(S_{iM}V_{jM}-\E\left[S_{iM}V_{jM}\right]\right)\right)^{2}>\epsilon\right) \\
&\leq\frac{1}{M^{2}\epsilon^{2}}\E\left[\left(\sum_{i=1}^{M}\sum_{j\in\mathcal{N}_{g(i)}^{G}}\frac{1}{M^G_{g(i)}}\left(S_{iM}V_{jM}-\E\left[S_{iM}V_{jM}\right]\right)\right)^{2}\right] \\
&=\frac{1}{M^{2}\epsilon^{2}}Var\left(\sum_{i=1}^{M}\sum_{j\in\mathcal{N}_{g(i)}^{G}}\frac{1}{M^G_{g(i)}}S_{iM}V_{jM}\right) =\frac{1}{M^{2}\epsilon^{2}}Var\left(\sum_{g}\sum_{i,j\in\mathcal{N}_{g}^{G}}\frac{1}{M^G_{g}}S_{iM}V_{jM}\right) \\
&\leq C\frac{1}{M^{2}\epsilon^{2}}\sum_{g}\sum_{g^{\prime}}\sum_{i,j\in\mathcal{N}_{g}^{G}}\sum_{k,l\in\mathcal{N}_{g^{\prime}}^{G}}\left(A_{ik}+A_{il}+A_{jk}+A_{jl}\right)
=o(1)
\end{align*}

\bigskip
\noindent
\textbf{Proof of Proposition \ref{lem:OWFE}:}
Define the (infeasible) terms in the following way:
\begin{align*}
\alpha_{gM} & :=\frac{1}{M_{g}^{G}}\sum_{i\in\mathcal{N}_{g}^{G}}y_{iM}(0)\\
e_{iM}(0) & :=y_{iM}(0)-\alpha_{g(i)M}\\
e_{iM}(1) & :=y_{iM}(1)-\alpha_{g(i)M}-\tau_{g(i)M}
\tag{\stepcounter{equation}\theequation}
\end{align*}

Since $Y_{iM}:=X_{iM}y_{iM}(1)+\left(1-X_{iM}\right)y_{iM}(0)$,
\begin{align*}
Y_{iM} & =e_{iM}(1)X_{iM}+e_{iM}(0)\left(1-X_{iM}\right)+\alpha_{g(i)M}+\tau_{g(i)M}X_{iM}
\tag{\stepcounter{equation}\theequation}
\end{align*}

Substituting this expression into $\hat{\tau}_{FE}$,
\begin{align*}
\hat{\tau}_{OWFE} & =\frac{\sum_{i=1}^M \left(e_{iM}(1)X_{iM}+e_{iM}(0)\left(1-X_{iM}\right)+\alpha_{g(i)M}+\tau_{g(i)M}X_{iM}\right)\left(X_{iM}-\bar{X}_{g(i)M}\right)}{\sum_{i=1}^M X_{iM}\left(X_{iM}-\bar{X}_{g(i)M}\right)}\\
 & =\frac{\sum_{i=1}^M \left(\left(e_{iM}(1)+\tau_{g(i)M}\right)X_{iM}+e_{iM}(0)\left(1-X_{iM}\right)\right)\left(X_{iM}-\bar{X}_{g(i)M}\right)}{\sum_{i=1}^M X_{iM}\left(X_{iM}-\bar{X}_{g(i)M}\right)}
 \tag{\stepcounter{equation}\theequation}
\end{align*}
by noticing that:
\begin{align*}
\sum_{i=1}^M \alpha_{g(i)M}\left(X_{iM}-\bar{X}_{g(i)M}\right) & =\sum_{g=1}^G\sum_{i\in\mathcal{N}_{g}^{G}} \alpha_{g(i)M}\left(X_{iM}-\frac{1}{\bar{N}_{gM}}\sum_{i\in\mathcal{N}_{g}^{G}} X_{iM}\right)\\
 & =\sum_{g=1}^G\alpha_{gM}\left(\sum_{i\in\mathcal{N}_{g}^{G}} X_{iM}-\sum_{j\in\mathcal{N}_{g}^{G}}R_{jM}X_{jM}\right)=0.
 \tag{\stepcounter{equation}\theequation}
\end{align*}

Convergence of $\hat{\tau}_{OWFE}$ is immediate from applying Lemma \ref{lem:OWclusLLN}.
Every cluster has independent assignment probability of $A_{gM}$.
To evaluate the denominator, observe that, due to Lemma \ref{lem:mwclusW}:
\begin{align*}
\E&\left[X_{iM}\bar{X}_{g(i)M}\right]	=\E\left[X_{iM}\frac{1}{M_{gM}^{G}}\left(\sum_{j\in\mathcal{N}_{g(i)}^{G}}X_{jM}\right)\right] \\
	&=\frac{1}{M_{g}^{G}}\sum_{i\in\mathcal{N}_{g(i)}^{G}}\E\left[X_{iM}X_{jM}\right]\\
	&=\frac{1}{M_{g}^{G}}\E\left[X_{iM}^{2}\right]+\frac{1}{M_{g}^{G}}\sum_{i\in\mathcal{N}_{(g(i),h(i))}^{G\cap H}\backslash\{i\}}\E\left[X_{iM}X_{jM}\right]+\frac{1}{M_{g}^{G}}\sum_{i\in\mathcal{N}_{g(i)}^{G}\backslash\mathcal{N}_{(g(i),h(i))}^{G\cap H}} \E\left[X_{iM}X_{jM}\right]\\
	&=\frac{1}{M_{g}^{G}}\mu_{A}\mu_{B}+\frac{1}{M_{g}^{G}}\sum_{i\in\mathcal{N}_{(g(i),h(i))}^{G\cap H}\backslash\{i\}}\left(\mu_{A}^{2}+\sigma_{A}^{2}\right)\left(\mu_{B}^{2}+\sigma_{B}^{2}\right)+\frac{1}{M_{g}^{G}}\sum_{i\in\mathcal{N}_{g(i)}^{G}\backslash\mathcal{N}_{(g(i),h(i))}^{G\cap H}}\left(\mu_{A}^{2}+\sigma_{A}^{2}\right)\mu_{B}^{2}\\
	&=\frac{1}{M_{g(i)}^{G}}\mu_{A}\mu_{B}+\frac{M_{(g(i),h(i))}^{G\cap H}-1}{M_{g(i)}^{G}}\left(\mu_{A}^{2}+\sigma_{A}^{2}\right)\left(\mu_{B}^{2}+\sigma_{B}^{2}\right)+\frac{M_{g(i)}^{G}-M_{(g(i),h(i))}^{G\cap H}}{M_{g(i)}^{G}}\left(\mu_{A}^{2}+\sigma_{A}^{2}\right)\mu_{B}^{2}\\
	&=\frac{1}{M_{g(i)}^{G}}\left(\mu_{A}\mu_{B}+\left(M_{(g(i),h(i))}^{G\cap H}-1\right)\left(\mu_{A}^{2}+\sigma_{A}^{2}\right)\left(\mu_{B}^{2}+\sigma_{B}^{2}\right)+\left(M_{g(i)}^{G}-M_{(g(i),h(i))}^{G\cap H}\right)\left(\mu_{A}^{2}+\sigma_{A}^{2}\right)\mu_{B}^{2}\right).
\end{align*}

By imposing $A_{g},B_{h}\in\{0,1\}$ so that $\mu_{A}^{2}+\sigma_{A}^{2}=\mu_{A}$ as $\sigma_{A}^{2}=\mu_{A}\left(1-\mu_{A}\right)$, 
\begin{align*}
\E&\left[X_{iM}\left(X_{iM}-\bar{X}_{g(i)M}\right)\right]	=\mu_{A}\mu_{B}-\frac{1}{M_{g(i)}^{G}}\left(\mu_{A}\mu_{B}+\left(M_{(g(i),h(i))}^{G\cap H}-1\right)\mu_{A}\mu_{B}+\left(M_{g(i)}^{G}-M_{(g(i),h(i))}^{G\cap H}\right)\mu_{A}\mu_{B}^{2}\right) \\
	&=\mu_{A}\mu_{B}-\frac{\mu_{A}\mu_{B}}{M_{g(i)}^{G}}\left(M_{(g(i),h(i))}^{G\cap H}+\left(M_{g(i)}^{G}-M_{(g(i),h(i))}^{G\cap H}\right)\mu_{B}\right)\\
	&=\mu_{A}\mu_{B}\left(1-\frac{M_{(g(i),h(i))}^{G\cap H}+\left(M_{g(i)}^{G}-M_{(g(i),h(i))}^{G\cap H}\right)\mu_{B}}{M_{g(i)}^{G}}\right).
\end{align*}

Next, we proceed to the numerator:
\begin{align*}
\sum_{i=1}^M &  \left(\left(e_{iM}(1)+\tau_{g(i)M}\right)X_{iM}+e_{iM}(0)\left(1-X_{iM}\right)\right)\left(X_{iM}-\bar{X}_{g(i)M}\right)\\
 & =\sum_{i=1}^M \tau_{g(i)M}X_{iM}\left(X_{iM}-\bar{X}_{g(i)M}\right)+\sum_{i=1}^M \left(e_{iM}(1)-e_{iM}(0)\right)X_{iM}\left(X_{iM}-\bar{X}_{g(i)M}\right)\\
 & \quad+\sum_{i=1}^M e_{iM}(0)\left(X_{iM}-\bar{X}_{g(i)M}\right).
 \tag{\stepcounter{equation}\theequation}
\end{align*}

Taking expectations of the final term, $\sum_{i=1}^M\E\left[ e_{iM}(0)\left(X_{iM}-\bar{X}_{g(i)M}\right)\right]  =0 $ is immediate. 
Using previous results on the first expectation,
\begin{align*}
\sum_{i=1}^M & (\tau_{g(i)M}+ e_{iM}(1) - e_{iM}(0) )\E\left[X_{iM}\left(X_{iM}-\bar{X}_{g(i)M}\right)\right]\\
 & = \sum_{i=1}^M  (y_{iM}(1) - y_{iM}(0) )\E\left[X_{iM}\left(X_{iM}-\bar{X}_{g(i)M}\right)\right].
 \tag{\stepcounter{equation}\theequation}
\end{align*}

\noindent

The following lemma is used to derive the two-way fixed effects estimand.
\begin{lemma} \label{lem:TWFE2}
\footnotesize
\begin{align*}
\sum_{i=1}^M&\E\left[\tilde{X}_{iM}Y_{iM}\right] =\mu_{A}\mu_{B}\left(\sum_{c}M_{c}^{G\cap H}\tau_{cM}-\sum_{g=1}^G\sum_{c\in\mathcal{M}_{g}^{G}}\tau_{cM}\frac{\left(M_{c}^{G\cap H}\right)^{2}}{M_{g}^{G}}-\sum_{h=1}^H\sum_{c\in\mathcal{M}_{h}^{H}}\tau_{cM}\frac{\left(M_{c}^{G\cap H}\right)^{2}}{M_{h}^{H}}+\frac{1}{M}\sum_{c}\tau_{cM}\left(M_{c}^{G\cap H}\right)^{2}\right)\\
 & \quad+\frac{1}{M}\mu_{A}^{2}\mu_{B}^{2}\left(M\sum_{c}M_{c}^{G\cap H}\tau_{cM}-\sum_{g=1}^GM_{g}^{G}\sum_{c\in\mathcal{M}_{g}^{G}}M_{c}^{G\cap H}\tau_{cM}-\sum_{h=1}^HM_{h}^{H}\sum_{c\in\mathcal{M}_{h}^{H}}M_{c}^{G\cap H}\tau_{cM}+\sum_{c}\left(M_{c}^{G\cap H}\right)^{2}\tau_{cM}\right)\\
 & \quad+\mu_{A}\mu_{B}^{2}\left(\frac{1}{M}\sum_{g=1}^G\sum_{c\in\mathcal{M}_{g}^{G}}\tau_{cM}\left(M_{g}^{G}M_{c}^{G\cap H}-\left(M_{c}^{G\cap H}\right)^{2}\right)-\sum_{g=1}^G\sum_{c\in\mathcal{M}_{g}^{G}}\tau_{cM}\left(M_{c}^{G\cap H}-\frac{\left(M_{c}^{G\cap H}\right)^{2}}{M_{g}^{G}}\right)\right)\\
 & \quad+\mu_{A}^{2}\mu_{B}\left(\frac{1}{M}\sum_{h=1}^H\sum_{c\in\mathcal{M}_{g}^{G}}\tau_{cM}\left(M_{h}^{H}M_{c}^{G\cap H}-\left(M_{c}^{G\cap H}\right)^{2}\right)-\sum_{h=1}^H\sum_{c\in\mathcal{M}_{h}^{H}}\tau_{cM}\left(M_{c}^{G\cap H}-\frac{\left(M_{c}^{G\cap H}\right)^{2}}{M_{h}^{H}}\right)\right)
 \tag{\stepcounter{equation}\theequation}
\end{align*}
\normalsize
\begin{align*}
\sum_{i=1}^M\E\left[\tilde{X}_{iM}X_{iM}\right] & =\mu_{A}\mu_{B}\left(M-\sum_{g=1}^G\sum_{c\in\mathcal{M}_{g}^{G}}\frac{\left(M_{c}^{G\cap H}\right)^{2}}{M_{g}^{G}}-\sum_{h=1}^H\sum_{c\in\mathcal{M}_{h}^{H}}\frac{\left(M_{c}^{G\cap H}\right)^{2}}{M_{h}^{H}}+\frac{1}{M}\sum_{c}\left(M_{c}^{G\cap H}\right)^{2}\right)\\
 & \quad-\mu_{A}\mu_{B}^{2}\left(M-\sum_{g=1}^G\sum_{c\in\mathcal{M}_{g}^{G}}\frac{\left(M_{c}^{G\cap H}\right)^{2}}{M_{g}^{G}}-\frac{1}{M}\left(\sum_{g=1}^G\left(M_{g}^{G}\right)^{2}-\sum_{c}\left(M_{c}^{G\cap H}\right)^{2}\right)\right)\\
 & \quad-\mu_{A}^{2}\mu_{B}\left(M-\sum_{h=1}^H\sum_{c\in\mathcal{M}_{h}^{H}}\frac{\left(M_{c}^{G\cap H}\right)^{2}}{M_{h}^{H}}-\frac{1}{M}\left(\sum_{h=1}^H\left(M_{h}^{H}\right)^{2}-\sum_{c}\left(M_{c}^{G\cap H}\right)^{2}\right)\right)\\
 & \quad+\mu_{A}^{2}\mu_{B}^{2}\left(M-\frac{1}{M}\sum_{g=1}^G\left(M_{g}^{G}\right)^{2}-\frac{1}{M}\sum_{h=1}^H\left(M_{h}^{H}\right)^{2}+\frac{1}{M}\sum_{c}\left(M_{c}^{G\cap H}\right)^{2}\right)
 \tag{\stepcounter{equation}\theequation}
\end{align*}

Convergence occurs by applying Lemma \ref{lem:OWclusLLN} to the numerator and denominator of $\hat{\tau}_{FE}$ separately. 
\end{lemma}

\noindent
\textbf{Proof:}
Let $c:=(g,h)$ denote the intersection of clusters $g$ and $h$.
Previously, we defined the residuals $e_{iM}(d)$ with respect to the cluster. Now, we define it with respect to the cluster intersection $c$. 
To further ease notation, we use $c\in\mathcal{M}_{g}^{G}$ to denote that
intersection $c$ has the $G$ index of $g$. Then, $\sum_{c\in\mathcal{M}_{g}^{G}}M_{c}^{G\cap H}=M_{g}^{G}$. 
As before, we have:
\begin{align*}
\alpha_{cM} & :=\frac{1}{M_{c}^{G\cap H}}\sum_{i\in\mathcal{N}_{c}^{G\cap H}}y_{iM}(0)\\
\tau_{cM} & :=\frac{1}{M_{c}^{G\cap H}}\sum_{i\in\mathcal{N}_{c}^{G\cap H}}\left(y_{iM}(1)-y_{iM}(0)\right)\\
e_{iM}(0) & :=y_{iM}(0)-\alpha_{c(i)M}\\
e_{iM}(1) & :=y_{iM}(1)-\alpha_{c(i)M}-\tau_{c(i)M}
\end{align*}
\begin{align*}
Y_{iM} & =e_{iM}(1)X_{iM}+e_{iM}(0)\left(1-X_{iM}\right)+\alpha_{g(i)M}+\tau_{g(i)M}X_{iM}
\tag{\stepcounter{equation}\theequation}
\end{align*}

The estimator in TWFE is slightly different, because the residualization is different. 
First consider the expectation of the denominator $\sum_{i=1}^M\E\left[\tilde{X}_{iM}X_{iM}\right]$.
The assignment mechanism is where $X_{iM}=A_{g(i)M}B_{h(i)M}$, where
$A,B$ are independent with means $\mu_{A},\mu_{B}$ respectively.
This means that $\E\left[X_{iM}\right]=\mu_{A}\mu_{B}$.
\begin{align*}
\sum_{i=1}^M\E\left[\tilde{X}_{iM}X_{iM}\right] & =\sum_{i=1}^M\E\left[X_{iM}\left(X_{iM}-\frac{1}{M_{g}^{G}}\sum_{j\in\mathcal{N}_{g}^{G}}X_{jM}-\frac{1}{M_{h}^{H}}\sum_{j\in\mathcal{N}_{h}}X_{jM}+\frac{1}{M}\sum_{i=1}^MX_{iM}\right)\right]\\
 & =\sum_{i=1}^M\E\left[X_{iM}-\frac{1}{M_{g}^{G}}\sum_{j\in\mathcal{N}_{g}^{G}}X_{iM}X_{jM}-\frac{1}{M_{h}^{H}}\sum_{j\in\mathcal{N}_{h}}X_{iM}X_{jM}+\frac{1}{M}\sum_{j}X_{iM}X_{jM}\right]
 \tag{\stepcounter{equation}\theequation}
\end{align*}

For a given $i$, let $g=g(i)$:
\begin{align*}
\sum_{j\in\mathcal{N}_{g}^{G}}\E\left[X_{iM}X_{jM}\right] & =\sum_{j\in\mathcal{N}_{g}^{G}}\E\left[A_{gM}\right]\E\left[B_{h(i)M}B_{h(j)M}\right]\\
 & =\mu_{A}\left(\sum_{j\in\mathcal{N}_{c(i)}^{G\cap H}}\E\left[B_{h(i)M}B_{h(j)M}\right]+\sum_{j\in\mathcal{N}_{g}^{G}\backslash\mathcal{N}_{c(i)}^{G\cap H}}\E\left[B_{h(i)M}B_{h(j)M}\right]\right)\\
 & =\mu_{A}\left(\sum_{j\in\mathcal{N}_{c(i)}^{G\cap H}}\E\left[B_{h(i)M}\right]+\sum_{j\in\mathcal{N}_{g}^{G}\backslash\mathcal{N}_{c(i)}^{G\cap H}}\E\left[B_{h(i)M}\right]\E\left[B_{h(j)M}\right]\right)\\
 & =M_{c(i)}^{G\cap H}\mu_{A}\mu_{B}+\left(M_{g}^{G}-M_{c(i)}^{G\cap H}\right)\mu_{A}\mu_{B}^{2}
 \tag{\stepcounter{equation}\theequation}
\end{align*}

Using this in the larger sum,
\begin{align*}
\sum_{i=1}^M &\frac{1}{M_{g}^{G}}\sum_{j\in\mathcal{N}_{g}}\E\left[X_{iM}X_{jM}\right]  =\sum_{g=1}^G\sum_{i\in\mathcal{N}_{g}^{G}}\frac{1}{M_{g}^{G}}\left(M_{c(i)}^{G\cap H}\mu_{A}\mu_{B}+\left(M_{g}^{G}-M_{c(i)}^{G\cap H}\right)\mu_{A}\mu_{B}^{2}\right)\\
 & =\sum_{g=1}^G\sum_{i\in\mathcal{N}_{g}^{G}}\frac{M_{c(i)}^{G\cap H}}{M_{g}^{G}}\mu_{A}\mu_{B}+\sum_{g=1}^G\sum_{i\in\mathcal{N}_{g}^{G}}\left(1-\frac{M_{c(i)}^{G\cap H}}{M_{g}^{G}}\right)\mu_{A}\mu_{B}^{2}\\
 & =\sum_{g=1}^G\sum_{c\in\mathcal{M}_{g}^{G}}\frac{\left(M_{c}^{G\cap H}\right)^{2}}{M_{g}^{G}}\mu_{A}\mu_{B}+\sum_{g=1}^G\sum_{c\in\mathcal{M}_{g}^{G}}M_{c}^{G\cap H}\mu_{A}\mu_{B}^{2}-\sum_{g=1}^G\sum_{c\in\mathcal{M}_{g}^{G}}\frac{\left(M_{c}^{G\cap H}\right)^{2}}{M_{g}^{G}}\mu_{A}\mu_{B}^{2}\\
 & =\sum_{g=1}^G\sum_{c\in\mathcal{M}_{g}^{G}}\frac{\left(M_{c}^{G\cap H}\right)^{2}}{M_{g}^{G}}\mu_{A}\mu_{B}\left(1-\mu_{B}\right)+M\mu_{A}\mu_{B}^{2}\\
 & =\mu_{A}\mu_{B}^{2}\left(M-\sum_{g=1}^G\sum_{c\in\mathcal{M}_{g}^{G}}\frac{\left(M_{c}^{G\cap H}\right)^{2}}{M_{g}^{G}}\right)+\mu_{A}\mu_{B}\left(\sum_{g=1}^G\sum_{c\in\mathcal{M}_{g}^{G}}\frac{\left(M_{c}^{G\cap H}\right)^{2}}{M_{g}^{G}}\right)
 \tag{\stepcounter{equation}\theequation}
\end{align*}

Similarly,
\begin{equation}
\sum_{i=1}^M\frac{1}{M_{h}^{H}}\sum_{j\in\mathcal{N}_{h}^{H}}\E\left[X_{iM}X_{jM}\right]=\sum_{h=1}^H\sum_{c\in\mathcal{M}_{c}^{H}}\frac{\left(M_{c}^{G\cap H}\right)^{2}}{M_{h}^{H}}\mu_{A}\mu_{B}\left(1-\mu_{A}\right)+M\mu_{A}^{2}\mu_{B}
\end{equation}

Finally,
\begin{align*}
\sum_{j}\E\left[X_{iM}X_{jM}\right] & =\sum_{j\in\mathcal{N}_{g}^{G}}\E\left[X_{iM}X_{jM}\right]+\sum_{j\in\mathcal{N}_{h}^{H}}\E\left[X_{iM}X_{jM}\right] \\
&\quad -\sum_{j\in\mathcal{N}_{c}^{G\cap H}}\E\left[X_{iM}X_{jM}\right]+\sum_{j\notin\mathcal{N}_{g}^{G}\cup\mathcal{N}_{h}^{H}}\E\left[X_{iM}X_{jM}\right]
\tag{\stepcounter{equation}\theequation}
\end{align*}

Observe that:

\begin{align*}
\sum_{j\in\mathcal{N}_{c(i)}^{G\cap H}}\E\left[X_{iM}X_{jM}\right] & =M_{c(i)}^{G\cap H}\mu_{A}\mu_{B}\\
\sum_{i=1}^M\sum_{j\in\mathcal{N}_{c(i)}^{G\cap H}}\E\left[X_{iM}X_{jM}\right] & =\sum_{c}M_{c}^{G\cap H}\mu_{A}\mu_{B}
\tag{\stepcounter{equation}\theequation}
\end{align*}

Since $\sum_{j\notin\mathcal{N}_{g}^{G}\cup\mathcal{N}_{h}^{H}}\E\left[X_{iM}X_{jM}\right]=\left(M-M_{g(i)}^{G}-M_{h(i)}^{H}+M_{c(i)}^{G\cap H}\right)\mu_{A}^{2}\mu_{B}^{2}$,
\begin{align*}
\sum_{i=1}^M\sum_{j\notin\mathcal{N}_{g}^{G}\cup\mathcal{N}_{h}^{H}}\E\left[X_{iM}X_{jM}\right] & =\mu_{A}^{2}\mu_{B}^{2}\sum_{i=1}^M\left(M-M_{g(i)}^{G}-M_{h(i)}^{H}+M_{c(i)}^{G\cap H}\right)\\
 & =\mu_{A}^{2}\mu_{B}^{2}\left(M^{2}-\sum_{g=1}^G\left(M_{g}^{G}\right)^{2}-\sum_{h=1}^H\left(M_{h}^{H}\right)^{2}+\sum_{c}\left(M_{c}^{G\cap H}\right)^{2}\right)
 \tag{\stepcounter{equation}\theequation}
\end{align*}

Evaluate the first few terms:
\begin{align*}
\sum_{i=1}^M\sum_{j\in\mathcal{N}_{g(i)}^{G}}\E\left[X_{iM}X_{jM}\right] & =\sum_{g=1}^G\sum_{i\in\mathcal{N}_{g}^{G}}\left(M_{c(i)}^{G\cap H}\mu_{A}\mu_{B}+\left(n_{g}-M_{c(i)}^{G\cap H}\right)\mu_{A}\mu_{B}^{2}\right)\\
 & =\sum_{g=1}^G\sum_{c\in\mathcal{M}_{g}^{G}}\left(M_{c}^{G\cap H}\right)^{2}\mu_{A}\mu_{B}+\sum_{g=1}^G\sum_{c\in\mathcal{M}_{g}^{G}}M_{c}^{G\cap H}\left(M_{g}^{G}-M_{c}^{G\cap H}\right)\mu_{A}\mu_{B}^{2}\\
 & =\sum_{g=1}^G\sum_{c\in\mathcal{M}_{g}^{G}}\left(M_{c}^{G\cap H}\right)^{2}\mu_{A}\mu_{B}\left(1-\mu_{B}\right)+\sum_{g=1}^G\left(M_{g}^{G}\right)^{2}\mu_{A}\mu_{B}^{2}\\
 & =\mu_{A}\mu_{B}^{2}\left(\sum_{g=1}^G\left(M_{g}^{G}\right)^{2}-\sum_{g=1}^G\sum_{c\in\mathcal{M}_{g}^{G}}\left(M_{c}^{G\cap H}\right)^{2}\right)+\mu_{A}\mu_{B}\left(\sum_{g=1}^G\sum_{c\in\mathcal{M}_{g}^{G}}\left(M_{c}^{G\cap H}\right)^{2}\right)\\
 & =\mu_{A}\mu_{B}^{2}\left(\sum_{g=1}^G\left(M_{g}^{G}\right)^{2}-\sum_{c}\left(M_{c}^{G\cap H}\right)^{2}\right)+\mu_{A}\mu_{B}\left(\sum_{c}\left(M_{c}^{G\cap H}\right)^{2}\right)
 \tag{\stepcounter{equation}\theequation}
\end{align*}

Hence,
\begin{align*}
\sum_{i=1}^M &\frac{1}{n}\sum_{j}\E\left[X_{iM}X_{jM}\right]  =\frac{1}{M}\left(\mu_{A}\mu_{B}^{2}\left(\sum_{g=1}^G\left(M_{g}^{G}\right)^{2}-\sum_{c}\left(M_{c}^{G\cap H}\right)^{2}\right)+\mu_{A}\mu_{B}\left(\sum_{c}\left(M_{c}^{G\cap H}\right)^{2}\right)\right)\\
 & \quad+\frac{1}{M}\left(\mu_{A}^{2}\mu_{B}\left(\sum_{h=1}^H\left(M_{h}^{H}\right)^{2}-\sum_{c}\left(M_{c}^{G\cap H}\right)^{2}\right)+\mu_{A}\mu_{B}\left(\sum_{c}\left(M_{c}^{G\cap H}\right)^{2}\right)\right)\\
 & \quad-\frac{1}{M}\left(\sum_{c}\left(M_{c}^{G\cap H}\right)^{2}\mu_{A}\mu_{B}\right)+\frac{1}{M}\mu_{A}^{2}\mu_{B}^{2}\left(n^{2}-\sum_{g=1}^G\left(M_{g}^{G}\right)^{2}-\sum_{h=1}^H\left(M_{h}^{H}\right)^{2}+\sum_{c}\left(M_{c}^{G\cap H}\right)^{2}\right)
 \tag{\stepcounter{equation}\theequation}
\end{align*}

Putting these results together,
\begin{align*}
\sum_{i=1}^M &\E\left[\tilde{X}_{iM}X_{iM}\right] =M\mu_{A}\mu_{B}-\mu_{A}\mu_{B}^{2}\left(M-\sum_{g=1}^G\sum_{c\in\mathcal{M}_{g}^{G}}\frac{\left(M_{c}^{G\cap H}\right)^{2}}{M_{g}^{G}}\right)-\mu_{A}\mu_{B}\left(\sum_{g=1}^G\sum_{c\in\mathcal{M}_{g}^{G}}\frac{\left(M_{c}^{G\cap H}\right)^{2}}{M_{g}^{G}}\right)\\
 & \quad-\mu_{A}^{2}\mu_{B}\left(M-\sum_{h=1}^H\sum_{c\in\mathcal{M}_{h}^{H}}\frac{\left(M_{c}^{G\cap H}\right)^{2}}{M_{h}^{H}}\right)-\mu_{A}\mu_{B}\left(\sum_{h=1}^H\sum_{c\in\mathcal{M}_{h}^{H}}\frac{\left(M_{c}^{G\cap H}\right)^{2}}{M_{h}^{H}}\right)\\
 & \quad+\frac{1}{M}\left(\mu_{A}\mu_{B}^{2}\left(\sum_{g=1}^G\left(M_{g}^{G}\right)^{2}-\sum_{c}\left(M_{c}^{G\cap H}\right)^{2}\right)+\mu_{A}\mu_{B}\left(\sum_{c}\left(M_{c}^{G\cap H}\right)^{2}\right)\right)\\
 & \quad+\frac{1}{M}\left(\mu_{A}^{2}\mu_{B}\left(\sum_{h=1}^H\left(M_{h}^{H}\right)^{2}-\sum_{c}\left(M_{c}^{G\cap H}\right)^{2}\right)+\mu_{A}\mu_{B}\left(\sum_{c}\left(M_{c}^{G\cap H}\right)^{2}\right)\right)\\
 & \quad-\frac{1}{M}\left(\sum_{c}\left(M_{c}^{G\cap H}\right)^{2}\mu_{A}\mu_{B}\right)+\frac{1}{M}\mu_{A}^{2}\mu_{B}^{2}\left(n^{2}-\sum_{g=1}^G\left(M_{g}^{G}\right)^{2}-\sum_{h=1}^H\left(M_{h}^{H}\right)^{2}+\sum_{c}\left(M_{c}^{G\cap H}\right)^{2}\right),
 \tag{\stepcounter{equation}\theequation}
\end{align*}
which simplifies to the expression in the lemma.

Proceeding with the numerator, 
\begin{align*}
\sum_{i=1}^M\E\left[\tilde{X}_{iM}Y_{iM}\right] & =\sum_{i=1}^M\E\left[Y_{iM}X_{iM}-\frac{1}{M_{g}^{G}}\sum_{j\in\mathcal{N}_{g}^{G}}Y_{iM}X_{jM}-\frac{1}{M_{h}^{H}}\sum_{j\in\mathcal{N}_{h}^{H}}Y_{iM}X_{jM}+\frac{1}{M}\sum_{j}Y_{iM}X_{jM}\right]
\tag{\stepcounter{equation}\theequation}
\end{align*}

Let's look at the first term, using $Y_{iM}=e_{iM}(1)X_{iM}+e_{iM}(0)\left(1-X_{iM}\right)+\alpha_{g(i)M}+\tau_{g(i)M}X_{iM}$:
\begin{align*}
\sum_{i=1}^M\E\left[Y_{iM}X_{iM}\right] & =\sum_{i=1}^M\left(e_{iM}(1)+\tau_{c(i)M}+\alpha_{c(i)M}\right)\mu_{A}\mu_{B}\\
 & =\sum_{c}M_{c}^{G\cap H}\left(\tau_{cM}+\alpha_{cM}\right)\mu_{A}\mu_{B}.
 \tag{\stepcounter{equation}\theequation}
\end{align*}

For a given $i$, let $g=g(i)$:
\begin{align*}
\sum_{j\in\mathcal{N}_{g}^{G}}&\E\left[Y_{iM}X_{jM}\right] =\sum_{j\in\mathcal{N}_{g}^{G}}\E\left[\left(\left(e_{iM}(1)+\tau_{c(i)M}\right)X_{iM}+e_{iM}(0)\left(1-X_{iM}\right)+\alpha_{c(i)M}\right)X_{jM}\right]\\
 & =\sum_{j\in\mathcal{N}_{g}^{G}}\left(\left(e_{iM}(1)+\tau_{c(i)M}\right)\E\left[X_{iM}X_{jM}\right]+\left(e_{iM}(0)+\alpha_{c(i)M}\right)\mu_{A}\mu_{B}-e_{iM}(0)\E\left[X_{iM}X_{jM}\right]\right)\\
 & =\left(e_{iM}(1)+\tau_{c(i)M}-e_{iM}(0)\right)\sum_{j\in\mathcal{N}_{g}^{G}}\E\left[X_{iM}X_{jM}\right]+M_{g}^{G}\left(e_{iM}(0)+\alpha_{c(i)M}\right)\mu_{A}\mu_{B},
 \tag{\stepcounter{equation}\theequation}
\end{align*}
where $\sum_{j\in\mathcal{N}_{g}^{G}}\E\left[X_{iM}X_{jM}\right]=M_{c(i)}^{G\cap H}\mu_{A}\mu_{B}+\left(M_{g}^{G}-M_{c(i)}^{G\cap H}\right)\mu_{A}\mu_{B}^{2}$
was derived from before. 
Using the notation $\bar{e}_{cM}(w)=\frac{1}{M_{c}^{G\cap H}}\sum_{i\in\mathcal{N}_{c}^{G\cap H}}e_{iM}(w)$ in the larger sum, 
\begin{align*}
\sum_{i=1}^M&\frac{1}{M_{g}^{G}}\sum_{j\in\mathcal{N}_{g}^{G}}\E\left[Y_{iM}X_{jM}\right] \\
&=\sum_{g=1}^G\sum_{i\in\mathcal{N}_{g}^{G}}\frac{1}{M_{g}^{G}}\bigg(\left(e_{iM}(1)-e_{iM}(0)+\tau_{c(i)M}\right)\left(M_{c(i)}^{G\cap H}\mu_{A}\mu_{B}+\left(M_{g}^{G}-M_{c(i)}^{G\cap H}\right)\mu_{A}\mu_{B}^{2}\right)\\
&\quad+M_{g}^{G}\left(e_{iM}(0)+\alpha_{c(i)M}\right)\mu_{A}\mu_{B}\bigg)\\
 & =\sum_{g=1}^G\sum_{c\in\mathcal{M}_{g}^{G}}\left(\bar{e}_{cM}(1)-\bar{e}_{cM}(0)+\tau_{cM}\right)\left(\frac{\left(M_{c}^{G\cap H}\right)^{2}}{M_{g}^{G}}\mu_{A}\mu_{B}+M_{c}^{G\cap H}\mu_{A}\mu_{B}^{2}-\frac{\left(M_{c}^{G\cap H}\right)^{2}}{M_{g}^{G}}\mu_{A}\mu_{B}^{2}\right) \\
 &\quad +\sum_{g=1}^G\sum_{c\in\mathcal{M}_{g}^{G}}M_{c}^{G\cap H}\alpha_{cM}\mu_{A}\mu_{B}
 \tag{\stepcounter{equation}\theequation}
\end{align*}

Similarly,
\begin{align*}
\sum_{i=1}^M &\frac{1}{M_{h}^{H}}\sum_{j\in\mathcal{N}_{h}^{H}}\E\left[Y_{iM}X_{jM}\right]\\
&=\sum_{h=1}^H\sum_{c\in\mathcal{M}_{h}^{H}}\left(\bar{e}_{cM}(1)-\bar{e}_{cM}(0)+\tau_{cM}\right)\left(\frac{\left(M_{c}^{G\cap H}\right)^{2}}{M_{h}^{H}}\mu_{A}\mu_{B}+M_{c}^{G\cap H}\mu_{A}\mu_{B}^{2}-\frac{\left(M_{c}^{G\cap H}\right)^{2}}{M_{h}^{H}}\mu_{A}\mu_{B}^{2}\right) \\
&\quad+\sum_{h=1}^H\sum_{c\in\mathcal{M}_{h}^{H}}M_{c}^{G\cap H}\alpha_{cM}\mu_{A}\mu_{B}
\tag{\stepcounter{equation}\theequation}
\end{align*}

Finally,

\begin{align*}
\sum_{j}\E\left[Y_{iM}X_{jM}\right] & =\sum_{j\in\mathcal{N}_{g}^{G}}\E\left[Y_{iM}X_{jM}\right]+\sum_{j\in\mathcal{N}_{h}^{H}}\E\left[Y_{iM}X_{jM}\right]-\sum_{j\in\mathcal{N}_{c}^{G\cap H}}\E\left[Y_{iM}X_{jM}\right]+\sum_{j\notin\mathcal{N}_{g}^{G}\cup\mathcal{N}_{h}^{H}}\E\left[Y_{iM}X_{jM}\right]
\tag{\stepcounter{equation}\theequation}
\end{align*}

Using $\sum_{j\in\mathcal{N}_{c(i)}^{G\cap H}}\E\left[X_{iM}X_{jM}\right]=M_{c(i)}^{G\cap H}\mu_{A}\mu_{B}$
and $\sum_{i=1}^M\sum_{j\in\mathcal{N}_{c(i)}^{G\cap H}}\E\left[X_{iM}X_{jM}\right]=\sum_{c}\left(M_{c(i)}^{G\cap H}\right)^{2}\mu_{A}\mu_{B}$,
observe that:
\begin{align*}
\sum_{j\in\mathcal{N}_{c(i)}^{G\cap H}}\E\left[Y_{iM}X_{jM}\right] & =\sum_{j\in\mathcal{N}_{c(i)}^{G\cap H}}\E\left[\left(e_{iM}(1)X_{iM}+e_{iM}(0)\left(1-X_{iM}\right)+\alpha_{c(i)M}+\tau_{c(i)M}X_{iM}\right)X_{jM}\right]\\
 & =\sum_{j\in\mathcal{N}_{c(i)}^{G\cap H}}\left(\left(e_{iM}(1)+\tau_{c(i)M}-e_{iM}(0)\right)\E\left[W_{i}W_{j}\right]+\left(e_{iM}(0)+\alpha_{c(i)M}\right)\mu_{A}\mu_{B}\right)\\
 & =\left(\bar{e}_{c(i)M}(1)-\bar{e}_{c(i)M}(0)+\tau_{c(i)M}\right)M_{c(i)}^{G\cap H}\mu_{A}\mu_{B}+M_{c(i)}^{G\cap H}\left(\bar{e}_{c(i)M}(0)+\alpha_{c(i)M}\right)\mu_{A}\mu_{B}\\
 & =\left(\bar{e}_{c(i)M}(1)+\alpha_{c(i)M}+\tau_{c(i)M}\right)M_{c(i)}^{G\cap H}\mu_{A}\mu_{B}
 \tag{\stepcounter{equation}\theequation}
\end{align*}

\begin{align*}
\sum_{i=1}^M\sum_{j\in\mathcal{N}_{c(i)}^{G\cap H}}\E\left[Y_{iM}X_{jM}\right] & =\sum_{c}\left(\bar{e}_{cM}(1)+\alpha_{cM}+\tau_{cM}\right)\left(M_{cM}^{G\cap H}\right)^{2}\mu_{A}\mu_{B}
\tag{\stepcounter{equation}\theequation}
\end{align*}

Let's focus on the last term:
\begin{align*}
\sum_{j\notin\mathcal{N}_{g}^{G}\cup\mathcal{N}_{h}^{H}}\E\left[Y_{iM}X_{jM}\right] & =\left(M-M_{g(i)}^{G}-M_{h(i)}^{H}+M_{c(i)}^{G\cap H}\right)\E\left[Y_{iM}\right]\E\left[X_{jM}\right]\\
 & =\left(M-M_{g(i)}^{G}-M_{h(i)}^{H}+M_{c(i)}^{G\cap H}\right)\E\left[X_{iM}y_{iM}(1)+\left(1-X_{iM}\right)y_{iM}(0)\right]\mu_{A}\mu_{B}\\
 & =\left(M-M_{g(i)}^{G}-M_{h(i)}^{H}+M_{c(i)}^{G\cap H}\right)\left(y_{iM}(0)+\mu_{A}\mu_{B}\left(y_{iM}(1)-y_{iM}(0)\right)\right)\mu_{A}\mu_{B}
 \tag{\stepcounter{equation}\theequation}
\end{align*}

Hence,
\begin{align*}
\sum_{i=1}^M &\sum_{j\notin\mathcal{N}_{g}^{G}\cup\mathcal{N}_{h}^{H}}\E\left[Y_{iM}X_{jM}\right] \\
& =\mu_{A}\mu_{B}\sum_{i=1}^M\left(M-M_{g(i)}^{G}-M_{h(i)}^{H}+M_{c(i)}^{G\cap H}\right)\left(y_{iM}(0)+\mu_{A}\mu_{B}\left(y_{iM}(1)-y_{iM}(0)\right)\right)
\tag{\stepcounter{equation}\theequation}
\end{align*}

To simplify this expression, observe that:
\begin{align*}
\sum_{i=1}^My_{iM}(0) & =\sum_{c}M_{c}^{G\cap H}\alpha_{cM}\\
\sum_{i=1}^MM_{g(i)}^{G}y_{i}(0) & =\sum_{g=1}^GM_{g}^{G}\sum_{c\in\mathcal{M}_{g}^{G}}M_{c}^{G\cap H}\alpha_{cM}\\
\sum_{i=1}^MM_{c(i)}^{G\cap H}y_{i}(0) & =\sum_{c}\left(M_{c}^{G\cap H}\right)^{2}\alpha_{cM}
\tag{\stepcounter{equation}\theequation}
\end{align*}

\begin{align*}
\sum_{i=1}^M&\sum_{j\notin\mathcal{N}_{g}^{G}\cup\mathcal{N}_{h}^{H}}\E\left[Y_{iM}W_{jM}\right] =\mu_{A}\mu_{B}\bigg(M\sum_{c}M_{c}^{G\cap H}\alpha_{cM}\\
&\quad-\sum_{g=1}^GM_{g}^{G}\sum_{c\in\mathcal{M}_{g}^{G}}M_{c}^{G\cap H}\alpha_{cM}-\sum_{h=1}^HM_{h}^{H}\sum_{c\in\mathcal{M}_{h}^{H}}M_{c}^{G\cap H}\alpha_{cM}+\sum_{c}\left(M_{c}^{G\cap H}\right)^{2}\alpha_{cM}\bigg)\\
 & \quad+\mu_{A}^{2}\mu_{B}^{2}\bigg(M\sum_{c}M_{c}^{G\cap H}\tau_{cM}-\sum_{g=1}^GM_{g}^{G}\sum_{c\in\mathcal{M}_{g}^{G}}M_{c}^{G\cap H}\tau_{cM}-\sum_{h=1}^HM_{h}^{H}\sum_{c\in\mathcal{M}_{h}^{H}}M_{c}^{G\cap H}\tau_{cM} \\
 &\quad \quad +\sum_{c}\left(M_{c}^{G\cap H}\right)^{2}\tau_{cM}\bigg)
 \tag{\stepcounter{equation}\theequation}
\end{align*}

\begin{align*}
\sum_{i=1}^M&\sum_{j\in\mathcal{N}_{g}^{G}}\E\left[Y_{iM}X_{jM}\right] =\sum_{g=1}^G\sum_{i\in\mathcal{N}_{g}^{G}}\bigg(\left(e_{iM}(1)-e_{iM}(0)+\tau_{c(i)M}\right)\left(M_{c(i)}^{G\cap H}\mu_{A}\mu_{B}+\left(M_{g}^{G}-M_{c(i)}^{G\cap H}\right)\mu_{A}\mu_{B}^{2}\right)\\
&\quad+M_{g}^{G}\left(e_{iM}(0)+\alpha_{c(i)M}\right)\mu_{A}\mu_{B}\bigg)\\
 & =\sum_{g=1}^G\sum_{c\in\mathcal{M}_{g}^{G}}\left(\bar{e}_{cM}(1)-\bar{e}_{cM}(0)+\tau_{cM}\right)\left(\left(M_{c}^{G\cap H}\right)^{2}\mu_{A}\mu_{B}+M_{g}^{G}M_{c}^{G\cap H}\mu_{A}\mu_{B}^{2}-\left(M_{c}^{G\cap H}\right)^{2}\mu_{A}\mu_{B}^{2}\right)\\
 & \quad+\sum_{g=1}^G\sum_{c\in\mathcal{M}_{g}^{G}}M_{g}^{G}M_{c}^{G\cap H}\left(\alpha_{cM}+\bar{e}_{cM}(0)\right)\mu_{A}\mu_{B}.
 \tag{\stepcounter{equation}\theequation}
\end{align*}

Hence, by using $\bar{e}_{cM}(w)=0$ as before,
\begin{align*}
\sum_{i=1}^M&\frac{1}{M}\sum_{j}\E\left[Y_{iM}X_{jM}\right]  =\frac{1}{M}\mu_{A}\mu_{B}\left(M\sum_{c}M_{c}^{G\cap H}\alpha_{cM}+\sum_{c}\tau_{cM}\left(M_{c}^{G\cap H}\right)^{2}\right)\\
 & \quad+\frac{1}{M}\mu_{A}^{2}\mu_{B}^{2}\bigg(M\sum_{c}M_{c}^{G\cap H}\tau_{cM}-\sum_{g=1}^GM_{g}^{G}\sum_{c\in\mathcal{M}_{g}^{G}}M_{c}^{G\cap H}\tau_{cM} \\
 &\quad \quad -\sum_{h=1}^HM_{h}^{H}\sum_{m\in\mathcal{M}_{h}}M_{c}^{G\cap H}\tau_{cM}+\sum_{c}\left(M_{c}^{G\cap H}\right)^{2}\tau_{cM}\bigg)\\
 & \quad+\frac{1}{M}\mu_{A}\mu_{B}^{2}\sum_{g=1}^G\sum_{c\in\mathcal{M}_{g}^{G}}\tau_{cM}\left(M_{g}^{G}M_{c}^{G\cap H}-\left(M_{c}^{G\cap H}\right)^{2}\right) \\
 & \quad +\frac{1}{M}\mu_{A}^{2}\mu_{B}\sum_{h=1}^H\sum_{m\in\mathcal{M}_{g}}\tau_{cM}\left(M_{h}^{H}M_{c}^{G\cap H}-\left(M_{c}^{G\cap H}\right)^{2}\right).
 \tag{\stepcounter{equation}\theequation}
\end{align*}
Combine the expressions to obtain the result.

\bigskip
\noindent
\textbf{Proof of Proposition \ref{lem:TWFE}:}

Under the hypothesis of the lemma, $M=M_{G}M_{H}k$, so by applying Lemma \ref{lem:TWFE2} and simplifying terms,
\begin{align*}
\sum_{i=1}^M\E\left[\tilde{X}_{iM}X_{iM}\right] & =\mu_{A}\mu_{B}\left(1-\mu_{B}-\mu_{A}+\mu_{A}\mu_{B}\right)\left(M-Gk-Hk+k\right)
 \tag{\stepcounter{equation}\theequation}
\end{align*}
\begin{align*}
\sum_{i=1}^M\E\left[\tilde{X}_{iM}Y_{iM}\right] 
 & =\tau_{M}\mu_{A}\mu_{B}\left(1-\mu_{B}-\mu_{A}+\mu_{A}\mu_{B}\right)\left(M-Gk-Hk+k\right)
 \tag{\stepcounter{equation}\theequation}
\end{align*}

Then, the estimand reduces to:
\begin{align*}
\frac{\sum_{i=1}^M\E\left[\tilde{X}_{iM}Y_{iM}\right]}{\sum_{i=1}^M\E\left[\tilde{X}_{iM}X_{iM}\right]} & =\frac{\tau_{M}\mu_{A}\mu_{B}\left(1-\mu_{B}-\mu_{A}+\mu_{A}\mu_{B}\right)\left(M-Gk-Hk+k\right)}{\mu_{A}\mu_{B}\left(1-\mu_{B}-\mu_{A}+\mu_{A}\mu_{B}\right)\left(M-Gk-Hk+k\right)}=\tau_{M}
\tag{\stepcounter{equation}\theequation}
\end{align*}

Convergence occurs by decomposing (\ref{eqn:TWFE_equation}) with (\ref{eqn:TWFE_transform}) then applying Lemma \ref{lem:OWclusLLN} to the respective terms.

\normalsize
\bibliography{finitepop_cluster}

\end{document}